\documentclass[aps,prl,amsmath,onecolumn,amssymb,floatfix,superscriptaddress,
11pt,notitlepage,longbibliography,nobibnotes]{revtex4-2}
\usepackage{times}
\usepackage{graphicx}
\usepackage{amsmath, braket, amsfonts}
\usepackage{amssymb}
\usepackage{natbib}
\setcitestyle{super}
\usepackage{sidecap}
\usepackage{bm, color, ulem}
\usepackage{xcolor}
\usepackage{lipsum}
\usepackage{placeins}  %

\usepackage[utf8]{inputenc}
\usepackage{graphics}
\graphicspath{ {./Figures/} }
\usepackage{lineno}
\usepackage{comment}

\usepackage{textcomp,gensymb}
\usepackage{prettyref}
\newrefformat{fig}{Fig.~\ref{#1}}
\newrefformat{exfig}{Supplementary Fig.~\ref{#1}}
\newrefformat{extab}{Supplementary Table~\ref{#1}}

\newcommand{\beginsupplement}{%
        \setcounter{table}{0}
        \setcounter{figure}{0}
        \setcounter{page}{1}
        \renewcommand{\figurename}{Supplementary Figure}
        \renewcommand{\tablename}{Supplementary Table}
        \renewcommand{\thepage}{S\arabic{page}}
     }

\newcommand{\TJWatson}{T. J. Watson Laboratory of Applied Physics, California Institute of Technology, Pasadena, California 91125, USA}
\newcommand{\IQIM}{Institute for Quantum Information and Matter, California Institute of Technology, Pasadena, California 91125, USA}
\newcommand{\Physics}{Department of Physics, California Institute of Technology, Pasadena, California 91125, USA}
\begin{document}

\title{Hierarchy of Symmetry Breaking Correlated Phases in Twisted Bilayer Graphene}

\author{Robert Polski}
    \thanks{These authors contributed equally to this work}
    \affiliation{\TJWatson}
    \affiliation{\IQIM}
\author{Yiran Zhang}
    \thanks{These authors contributed equally to this work}
    \affiliation{\TJWatson}
    \affiliation{\IQIM}
    \affiliation{\Physics}
\author{Yang Peng}
    \affiliation{Department of Physics and Astronomy, California State University, Northridge, California 91330, USA}
\author{Harpreet Singh Arora}
    \affiliation{\TJWatson}
    \affiliation{\IQIM}
\author{Youngjoon Choi}
    \affiliation{\TJWatson}
    \affiliation{\IQIM}
    \affiliation{\Physics}
\author{Hyunjin Kim}
    \affiliation{\TJWatson}
    \affiliation{\IQIM}
    \affiliation{\Physics}
\author{Kenji Watanabe}
    \affiliation{National Institute for Materials Science, Namiki 1-1, Tsukuba, Ibaraki 305 0044, Japan}
\author{Takashi Taniguchi}
    \affiliation{National Institute for Materials Science, Namiki 1-1, Tsukuba, Ibaraki 305 0044, Japan}
\author{Gil Refael}
    \affiliation{\IQIM}
    \affiliation{\Physics}
\author{Felix von Oppen}
    \affiliation{Dahlem Center for Complex Quantum Systems and Fachbereich Physik, Freie Universit\"at Berlin, 14195 Berlin, Germany}
\author{Stevan Nadj-Perge}
    \email[Correspondence: ]{s.nadj-perge@caltech.edu}
    \affiliation{\TJWatson}
    \affiliation{\IQIM}

\date{\today}

\maketitle


\noindent{\bf Twisted bilayer graphene (TBG) near the magic twist angle of
$\sim$1.1\degree~exhibits a rich phase diagram. However, the interplay between 
different phases and their dependence on twist angle is still elusive. 
Here, we explore the stability of various TBG phases and demonstrate that 
superconductivity near filling of two electrons per moir\'e unit cell alongside 
Fermi surface reconstructions, as well as entropy-driven high-temperature phase 
transitions and linear-in-T resistance occur over a range of twist angles 
which extends far beyond those exhibiting correlated insulating phases. 
In the vicinity of the magic angle, we also find a 
metallic phase that displays a hysteretic anomalous Hall 
effect and incipient Chern insulating behaviour. Such a metallic phase can be 
rationalized in terms of the interplay between interaction-driven deformations 
of TBG bands leading to Berry curvature redistribution and Fermi surface 
reconstruction. Our results 
provide an extensive perspective on the hierarchy of correlated 
phases in TBG as classified by their robustness against deviations from the 
magic angle or, equivalently, their electronic interaction requirements.}

TBG is a highly tunable platform for exploring the effects of strong electronic
interactions and topological bands\cite{caoCorrelatedInsulatorBehaviour2018,
caoUnconventionalSuperconductivityMagicangle2018,
yankowitzTuningSuperconductivityTwisted2019,
luSuperconductorsOrbitalMagnets2019,
aroraSuperconductivityMetallicTwisted2020,
serlinIntrinsicQuantizedAnomalous2019,
sharpeEmergentFerromagnetismThreequarters2019,
caoStrangeMetalMagicAngle2020, stepanovCompetingZerofieldChern2020}. At the 
magic angle, i.e., when the strength of 
the interactions among electrons is maximized relative to their kinetic energy, 
pronounced signatures of correlated phases
emerge\cite{bistritzerMoireBandsTwisted2011, 
caoCorrelatedInsulatorBehaviour2018, 
caoUnconventionalSuperconductivityMagicangle2018}. 
Away from the magic angle, the effective interaction strength is reduced and 
the correlated phases are believed to disappear rapidly. 
However, despite the strong impact of the twist angle on the phase diagram of 
nearly-magic TBG, this dependence is still experimentally 
under-explored. Here we report systematic measurements on multiple devices 
covering a wide range of twist angles 
between 0.79\degree~ and 1.23\degree~ (see \prettyref{extab:devicetable} for an 
overview) and examine the overall impact of twist angle, 
and thus strength of interactions, on the phase diagram of TBG.

\begin{figure*}[ht]
    \centering
    \includegraphics[width=16cm]{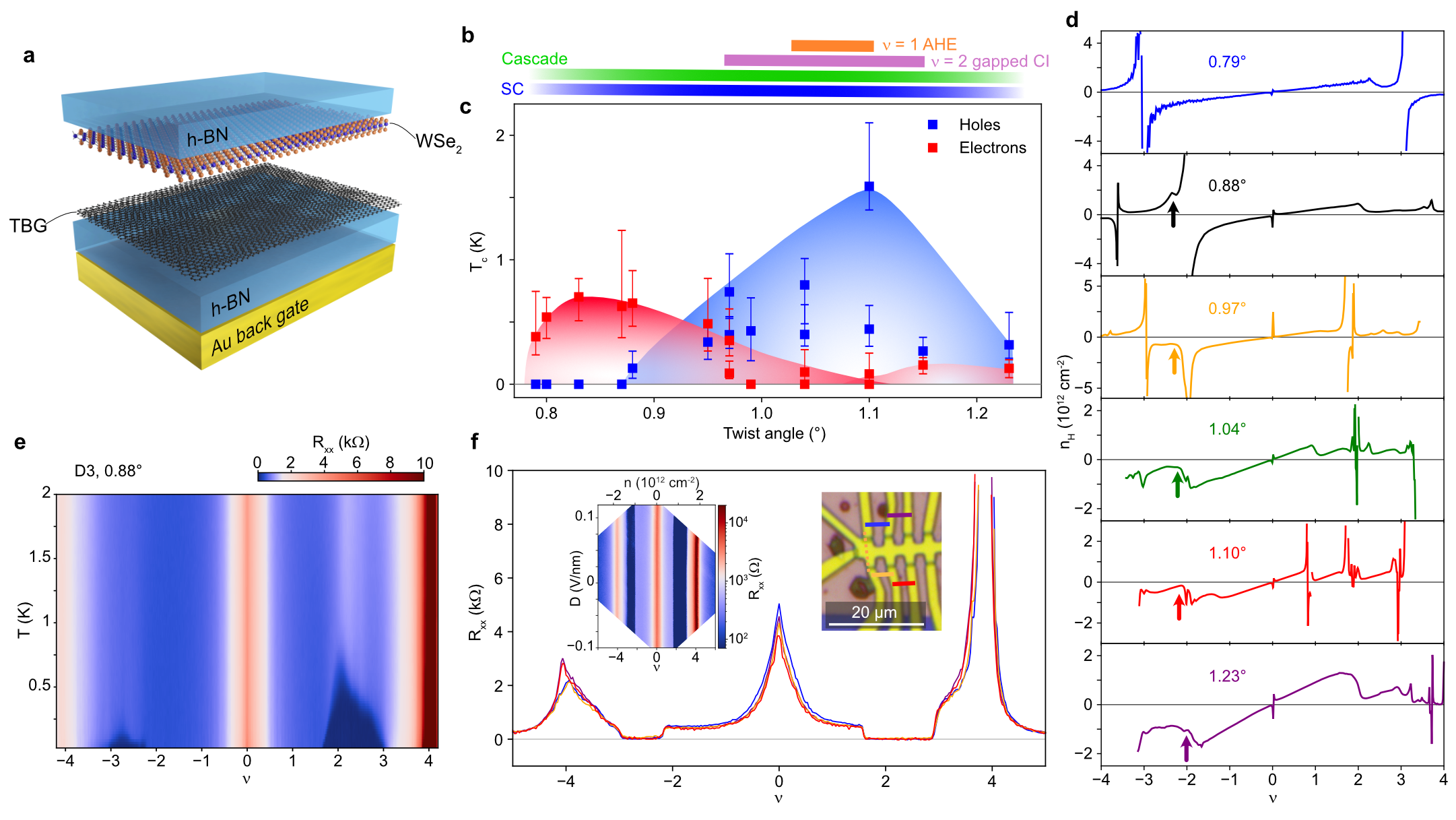}
    \caption{{\bf Overview of the symmetry-broken phases and SC in 
    TBG-WSe\textsubscript{2} for various twist angles.} {\bf a}, Schematic of 
    the hexagonal boron nitride (hBN)-encapsulated TBG-WSe\textsubscript{2} 
    structure with WSe\textsubscript{2} placed on top of TBG (see Methods for 
    details of fabrication, comparison to TBG devices without
    WSe\textsubscript{2}, and discussion of disorder). {\bf b, c}, Phase 
    diagram as a function of twist angle, indicating the regions which exhibit 
    the AHE due to ferromagnetism, $\nu$ = +2
    correlated insulators (CI), superconductivity (SC), and symmetry breaking 
    cascade effects as deduced from R$_{xx}$ peaks and Hall density resets. 
    The cascade and superconductivity start to disappear on 
    either side of the diagram, as indicated by the fading bar color. {\bf c}, 
    Critical temperatures T$_c$ of superconductivity for both holes and 
    electrons over a range of twist angles between 0.79\degree ~and 
    1.23\degree ~(squares indicate 50\% R$_n$ and the error bars 10\% 
    and 90\% R$_n$; for more details, see \prettyref{exfig:TandTcfigs}). 
    The gradient-filled domes are guides to the eye. {\bf d}, Hall density vs. 
    moiré filling factor. Flavor symmetry breaking correlations manifest as 
    Hall density resets (as seen clearly for 1.04\degree--1.23\degree ~on the 
    hole side, for instance, indicated by colored arrows) and occasionally as 
    singularities or hole-like regions (as seen at 1.10\degree ~and 
    0.97\degree ~on the electron side). {\bf e, f}, R$_{xx}$ values measured 
    for device D3, revealing the emerging hole-side superconductivity ({\bf e}) 
    and the uniformity of the sample over multiple contact pairs ({\bf f}). The 
    right inset of {\bf f} shows contact pairs corresponding to the colors in the main plot, and the left inset shows that the device behavior is independent of a $D$ field induced by a top gate in this device.}
    \label{fig:Fig1}
\end{figure*}

Figure \ref{fig:Fig1}a shows a schematic of the TBG-WSe\textsubscript{2} 
heterostructures (see \prettyref{exfig:opticalimages} 
and Methods for more details on device fabrication and the effects of 
WSe\textsubscript{2}). We find correlated insulators with well defined 
activation gaps for twist angles in the relatively narrow range of 
0.97\degree--1.15\degree, indicating 
that the addition of WSe\textsubscript{2} leaves the value of the magic angle 
unaffected (\prettyref{fig:Fig1}b). Unlike the correlated insulators, we find
that the cascade of high-temperature symmetry breaking 
transitions\cite{zondinerCascadePhaseTransitions2020, 
wongCascadeElectronicTransitions2020}  (discussed in more detail below) 
and superconductivity near $\nu$ = $\pm$2 (where $\nu$ is the number of 
electrons per unit cell) persist over a much wider range of 
twist angles (\prettyref{fig:Fig1}b,c; see also \prettyref{exfig:TandTcfigs} 
and Ref. \citenum{aroraSuperconductivityMetallicTwisted2020} for more data).
While all devices exhibit pronounced electron-hole asymmetry and a peak T$_c$ 
on the electron (hole) side which is shifted towards 
lower (higher) angles, superconductivity can be found well above 
($\theta$ = 1.23\degree, D2) and below ($\theta$ = 0.88\degree, D3) 
the magic angle for both negative and positive filling factors 
(see \prettyref{fig:Fig1}c). To the best of our knowledge, this 
is the largest reported range of twist angles exhibiting superconductivity 
for both electron and hole doping.

Importantly, the observed superconducting regions are consistently accompanied 
by Fermi surface reconstructions around $\nu$ = $\pm$2, as
manifested by a low-temperature reset in the Hall density. Consider, for 
example, the Hall density plots for the two lowest twist angles 
in \prettyref{fig:Fig1}d (blue and black curves). For 0.88\degree, hole-side 
SC~around $\nu$ = --2 has T$_c$ = 130 mK and is accompanied by the formation of 
a kink in the Hall density (black arrow), which is separate from
the van Hove singularity. At larger twist angles, the kink becomes a 
fully-developed Hall density reset to zero 
(colored arrows), corresponding to a more complete flavor symmetry 
breaking-induced Fermi surface reconstruction (see \prettyref{exfig:RxyT} for 
more data). In contrast, the device with twist angle 0.79\degree ~reveals a 
linear Hall density on the hole side that extends well beyond $\nu=-2$, 
ultimately reaching a van Hove 
singularity\cite{xieWeakfieldHallResistivity2020}. This signals the absence of 
an interaction-driven Fermi surface reconstruction. Interestingly, we also no 
longer find hole-side superconductivity for this twist angle. On the electron 
side, both twist angles
exhibit superconductivity and a kink in the Hall density due to Fermi surface 
reconstructions. 

We note that the addition of WSe\textsubscript{2}, while not changing the 
magic angle value, may help stabilize 
superconductivity\cite{aroraSuperconductivityMetallicTwisted2020} due to a 
reduction in disorder. We find in the best devices that four-point 
measurements almost perfectly overlap for different contact configurations 
(\prettyref{fig:Fig1}f), signaling high twist angle uniformity. Moreover, in
a dual-gated geometry no dependence on the 
displacement field is found (\prettyref{fig:Fig1}f, inset). 
However, we note that typical schemes of assessing disorder such as 
estimating the full width at 
half-maximum near charge neutrality that work well for monolayer graphene do 
not correlate with the superconducting T$_c$ or other disorder signatures in 
our samples (see methods and \prettyref{exfig:cnpdisorder}).

Our observations indicate that a fully flavor (i.e., spin and valley) symmetric
state strongly disfavors the formation of superconductivity. 
This rules out the simplest scenario for superconductivity based on 
electron-phonon coupling, which relies only on the local density of 
states\cite{qinCriticalMagneticFields2021}. 
Alternatively, and independently of the pairing mechanism, in the case of 
multiflavor
pairing, superconductivity and magnetism (i.e., flavor polarization) can be inherently
connected. This connection emerges from a simple energetic argument. If two
flavors pair, they could increase their condensation energy by exchanging  
particles with the other flavors, such that they maximize their density of states. Roughly, 
this is captured through a term akin to $\Delta^2 M$ in the free energy, but 
with M (and $\Delta$) being a matrix indicating the density of the various 
flavors on its diagonal, and correlations (pairing) 
in the off diagonal\cite{cherngSuperfluidityMagnetismMulticomponent2007}. 
This term in the free energy implies that strong flavor polarization 
(manifested in $M$) will generally increase the superconducting gap, and 
thus T$_c$, of a multiflavor superconductor. 
Conversely, a finite superconducting order parameter 
could also induce polarization. Our experimental observation that enhanced 
superconductivity occurs only in regions with prominent Hall density resets  is 
thus in line with multiflavor pairing. This could potentially reconcile
experiments with electron-phonon mechanisms of superconductivity, 
although we note that 
our results do not rule out unconventional mechanisms, based e.g., on 
flavor fluctuations\cite{youSuperconductivityValleyFluctuations2019} or the 
Kohn-Luttinger scenario\cite{gonzalezKohnLuttingerSuperconductivityTwisted2019}.

\begin{figure*}[ht]
    \centering
    \includegraphics[width=16cm]{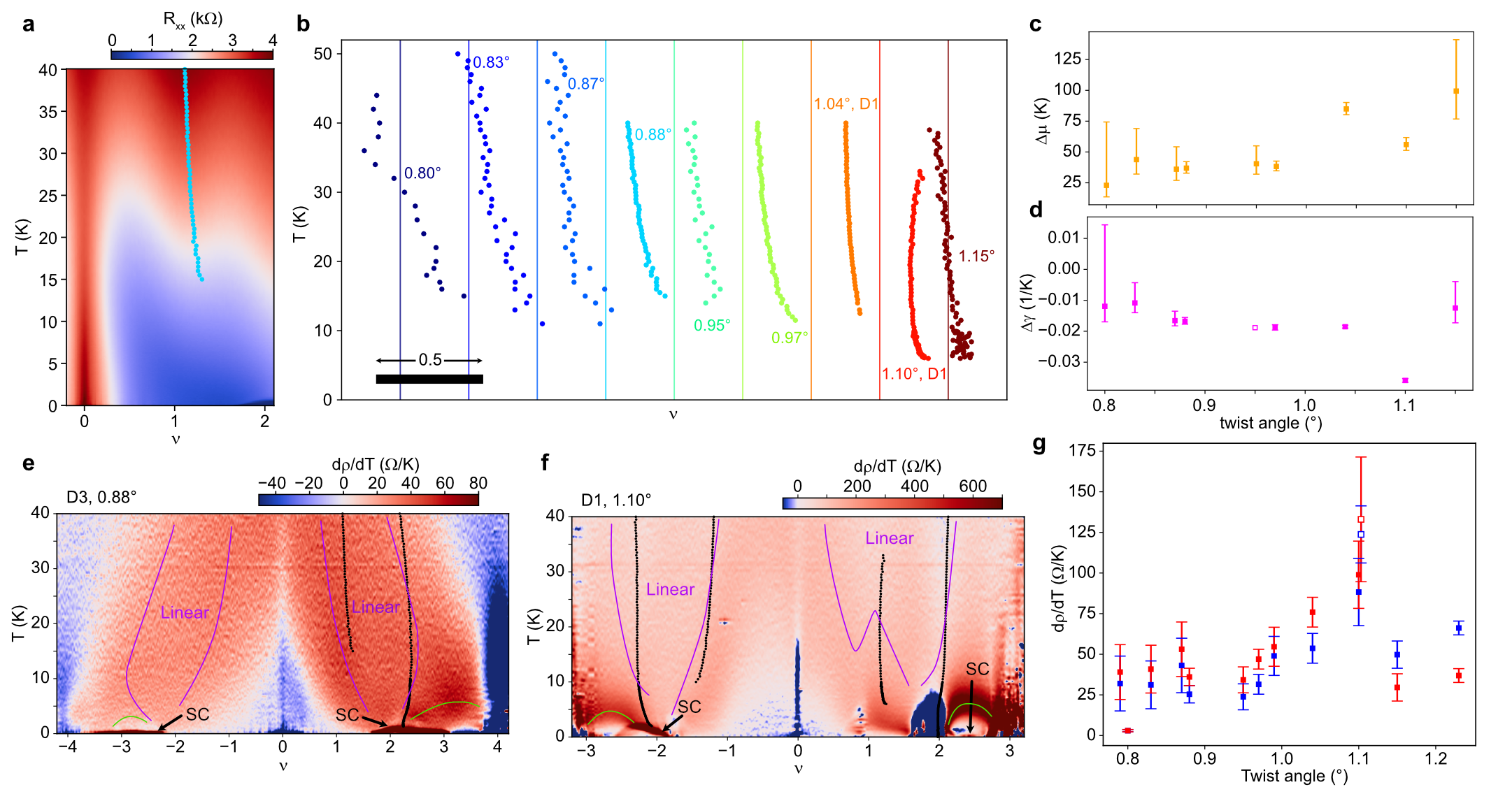}
    \caption{{\bf Pomeranchuk-like phase transitions near $\nu$ = +1 and linear $T$-dependence of resistivity.} 
    {\bf a}, Example of a peak fit at $\nu \approx 1$ for a device at twist angle 0.88\degree. {\bf b}, R$_{xx}$ peak 
    positions near filling factor $\nu \approx$ 1 as a function of temperature for devices with various 
    twist angles. The vertical colored lines represent the $\nu$ = 1 filling factor for the respectively 
    colored twist angle data. {\bf c, d}, Fit parameters $\Delta \mu$ and $\Delta \gamma$ representing change in the chemical 
    potential and specific heat (see Equation 1 in the Supplementary Information and Ref. \citenum{zondinerCascadePhaseTransitions2020}) 
    for the phase transition represented by the R$_{xx}$ peaks in {\bf b}. Error bars are 95\% confidence intervals (the hollow 
    square for 0.95\degree~ was set to the same $\Delta \gamma$ value as the 0.97\degree~ device due to the 
    unconstrained $\Delta \gamma$ value for the data points). {\bf e, f}, The derivative $\frac{d\rho}{dT}$ 
    (where $\rho$ = R$_{xx}$W/L is the resistance scaled by the width W and length L of the sample) for devices at 
    twist angles 0.88\degree~ and 1.10\degree, respectively. R$_{xx}$ peaks corresponding to Pomeranchuk-like transitions 
    are shown with black dots (corresponding to slight minima in $d\rho/dT$), superconductivity (SC) pockets are 
    shown with arrows, and the magenta lines are guides to the eye representing the approximate regions of T-linear 
    resistivity. The green lines reveal the inflection points under which low-temperature resistivity is 
    super-linear. {\bf g}, The resistivity slope $\frac{d\rho}{dT}$ for a range of twist angles, where red(blue) 
    is for electrons(holes). The values come from the average derivative over the area spanned by $1.5 < \nu < 1.8$ ($-2 < \nu < -1.6$) 
    for electrons (holes) and $15 < T < 38$ (error bars are the standard deviations). 
    Device D4, twist angle 1.10\degree~ is represented by hollow squares.}
    \label{fig:Fig2}
\end{figure*}

The principal features emerging at higher temperatures (above 5--10 K), 
such as the cascade of phase transitions between symmetry broken states near integer 
filling factors and the linear-in-T dependence of R$_{xx}$, are also present over a wide range of 
angles (\prettyref{fig:Fig2} and 
\prettyref{exfig:TandTcfigs}). In the case of $|\nu|\approx 1$, R$_{xx}$ peaks are associated 
with a Pomeranchuk-like phase transition\cite{saitoIsospinPomeranchukEffect2021,rozenEntropicEvidencePomeranchuk2021} 
between a flavor symmetric state near charge neutrality
and a symmetry broken phase with 
free local spin moments. The evolution of this phase boundary 
with temperature can be fit using a simple thermodynamic model including the entropy of localized spins\cite{rozenEntropicEvidencePomeranchuk2021}. 
The main parameters entering this model are the shift in the chemical potential $\Delta \mu$ due to the cascade transition and the  change in specific 
heat $\Delta \gamma$ between the two phases.

Our data suggest that this description works reasonably well over the entire 
range of angles investigated here (\prettyref{fig:Fig2}a-d). Moreover, $\Delta \gamma$ 
appears to be roughly constant except for device D1 at 1.10\degree~ right at 
the magic angle, perhaps indicating  additional correlation effects emerging for this 
angle. We note that this device also 
exhibits a metallic anomalous Hall phase near $\nu$ = 1 at low temperatures, 
discussed below. The slowly increasing $\Delta \mu$ signals stronger shifts of the 
bands as the twist angle is increased, which is consistent with the reduction 
of the moir\'e length scale and thus stronger
electronic interactions. 
The appearance of superconductivity, high-temperature symmetry breaking 
cascade transitions, and in particular Pomeranchuk-like transitions over similar twist 
angle ranges (see \prettyref{exfig:TandTcfigs} and 
Ref. \citenum{aroraSuperconductivityMetallicTwisted2020} to see the fading cascade at the limits of twist angles studied here) suggests a possible connection between these instabilities and points to similarities between TBG and heavy-fermion systems\cite{songMATBGTopologicalHeavy2021} which also show rich phase diagrams exhibiting similar phases\cite{stewartHeavyfermionSystems1984, lonzarichNewMicroscopicFramework2016, continentinoPomeranchukEffectUnstable2004}. 

The twist angle dependence of the linear-in-T resistivity behavior is shown in \prettyref{fig:Fig2}e-g. A linear temperature dependence of the resistivity can be due to electron-phonon scattering, at least at higher  
temperatures (above 5--10K)\cite{wuPhononinducedGiantLinearin2019, polshynLargeLinearintemperatureResistivity2019, sarmaStrangeMetallicityMoir2022}. 
Alternatively, this dependence has also been associated with strange metal behavior due to its 
onset at low temperatures and
its strength near the $|\nu|$ = 2 correlated phases\cite{caoStrangeMetalMagicAngle2020, jaouiQuantumCriticalBehavior2022}. Our data show that the 
linear-in-T behavior is qualitatively similar in devices away from  
(\prettyref{fig:Fig2}e; see \prettyref{exfig:drhodTangles} for more examples) and close to the  magic angle (\prettyref{fig:Fig2}f).
Both cases exhibit broad regions of linear-in-T behaviour fanning out from 
approximately $|\nu|$ = 2 (delineated by magenta lines in \prettyref{fig:Fig2}e,f). At lower 
$|\nu|$, this region is bordered by a broad region near charge neutrality, where the increase in resistance is quadratic (consistent with expectations for Fermi liquid behaviour). At 
higher $|\nu| > 2$,
there is a region where the temperature dependence of the resistivity crosses 
over from strongly super-linear to sub-linear as temperature increases. The 
intermediate inflection points, shown as green lines in \prettyref{fig:Fig2}e,f,
appear to be intertwined with other TBG phases as they occasionally touch 
the superconducting domes (both for electrons and holes at 0.88\degree~ 
as well as for holes at 1.10\degree) or
onset near $\nu \approx 2$, when a correlated insulating gap is present. This 
observation contrasts with the suggestion that the entire superconducting dome 
emerges below a linear-in-T phase\cite{jaouiQuantumCriticalBehavior2022}. Note 
that both the magnitude of the T dependence 
(as measured by the slope near the green lines) and the linear-in-T slope 
measured at higher temperatures (within the magenta regions) are enhanced 
near the magic angle (\prettyref{fig:Fig2}g). This is to be expected, as the 
Fermi velocity is minimized for this angle and could be even further 
reduced by interaction effects causing band flattening at nonzero 
filling\cite{goodwinHartreeTheoryCalculations2020}. 
Finally we note that, 
in general, the linear-in-T slope peaks around the same angle value for both 
electrons and holes. This contrasts with the observed doping asymmetry of the 
twist angles at which the superconducting $T_c$ becomes maximal, further 
highlighting the differences between high- and low-temperature symmetry 
breaking phenomena.

\begin{figure*}[ht]
    \centering
    \includegraphics[width=16cm]{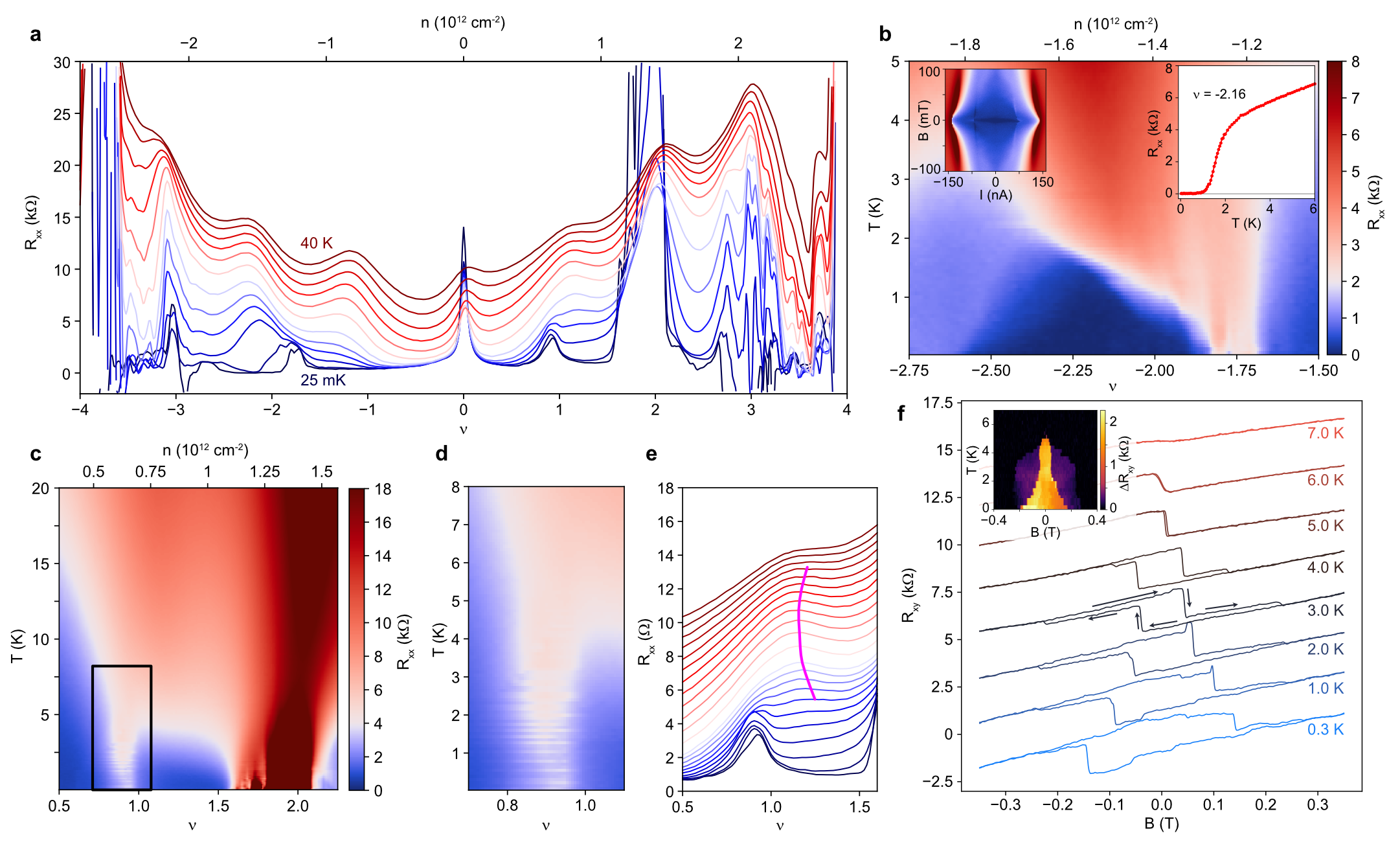}
    \caption{{\bf Superconductivity and correlated states 
    for a magic-angle TBG-WSe\textsubscript{2} device (D1).} {\bf a}, Line cuts of R$_{xx}$ versus
      filling factor $\nu$ for a range of temperatures (shown is
      25 mK, then 1, 3, 5, and 7 K, followed by every 5 K from 10 to 40 K).
      Superconductivity, correlated insulators, and orbital magnetism (incipient
      Chern insulator) emerge below 10K, whereas the symmetry breaking cascade
      transitions survive up to 40 K. {\bf b}, R\textsubscript{xx} versus
      temperature and filling factor on the hole side, showing a superconducting
      dome around $\nu$= --2. Insets, temperature dependence of
      R$_{xx}$ at $\nu$ = --2.16, showing the SC transition and
      Fraunhofer-like interference pattern ($\nu$ = --2.3). {\bf c}, R\textsubscript{xx} versus
      temperature and filling factor focusing around $\nu$ = 1. {\bf d}, Zoom
      into the black box in {\bf c}, revealing the switching behavior of the
      resistance, representative of the ferromagnetic state. {\bf e}, Dense
      line-cut plots for the area around $\nu$ = 1, showing that the evolution of
      R$_{xx}$ maxima above 10 K (magenta line) are distinct from the ferromagnetic peak. These peaks mark the T-dependent cascade transition
      previously studied around filling factor one in
      TBG\cite{rozenEntropicEvidencePomeranchuk2021,
        saitoIsospinPomeranchukEffect2021}. {\bf f}, Temperature dependence of R\textsubscript{xy}
      versus B at $\nu$ = 0.90 from 300 mK to 7 K. Successive curves are offset by
      2.2 k$\Omega$, and the arrows indicate the B field sweep directions. The inset
      demonstrates $\Delta$R$_{xy}$ = R$_{xy}^{B_\uparrow}$ -- R$_{xy}^{B_\downarrow}$
      versus B field and temperature at the same filling factor. }
    \label{fig:Fig3}
\end{figure*}

Now we focus on a magic-angle device with $\theta$ = 1.10\degree~ (\prettyref{fig:Fig3}). 
Upon cooldown, this device exhibits clear R$_{xx}$ peaks below T $\approx$ 40 K at every 
integer 0 $< |\nu| <$ 4. As temperature is lowered further, a correlated insulator (CI) 
develops near $\nu$ = +2, while other resistance peaks remain metallic or disappear gradually 
(\prettyref{fig:Fig3}a). For hole doping (\prettyref{fig:Fig3}b), the corresponding 
superconducting dome near $\nu$ = --2 reaches a maximal transition temperature of
T$_c = 1.6$~K, featuring vanishing longitudinal resistance R\textsubscript{xx} and a Fraunhofer-like pattern in line with previously 
reported hBN-encapsulated, high-quality magic-angle TBG
devices\cite{caoUnconventionalSuperconductivityMagicangle2018,
  yankowitzTuningSuperconductivityTwisted2019,
  luSuperconductorsOrbitalMagnets2019,
  saitoIndependentSuperconductorsCorrelated2020,
  stepanovUntyingInsulatingSuperconducting2020} (see also Methods for more detailed information).
  
In contrast to the features at $\nu$ = $\pm 2$, the evolution of R$_{xx}$
near $\nu$ = $\pm 1$ is more subtle. In the temperature range
6 K $<$ T $<$ 40 K, the R$_{xx}$ peaks near both 
$\nu$ = +1 and $\nu$ = --1  evolve
towards lower filling factors $|\nu|$ as temperature increases,
following the phase boundary discussed above (\prettyref{fig:Fig3}a,e). 
However, at lower temperatures, T $<$ 6 K, the two peaks show distinctly 
different behavior. While the hole-side peak completely disappears, reflecting
simple metallic behavior from charge neutrality to $\nu \approx$ --2, the peak near $\nu$ = 1 
gradually gives way to another peak emerging in the filling range 
0.8 $<\nu<$ 0.95  that persists to the lowest temperatures 
(\prettyref{fig:Fig3}c). Careful inspection reveals that R\textsubscript{xx} 
exhibits switching behavior around the peak, discontinuous resistance 
changes between the same sweep across $\nu$ taken at slightly different 
temperatures (\prettyref{fig:Fig3}d) presumably due to switching of domains. 
Further measurements in the $\nu$ and temperature range of this low-temperature 
R$_{xx}$ peak reveal an anomalous Hall effect (AHE). Figure \ref{fig:Fig3}f 
shows hysteresis loops in the Hall resistance, R$_{xy}$, for $\nu$ = 0.9 as  
measured from 0.3 K to 7 K. The loop has a coercive field of up to about 150 mT 
and is centered about zero magnetic field. The jump in resistance
$\Delta$R$_{xy}$ = R$_{xy}^{B_\uparrow}$ -- R$_{xy}^{B_\downarrow}$ reaches a 
maximal value of 2.5 k$\Omega$, significantly smaller than the resistance 
quantum, which persists until the Curie temperature $\sim$ 5 K
(\prettyref{fig:Fig3}f inset).  Signatures of an AHE phase are also observed at 
twist angles of 1.04\degree~ and 0.99\degree, 
as shown in \prettyref{exfig:AHEobs}.

\begin{figure*}[ht]
    \centering
    \includegraphics[width=16cm]{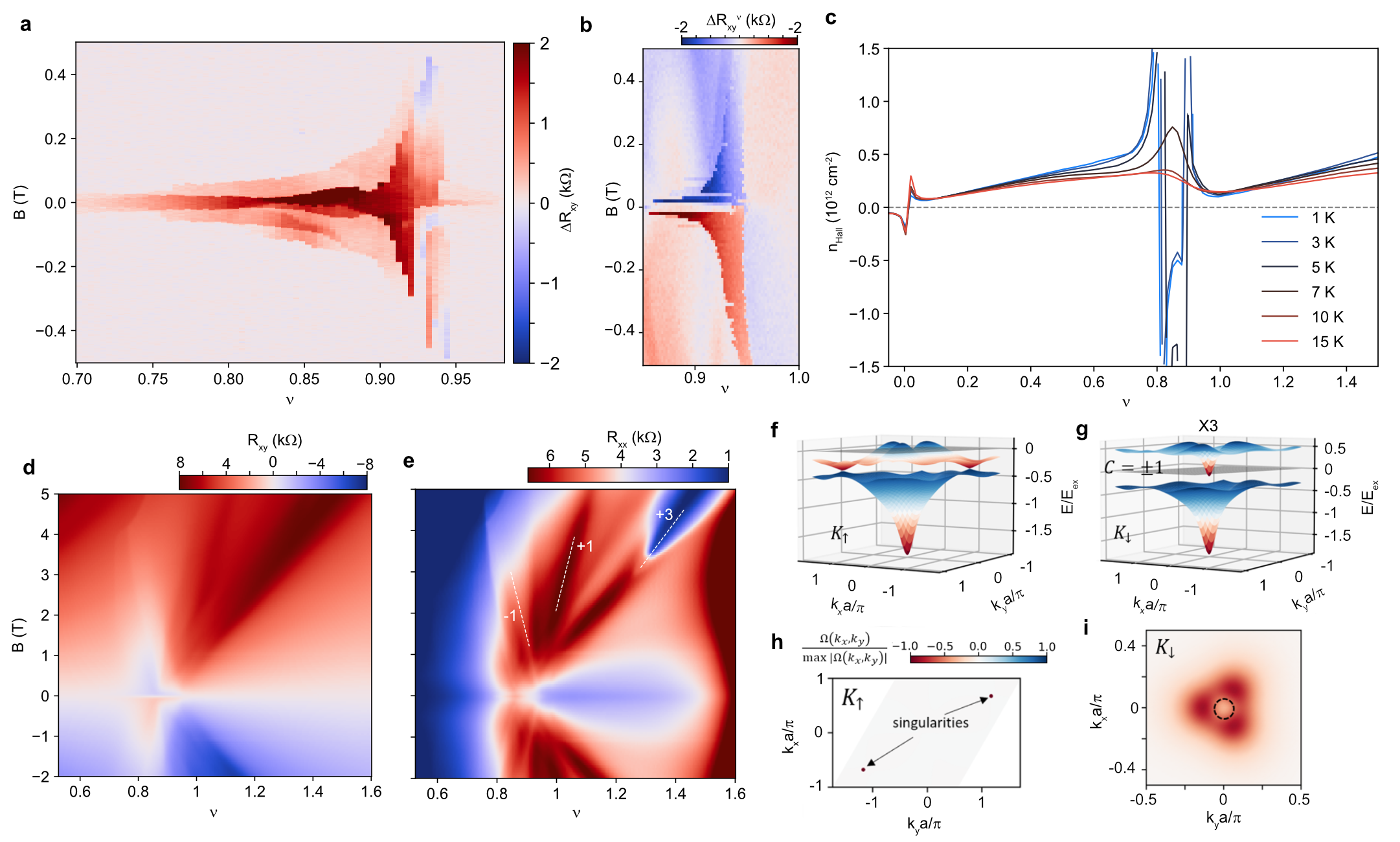}
    \caption{{\bf Anomalous Hall effect and zero-field $C$ = --1 correlated states
        near $\nu$ = 1.} {\bf a},
      $\Delta$R$_{xy}$ vs. B and $\nu$ around filling factor 1 measured at
      1.5K. The divergence of the coercive field happens around $\nu$ = 0.95. {\bf b}, Hall density difference as a result of sweeping $\nu$ while holding magnetic field constant, $\Delta$R$_{xy}^\nu$ = R$_{xy}^{\nu\uparrow}$ -- R$_{xy}^{\nu\downarrow}$.
      {\bf c}, Hall density versus $\nu$ measured at 0.5T (the value shown is
      antisymmetrized [R$_{xy}$(0.5T) -- R$_{xy}$(--0.5T)]/2) with temperature
      from 1K to 15K. n$_{Hall}$ changes sign when entering the anomalous Hall
      region in ($\nu$, T, B) parameter space, due to the AHE negative offset to R$_{xy}$ as a result of the zero-field symmetry breaking. However,
      the slope of the R$_{xy}$ vs. B measurements remain positive, as seen in {\bf d} and \prettyref{fig:Fig3}f, meaning the overall behavior is still electron-like.
      {\bf d, e}, Magnetic field and $\nu$ dependence of R$_{xy}$
      ({\bf d}) and R$_{xx}$ ({\bf e}) around filling factor 1
      measured at 3K. In addition to the $C$ = +3 Chern insulator developed at high magnetic field,
      features associated with $C$ = $\pm$1 also emerge from filling factor 1, as
      labeled in {\bf e}. The origin of the $C$ = --1 line is at $\nu$ = 0.94,
      indicating that an incipient Chern insulator $C$ = --1 is responsible for
      anomalous Hall signal observed around $\nu$ = 1.
      {\bf f, g}, The band structure obtained for each spin and mini-valley (K, K') flavor from the 10-band model for the case of broken C$_2\mathcal{T}$ symmetry at $\nu\simeq$ 0.81. The K$_\uparrow$ flavor, which is nearly filled, preserves the Dirac-like band structure ({\bf f}), whereas the other three flavors have a C$_2\mathcal{T}$-broken mass ({\bf g}). The gray planes represent the chemical potential. Here we assume the mass term is invariant under $\mathcal{T}$ operation, which gives overall Chern number $C$ = $\pm1$.
      {\bf h, i}, Berry curvature $\Omega_{k_x,k_y}$ for the conduction flat band in the K$_\uparrow$ flavor ({\bf h}) and for K$_\downarrow$ ({\bf i}), where the Berry curvature is concentrated above the $\Gamma$ pocket. The Fermi surface is plotted as a dotted circle. The other two flavors, from the opposite valley K', have the same Berry curvature as {\bf i} but opposite sign.}
    \label{fig:Fig4}
\end{figure*}

\begin{figure}[h]
    \includegraphics[width=16cm]{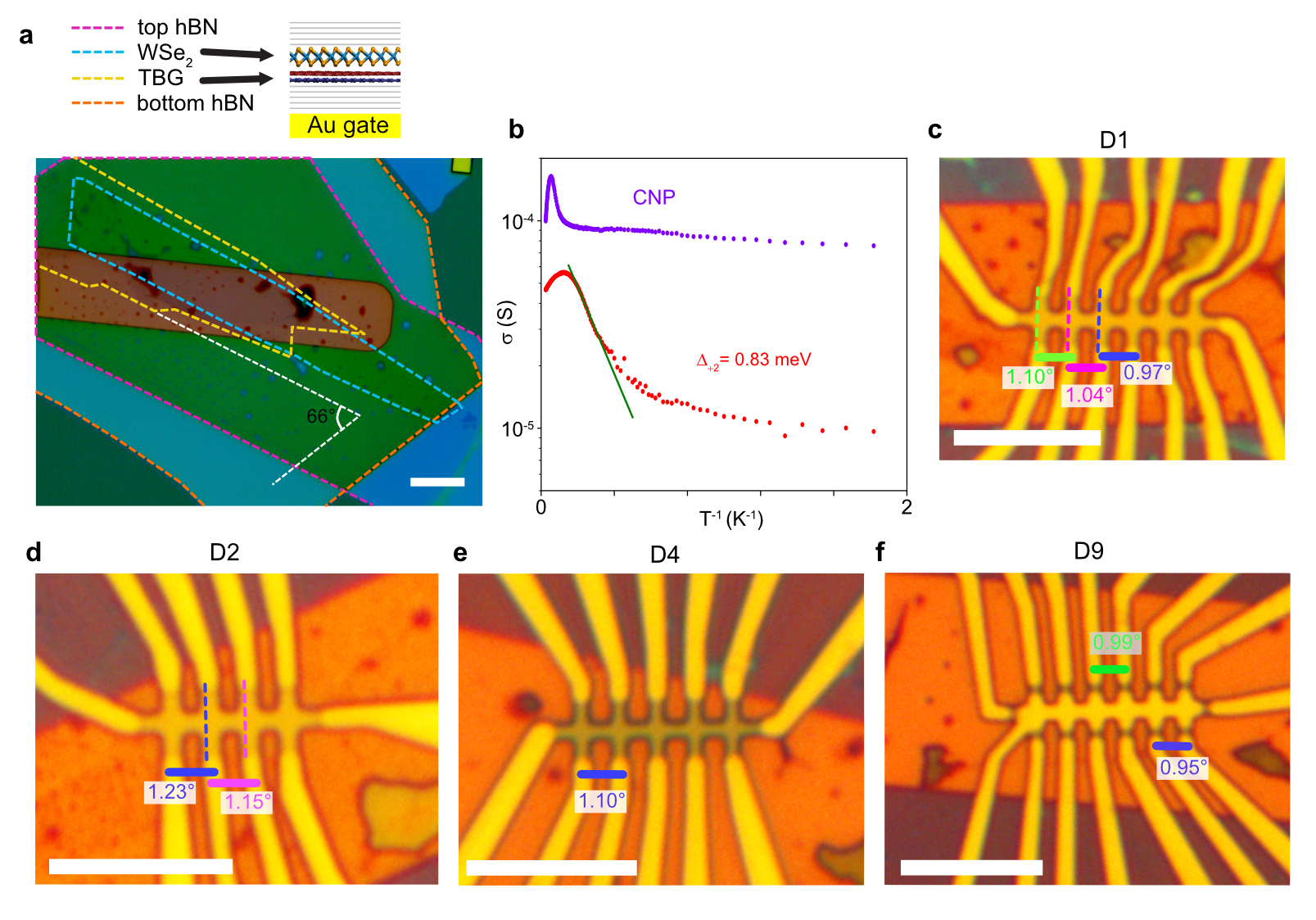}
    \centering
    \caption{{\bf Optical images of devices and correlated gap
        analysis.} {\bf a}, An optical image of D1 TBG-WSe\textsubscript{2}
      heterostructure with different layers delineated by dashed lines with
      different colors. The angle between graphene and bottom hBN edges is
      $\sim$66\degree, showing no obvious alignment. {\bf b}, Conductance versus 1/T
      for charge neutrality and partial filling factor $\nu$ = 2 from the data in \prettyref{fig:Fig3}a. The green line and the gap value shown are extracted from the
      activation fit, to the form $\sigma \propto e^{-\Delta/2k_B T}$.
      In contrast, the conductance at charge neutrality shows a much smaller variation in
      temperature without a clear activated gap. {\bf c-f}, Optical images of devices used in the study. Bold colored lines show contact pairs used for R$_{xx}$ measurements (and corresponding measured twist angles), while dashed colored lines show contact pairs used for R$_{xy}$ measurements. White scale bars represent 10 $\mu$m. For images of other devices in \prettyref{extab:devicetable}, see Ref. \citenum{aroraSuperconductivityMetallicTwisted2020}.}
\label{exfig:opticalimages}
\end{figure}

\begin{figure}[hbt]
    \includegraphics[width=10.5cm]{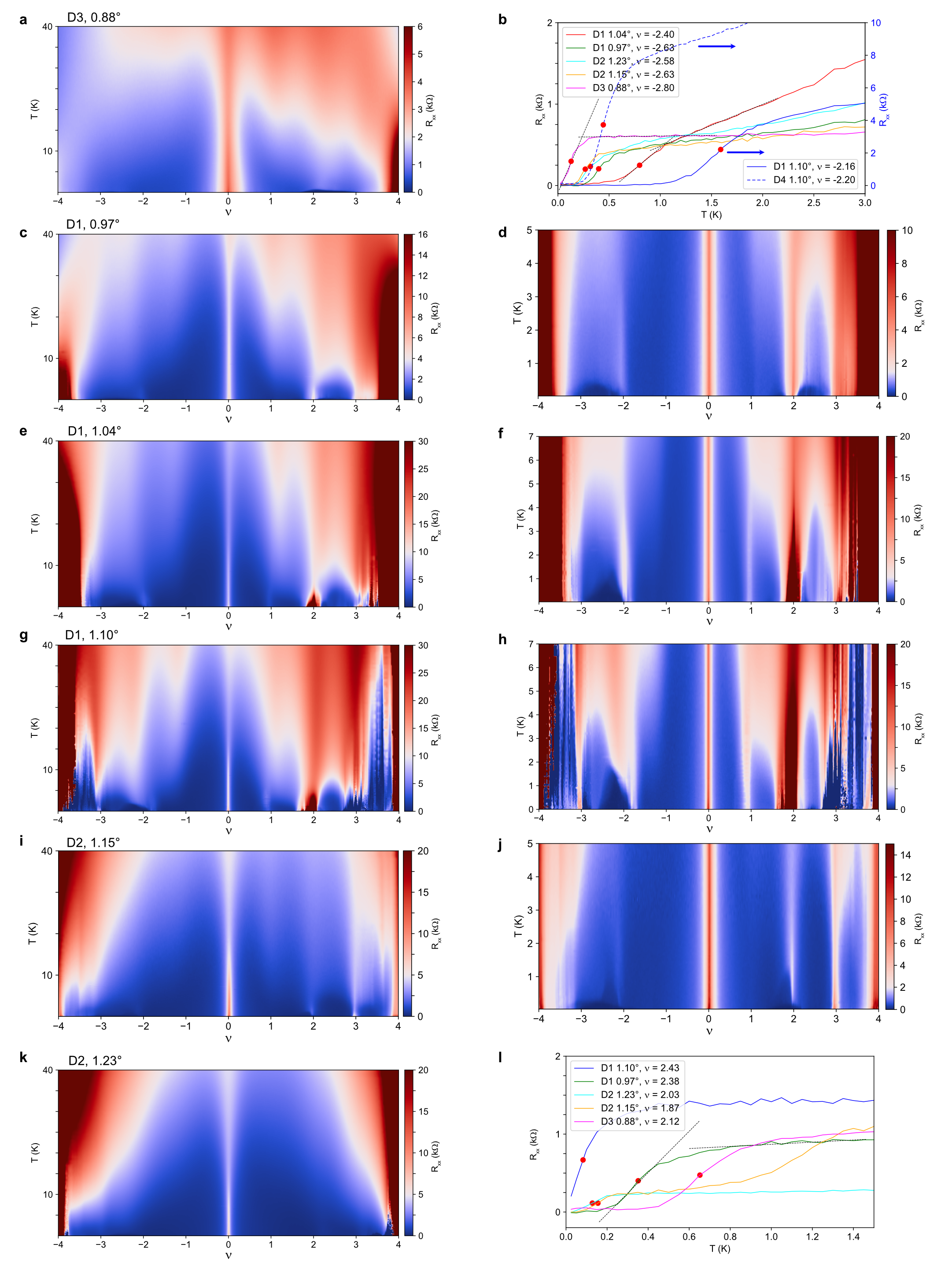}
    \centering
    \caption{{\bf Temperature dependence in a collection of devices.} Figures in
      the left column show the high-temperature (to 40 K) R$_{xx}$ vs. the moiré
      filling factor for a number of devices across the angle range. The
      corresponding figures in the right column reduce the temperature range to
      emphasize the correlated insulators and superconductivity appearing at low
      temperature. {\bf b} shows extracted T$_c$ on the hole side for each of
      the angles represented here. The resistance was much larger for device D1
      at 1.10\degree, as well as another device D4 at the same twist angle, so
      the y-axis for these curves is on the right of the plot. The extreme
      sensitivity of the correlations and superconductivity to the twist angle,
      cleanliness, homogeneity, and other factors occasionally results in
      different T$_c$ values for devices of the same angle. {\bf l}, The
      electron-side superconductivity T$_c$ values. T$_c$ was derived from 50\%
      R$_n$, where R$_n$ is defined as the intersection of line fits to the
      highly sloped region and the normal region just above the transition (line
      fits shown for selected curves as dashed black lines). The same method was
      used to determine the error-bars at 10\% and 90\% R$_n$ in
      \prettyref{fig:Fig1}c.}
\label{exfig:TandTcfigs}
\end{figure}

\begin{figure}[ht]
    \includegraphics[width=13cm]{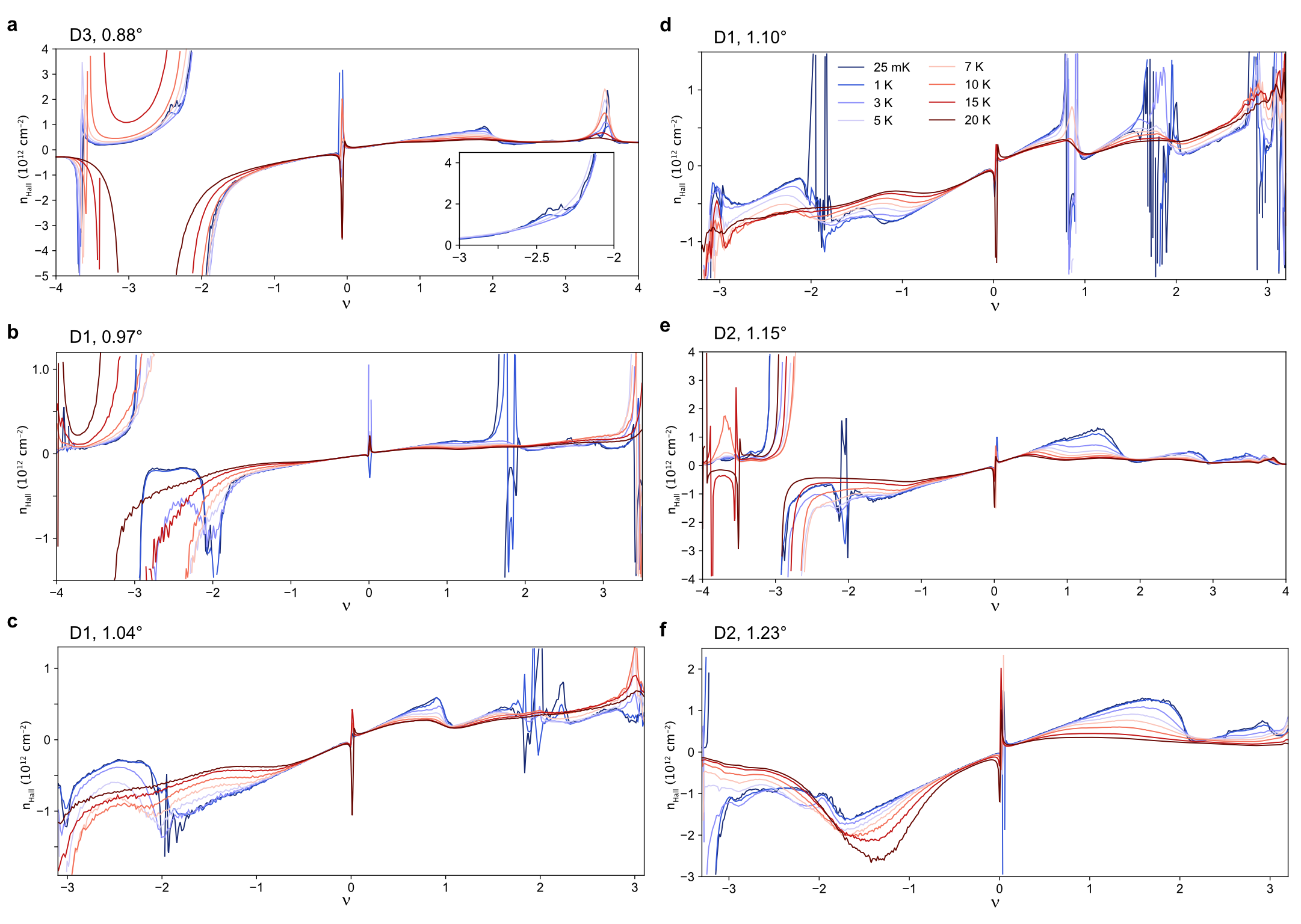}
    \centering
    \caption{{\bf Cascade behavior Hall density measurements at select
        temperatures.} More detailed Hall density data for the mentioned devices,
      taken at the the range of temperatures displayed in {\bf b}. The
      only deviation from these panels is in {\bf a}, where the
      lowest-temperature curve was measured at 50 mK instead of 25 mK. The inset of
      {\bf a} displays the $\nu \approx$ --2 region for the 0.88\degree
      ~device as it evolves at temperatures up to 5 K, revealing the appearance
      of the feature mentioned in the main text and \prettyref{fig:Fig1}d, which
      corresponds with the onset of hole-side superconductivity. This small
      feature, indicative of the onset of a cascade, survives to only around 2
      K. The rest of the panels reveal clear cascades, where the Hall density
      returns to near 0, at both $\nu$ = --2 and +2 at low temperatures.}
\label{exfig:RxyT}
\end{figure}

\begin{figure}[hb]
    \includegraphics[width=8cm]{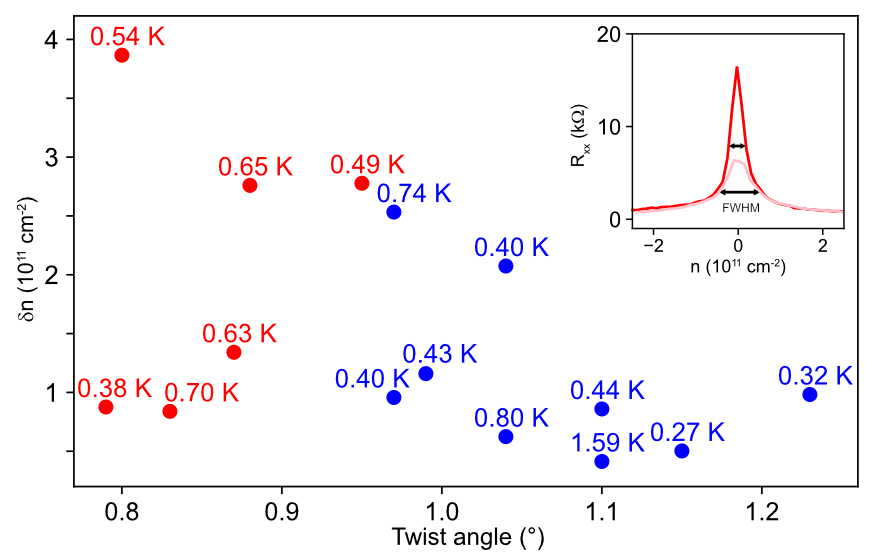}
    \centering
    \caption{{\bf Charge-neutrality point disorder measurement.} The disorder in monolayer graphene is commonly measured using the width of the charge neutrality peak $\delta n$. The full-width half-max used for this plot is shown in the inset for the 1.10\degree~ data point with T$_c$ of 1.59 K (red) and the 1.23\degree~ point (pink). However, we did not observe a correlation between superconducting T$_c$ and the disorder measured using this method for TBG. Listed next to each data point is the maximum T$_c$ measured for the twist angle, and the color corresponds to whether the superconductivity was on the hole (blue) or electron (red) side.}
\label{exfig:cnpdisorder}
\end{figure}

\begin{figure}[hbt]
    \includegraphics[width=10.5cm]{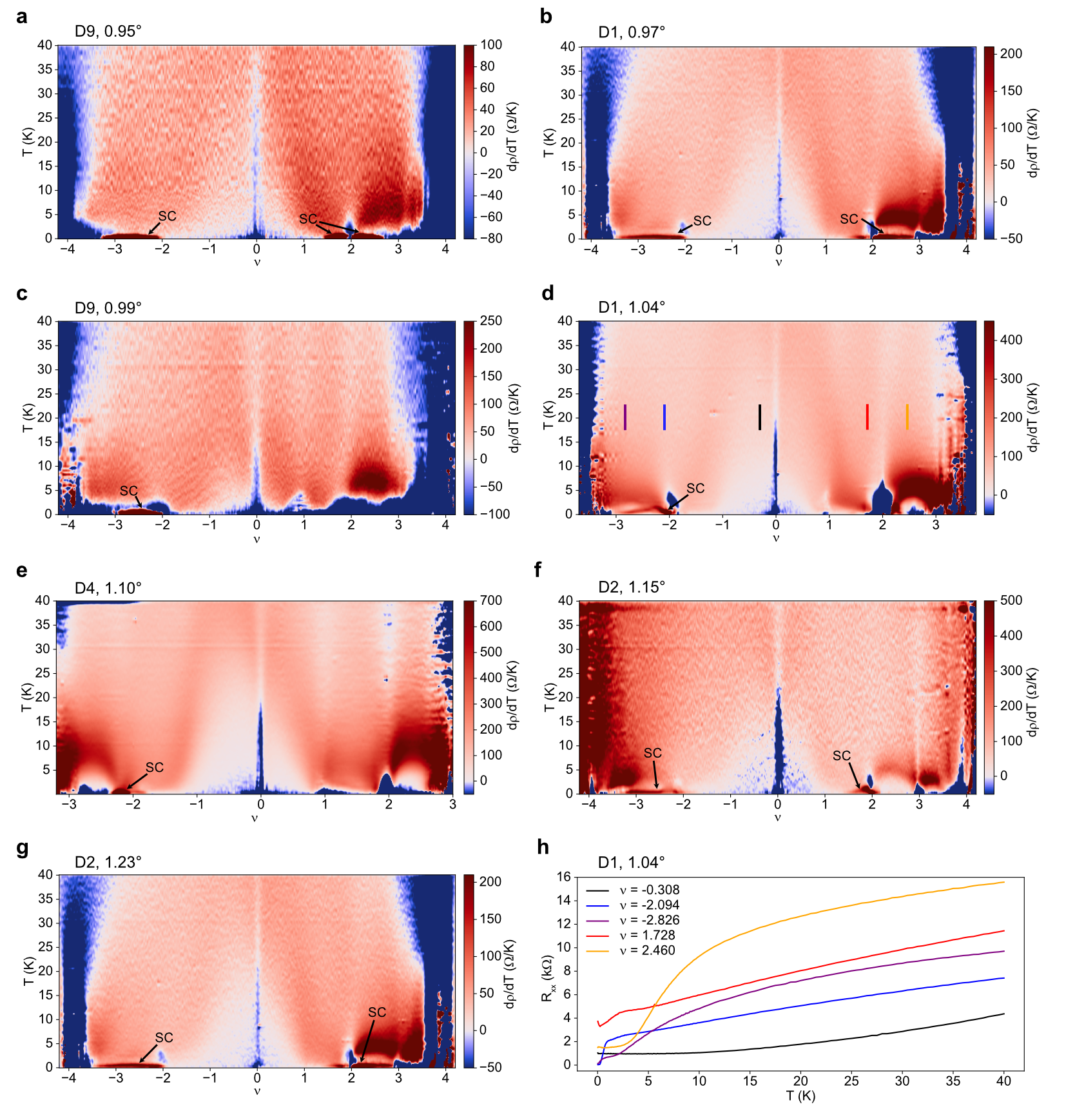}
    \centering
    \caption{{\bf d$\rho$/dT diagrams for devices of different angles.} {\bf a-g}, Derivatives of resistivity with respect to temperature up to 40 K for twist angles from 0.95\degree~ to 1.23\degree. {\bf h}, Selected linecuts of R$_{xx}$ vs. temperature for the 1.04\degree~ twist angle (filling factor values also shown as coloured lines in {\bf d}). The linecuts show the broad positive-curvature zone near charge neutrality (black line), the linear resistivity that persists down to a few Kelvin (although often blocked by a correlated insulator, superconducting, or other symmetry-broken state at low temperatures) near $|\nu|$ = 2 (red and blue lines), and the super-linear low temperature to sub-linear high temperature states at $|\nu| >$ 2 (purple and orange lines), which have large transition regions that prevent linear behavior until high temperatures. The curve at $\nu$ = --2.826 (purple) shows an example where the higher-temperature positive-curvature zone is seen as the superconducting dome is phasing out (small T$_c$).}
    \label{exfig:drhodTangles}
\end{figure}

Importantly, the observed AHE phase appears well below filling 
factor $\nu$ = 1, existing in the range 0.7 $<\nu<$ 1, with the
maximal $\Delta$R$_{xy}$ occurring near $\nu$ = 0.88
(\prettyref{fig:Fig4}a). Upon approaching $\nu$ = 0.95, the filling at which
R$_{xx}$ peaks, the coercive field diverges accompanied by a sudden
decrease in $\Delta$R$_{xy}$. Then signatures of small hysteresis loops in the 
opposite direction appear (see \prettyref{exfig:hysteresis} for a more 
clear example). Additionally, hysteresis is observed when sweeping $\nu$ in 
opposite directions (\prettyref{fig:Fig4}b) and holding the  magnetic field constant.

Measurements of R$_{xy}$ at elevated temperatures and over a wider
doping range further reveal the unusual nature of the observed AHE phase 
(\prettyref{fig:Fig4}c,d). Surprisingly, within the range of $\nu$ exhibiting 
the AHE and up to magnetic fields greater than 1 T, the sign of R$_{xy}$ is 
opposite to that of the surrounding doping range. Also, R$_{xy}$ changes sign 
when the temperature reaches $\sim$5 K, consistent with the measured AHE Curie 
temperature. While naively this behaviour might be due to a change of carrier 
type from electrons to holes, the measurements show a linear increase in 
R$_{xy}$ with increasing magnetic field throughout the entire doping range 
(\prettyref{fig:Fig3}f and \prettyref{fig:Fig4}e), consistent with dominant 
electron conduction.

The observation of hysteresis signals the emergence of an orbital 
ferromagnetic phase that arises from a band carrying nonzero Chern number 
$C$\cite{zhuVoltageControlledMagneticReversal2020, 
polshynElectricalSwitchingMagnetic2020}. 
A finite Chern number is also expected to result in R$_{xy}$ and R$_{xx}$ 
features that follow the Streda formula in an out-of-plane magnetic field B = 
$\frac{h}{Ce}n$. We observe clear maxima and minima in R$_{xx}$ approximately 
following the Streda formula at fields less than 3T (\prettyref{fig:Fig4}e), and 
specifically point to the R$_{xx}$ minimum following 
$C=$ --1 that extrapolates to $\nu$ = 0.95 at zero field, near where the 
coercivity diverges. The 
low-field features disappear by B = 3--4 T, where a finite-field Chern 
insulating phase corresponding to $C$ = 3 takes 
over. The observed switch in Chern number indicates the competing nature of 
these phases\cite{stepanovCompetingZerofieldChern2020, 
lianTwistedBilayerGraphene2021}.

However, our observations stand in contrast to discussions of zero-field Chern 
insulators, where the chemical potential falls into an insulating 
gap\cite{stepanovCompetingZerofieldChern2020,lianTwistedBilayerGraphene2021, 
linProximityinducedSpinorbitCoupling2021, groverImagingChernMosaic2022}.
Throughout the entire doping range, the AHE phase observed here 
appears metallic.
This is implied by the electron-like Hall resistivity and
the finite R$_{xx}$ peak at 0 T that is small compared to $h/e^2$
and increases with temperature. This contrasts with the expected vanishing 
of R$_{xx}$ for a gapped bulk with gapless edge
channels\cite{serlinIntrinsicQuantizedAnomalous2019} or a resistance $\sim h/e^2$ that can originate from domain walls extending between the
contacts\cite{polshynTopologicalChargeDensity2021}. We note that while disorder 
effects and the presence of domains observed in Ref. 
\cite{groverImagingChernMosaic2022} can, to some extent, explain 
the absence of a gap and imperfect quantization, local 
compressibility\cite{zondinerCascadePhaseTransitions2020, 
yuCorrelatedHofstadterSpectrum2021}
and scanning tunneling spectroscopy
measurements\cite{nuckollsStronglyCorrelatedChern2020, 
choiCorrelationdrivenTopologicalPhases2021} 
typically do not resolve a gapped state near $\nu=1$.
In the following, we present a scenario which would be consistent with the 
metallicity as well as the electron-like behaviour of R$_{xy}$ in the AHE phase 
seen in \prettyref{fig:Fig3}f and \prettyref{fig:Fig4}e. Our scenario relies on 
the fact that strong interactions can heavily deform the TBG bands, such that 
the $\Gamma$ point of the mini-Brillouin zone can be inverted, as reported by 
local spectroscopy 
measurements\cite{choiInteractiondrivenBandFlattening2021}. 

To model TBG, we employ a ten-band 
model\cite{poFaithfulTightbindingModels2019} that includes short-range Coulomb 
interactions\cite{choiElectronicCorrelationsTwisted2019}. First we show that this 
model, which uses a Hartree-Fock approximation, can capture the existence 
of the 
symmetry breaking cascade\cite{zondinerCascadePhaseTransitions2020,
  wongCascadeElectronicTransitions2020} and TBG band structure
deformations\cite{goodwinHartreeTheoryCalculations2020,
  ceaBandStructureInsulating2020, xieWeakfieldHallResistivity2020,
  choiInteractiondrivenBandFlattening2021} (see Supplementary Information for further discussion). 
Flavor-resolved mean-field band structures at $\nu$ = 0.81
are shown in \prettyref{fig:Fig4}f,g. Here, the symmetry breaking cascade
occurs well before $\nu$ reaches 1, with one of the spin-valley flavor bands 
being almost filled while the other three develop a gap and are shifted back 
to the vicinity of the charge neutrality point. Focusing on the three gapped 
bands, one can obtain a total Chern 
number of $C$ = $\pm$ 3 or $C$ = $\pm$1\cite{lianTwistedBilayerGraphene2021}, 
depending on 
the exact symmetry breaking mechanism, when the 
chemical potential is within the gap. For example, broken $\mathcal{T}$ 
symmetry naturally leads to $C$ = $\pm$3, and broken C$_2$ symmetry 
can give rise to $C$ = $\pm$1 phases\cite{nuckollsStronglyCorrelatedChern2020}.

Now consider the case displayed in \prettyref{fig:Fig4}g where the chemical 
potential for three flavors touches the bottom of the 
inverted electron pocket at the $\Gamma$ point. Despite being slightly electron 
doped, the three flavors still contribute to the anomalous Hall conductance. 
Since the Berry curvature of the upper band is small near the bottom of the 
inverted electron pocket, as shown in 
\prettyref{fig:Fig4}i, the Chern number remains approximately conserved. The 
consistently positive slope $\frac{dR_{xy}}{dB}$ of the Hall resistivity arises
from the electron-like bands of the barely filled flavors, while the 
apparent hole-like sign of $R_{xy}$ originates from the negative offset 
caused by the anomalous Hall effect. The experimentally observed hysteresis, in 
this scenario, would still be explained by orbital 
ferromagnetism\cite{zhuVoltageControlledMagneticReversal2020}. Due the metallic 
nature of the system, the Streda formula with $C=-1; \nu=1$ at $B=0$~T is only 
approximately satisfied. We note, that while the mean-field 
calculations presented here successfully capture a possible metallic AHE phase, 
other ground states with similar characteristics may also be 
possible\cite{kwanKekulSpiralOrder2021, 
lianTwistedBilayerGraphene2021, shavitTheoryCorrelatedInsulators2021, 
xieWeakfieldHallResistivity2020} and are hard to rule out based on our data. 
Finally similar scenario to the one proposed here can also explain AHE 
phases observed recently near $\nu=\pm2$\cite{tsengAnomalousHallEffect2022, linSpinorbitDrivenFerromagnetism2022}.
 
Our results show that the robustness to deviations of the twist angle from 
the magic angle divides TBG phases into two categories.  Superconductivity, 
cascade transitions, as well as the linear-in-T dependence of the 
resistivity are robust over a wide range of twist angles, spanning at least  
$0.8\degree \lesssim \theta \lesssim 1.23\degree$. Moreover, the cascade 
transitions near $\nu=\pm2$ appear to be a necessary prerequisite for the 
appearance of superconductivity, 
implying close relations between these two phases. In contrast, the 
correlated insulating and orbital ferromagnetic states
require a more subtle interplay of strong interactions, kinetic energy 
scales, and possible breaking of spatial symmetries. Due to this sensitivity, 
these phases appear in a more immediate vicinity of the magic angle, 
where the close competition between various phases can result in 
differing behaviour of devices with the same twist angle 
(e.g., see \prettyref{exfig:landaufan1p10}). 
This hierarchy of TBG phases will guide future theoretical frameworks aiming 
to explain the rich phenomenology of TBG and related structures.

\section{Methods}
\textbf{Device Fabrication:} The devices were fabricated using a ‘cut and stack’ method, 
in which graphene flakes were separated into two pieces using a sharp Platinum-Iridium tip; 
this prevents unwanted twisting and strain during tearing while allowing more control over 
the flake size and shape. First,
a thin hBN flake (10--30 nm) is picked up using a propylene carbonate film (PC)
previously placed on a polydimethylsiloxane (PDMS) stamp. Then the hBN is used
to pick up an exfoliated monolayer of WSe\textsubscript{2} (commercial source, HQ
graphene) before approaching the graphene. After picking up the first half of
the graphene flake, the transfer stage is rotated by approximately 1.1--1.3\degree
~(overshooting the target angle slightly), and then the second half is picked up, 
forming the twisted bilayer. Care was taken to approach and pick up
each stacking step slowly. In the last step, a thicker hBN (30--70 nm) is picked
up, and the whole stack is dropped on a predefined local gold back-gate at
150\degree ~C while the PC is released at 170\degree ~C. The PC is then cleaned
off with N-Methyl-2-Pyrrolidinone (NMP). The final geometry is defined by dry
etching with a CHF\textsubscript{3}/O\textsubscript{2} plasma and deposition of
ohmic edge contacts (Ti/Au, 5 nm/100 nm).

\textbf{Measurements:} All measurements were performed in a dilution refrigerator (Oxford Triton) with
a base temperature of $\sim$25 mK, using standard low-frequency lock-in
amplifier techniques. Unless otherwise specified, measurements are taken at
the base temperature. Frequencies of the lock-in amplifiers (Stanford
Research, models 830 and 865a) were kept in the range of 7--20 Hz in order to
measure the device's DC properties and the AC excitation was kept $<$5 nA (most 
measurements were taken at 0.5--1 nA to preserve the linearity of the system and 
avoid disturbing the fragile states at low temperatures). Each of the DC fridge 
lines pass through cold filters, including 4 Pi filters that filter out a range 
from $\sim$80 MHz to $>$10 GHz, as well as a two-pole RC low-pass filter.

\begin{figure}[hbt]
    \includegraphics[width=16cm]{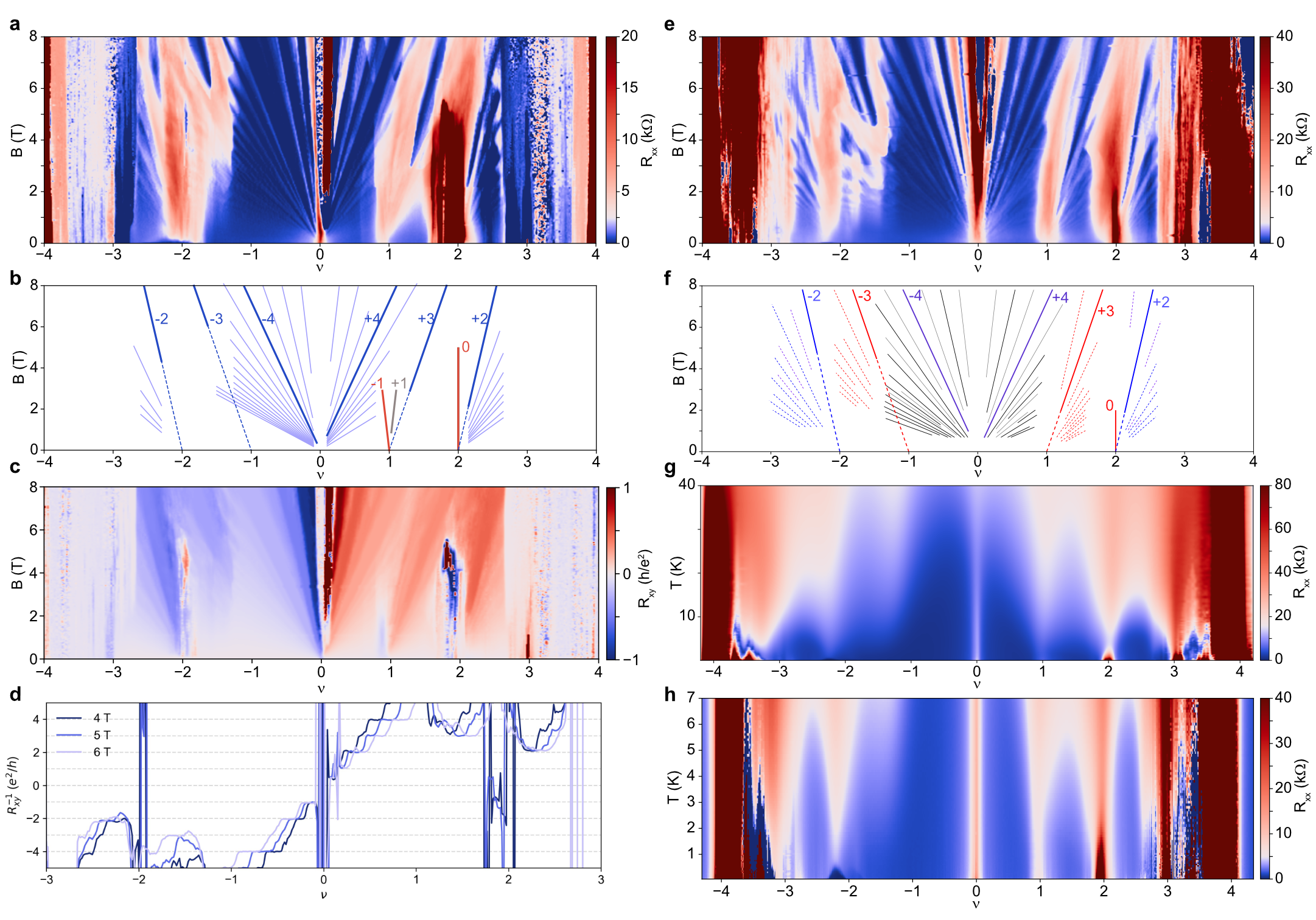}
    \centering
    \caption{{\bf R$_{xx}$ Landau Fans and comparison of
        D1, 1.10\degree and D4, 1.10\degree} {\bf a}, {\bf c}, R$_{xx}$ and
      R$_{xy}$ versus filling factor and magnetic field up to 8 T for D1. {\bf
        b}, Schematics showing correlated Chern insulators (bold blue lines) and
      zero-field competing Chern insulators (red lines) at the magic angle. {\bf d},
      Hall conductance showing well-quantized Chern insulators emanating from charge neutrality
      (broadest plateaus at $C$ = $\pm$4), $\nu$ = $\pm$1 ($C$ = $\pm$ 3) and $\nu$ = $\pm$2 ($C$ = $\pm$
      2). {\bf e, f}, Landau fan of D4 and schematic of visible Landau levels along with correlated Chern insulators (bold lines). Notice the fan around charge neutrality does not show the usual clear 4-fold degeneracy preference represented by a wider Landau level plateau in {\bf a}, and the fan emanating from $\nu$ = 1 persists to lower fields. This variance in magnetic field dependence reveals the sensitivity of the symmetry-broken states near 1.10\degree, particularly near $\nu$ = 1. {\bf g, h}, T-dependence of D4 at high and low temperatures, respectively. Contrast this with the T-dependence of D1 in \prettyref{fig:Fig3} and \prettyref{exfig:TandTcfigs}.}
\label{exfig:landaufan1p10}
\end{figure}

\begin{figure}[p]
    \includegraphics[width=16cm]{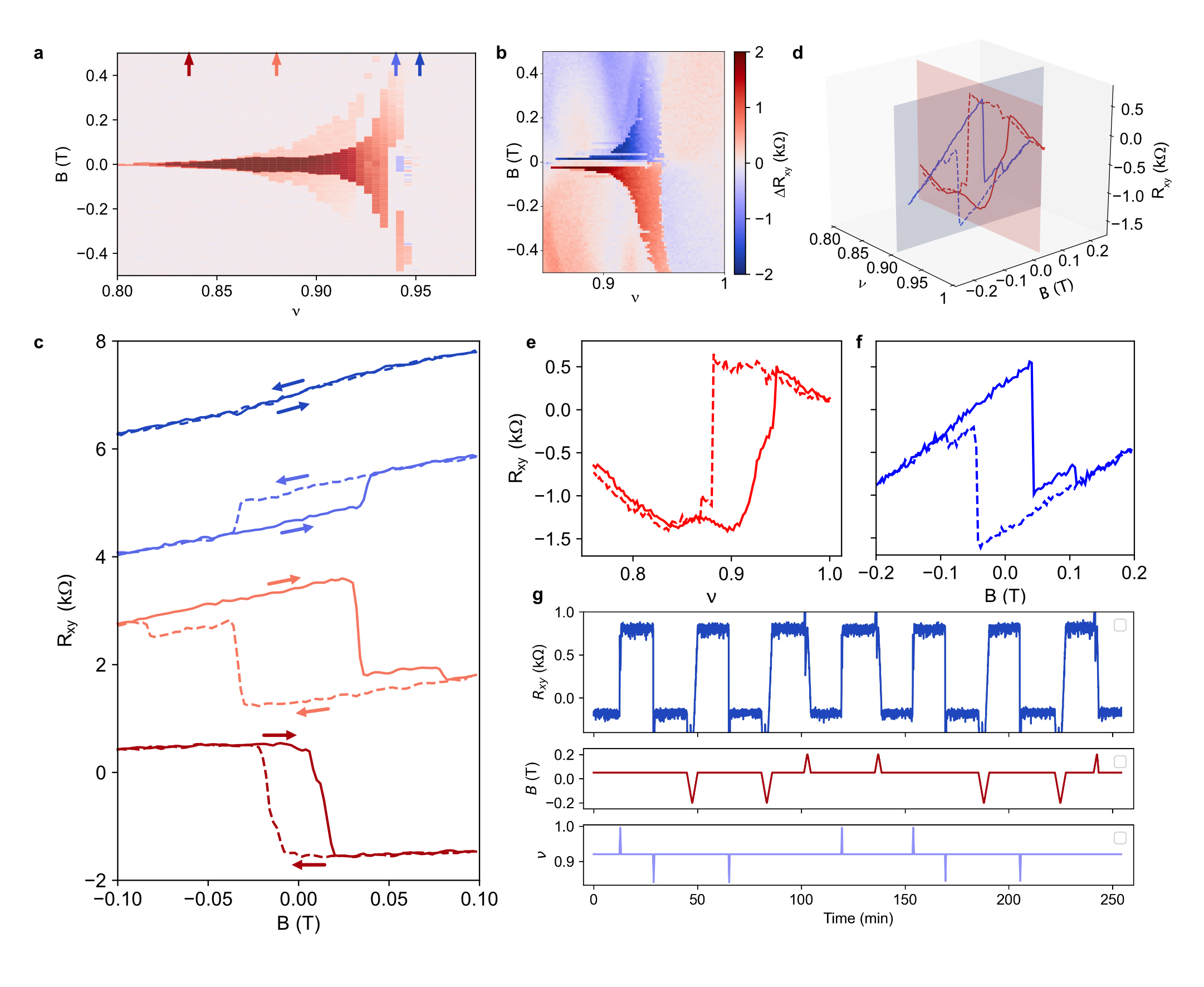}
    \centering
    \caption{{\bf Characterization of switching behaviour in D1 near $\nu=1$.} {\bf a}, $\Delta$R$_{xy}$ 
    (forward field sweep minus backward sweep) vs. B and
      $\nu$ around filling factor 1 measured at 3.5 K. {\bf b}, $\Delta$R$_{xy}$ 
      (forward sweep in $\nu$ minus backward sweep)
      between trace and retrace as the density is swept using gates, with gate
      sweeping taken at fixed magnetic field. {\bf c}, Hysteresis loops measured
      at filling factors marked by arrows in {\bf a}. {\bf d}--{\bf f},
      Hysteresis loops as a function of $\nu$ and B. {\bf d} shows the 3D perspective 
      of {\bf e} ($\nu$ sweep forward is solid, backward is dashed) and {\bf f} 
      (B sweep forward is solid, backward sweep is dashed). The density sweep in 
      {\bf e} was measured at 30 mT, after cycling to 200 mT to align the domains. {\bf g}, 
      Pulses of B and
      $\nu$ showing reproducible switching of magnetic state, with bit-like
      switching of R$_{xy}$.}
\label{exfig:hysteresis}
\end{figure}

\begin{figure}[ht]
    \includegraphics[width=14.5cm]{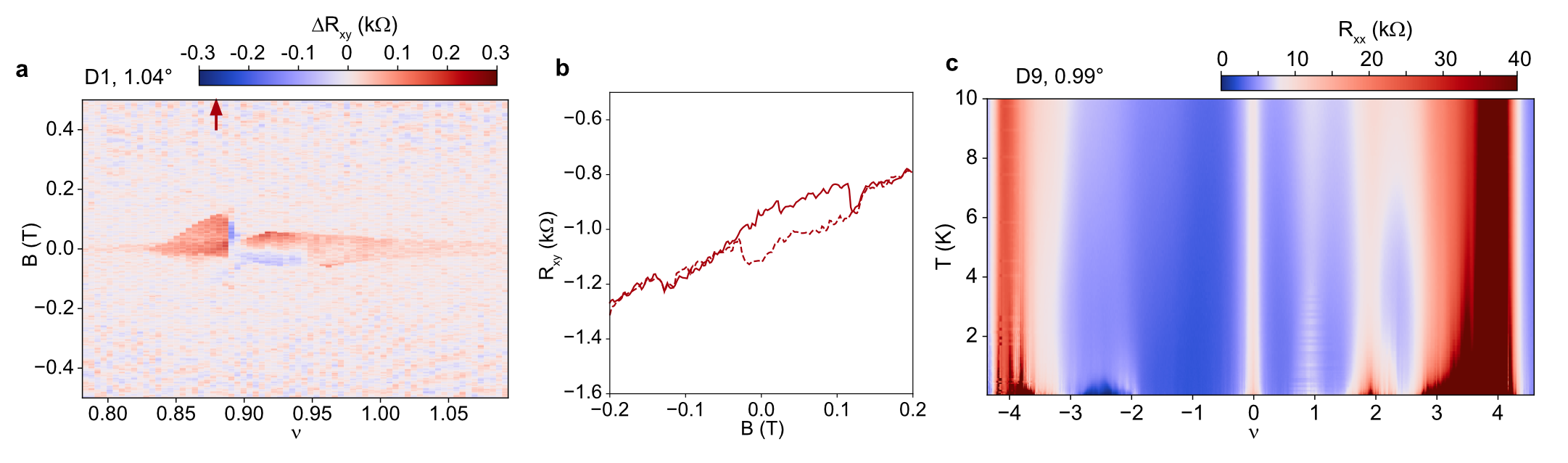}
    \centering
    \caption{{\bf $\nu=1$ anomalous hall effect slightly away from magic
        angle.} {\bf a}, $\Delta$R$_{xy}$ vs. B and $\nu$ around filling factor
      1 measured with twist angle 1.04\degree ~in device D1 at 1.5K. {\bf b},
      Line cut of R$_{xy}$ versus B at $\nu$ = 0.87 (red arrow) for the same 
      device. {\bf c}, Temperature dependence of another device D9, twist 
      angle 0.99\degree, showing evidence of switching behavior (similar to that 
      seen in \prettyref{fig:Fig3}e) and therefore possible ferromagnetism 
      near $\nu$ = 1. Bad contacts prevented us from measuring R$_{xy}$ data in this device.}
\label{exfig:AHEobs}
\end{figure}

\textbf{Similarities to hBN-encapsulated samples:} Here we briefly discuss similarities 
between devices studied here and hBN-encapsulated devices. While we showed previously that 
WSe\textsubscript{2} induces some amount of
spin-orbit coupling\cite{aroraSuperconductivityMetallicTwisted2020}, it does not change 
the magic angle significantly since both the most
prominent correlated insulating states at $\nu$ = +2 and highest T$_c$-superconductivity 
are observed near 1.1\degree. Moreover, the trends of metallic resistance peaks at
high temperature (\prettyref{exfig:TandTcfigs}) and the gaps between the flat and 
dispersive bands are consistent with hBN-encapsulated 
devices\cite{polshynLargeLinearintemperatureResistivity2019}. Note that the samples 
showing similar full-filling gaps likely have the same ratio between the tunneling 
energies on AA sites and AB sites (often represented as w$_{AA}$/w$_{AB}$ in the continuum
model)\cite{namLatticeRelaxationEnergy2017}. In this context, our main finding is 
that adding a WSe$_2$ layer decreases the amount of twist angle disorder on 
the hundreds-of-nanometer to $\micro$m scale in devices, 
as confirmed separately by STM measurements\cite{choiCorrelationdrivenTopologicalPhases2021}. 
This finding is also consistent with our four point
measurements for some of our best devices (see for example \prettyref{fig:Fig1}f). Another sign 
of decreased disorder is found in the well developed R$_{xy}$ plateaus at relatively 
low magnetic fields in our devices\cite{aroraSuperconductivityMetallicTwisted2020}
(see also \prettyref{exfig:landaufan1p10}).
This may explain our observation, in part, of superconductivity over a large angle range although the 
spin-orbit coupling may still play a role, particularly in the electron-side 
superconductivity away from the magic angle where hBN encapsulated TBG data is lacking and 
comparison is not possible.

\textbf{Disorder in TBG-WSe\textsubscript{2} devices:} In all TBG devices studied so far, 
it appears that there are significant device-to-device 
variations that are often associated with disorder. In addition to 
disorder that is intrinsic to graphene (such as charge disorder originating from residual 
polymers and other impurities, disordered edges, strain from wrinkles or bubbles, 
strain from the substrate or back gate), in TBG twist angle disorder is believed to
play an important role. As previously reported, it generates domains and gradual twist-angle 
shifts on length scales from 100 nm to a micron\cite{uriMappingTwistangleDisorder2020, 
choiCorrelationdrivenTopologicalPhases2021}. In 
this context, characterizing TBG disorder through transport measurements is somewhat 
more elusive as transport averages over device length scales (a few µm). It is important to 
emphasize that measurements of disorder that are typically used in a single layer of graphene 
and commonly detected through broadening of of charge-neutrality peaks in 
longitudinal or Hall resistance (full width at half max---FWHM---of R$_{xx}$), do not correlate 
well with superconductivity or other features that may point towards disorder in the TBG samples 
(\prettyref{fig:Fig1}g). 

\begin{figure}[hb]
    \includegraphics[width=15.5cm]{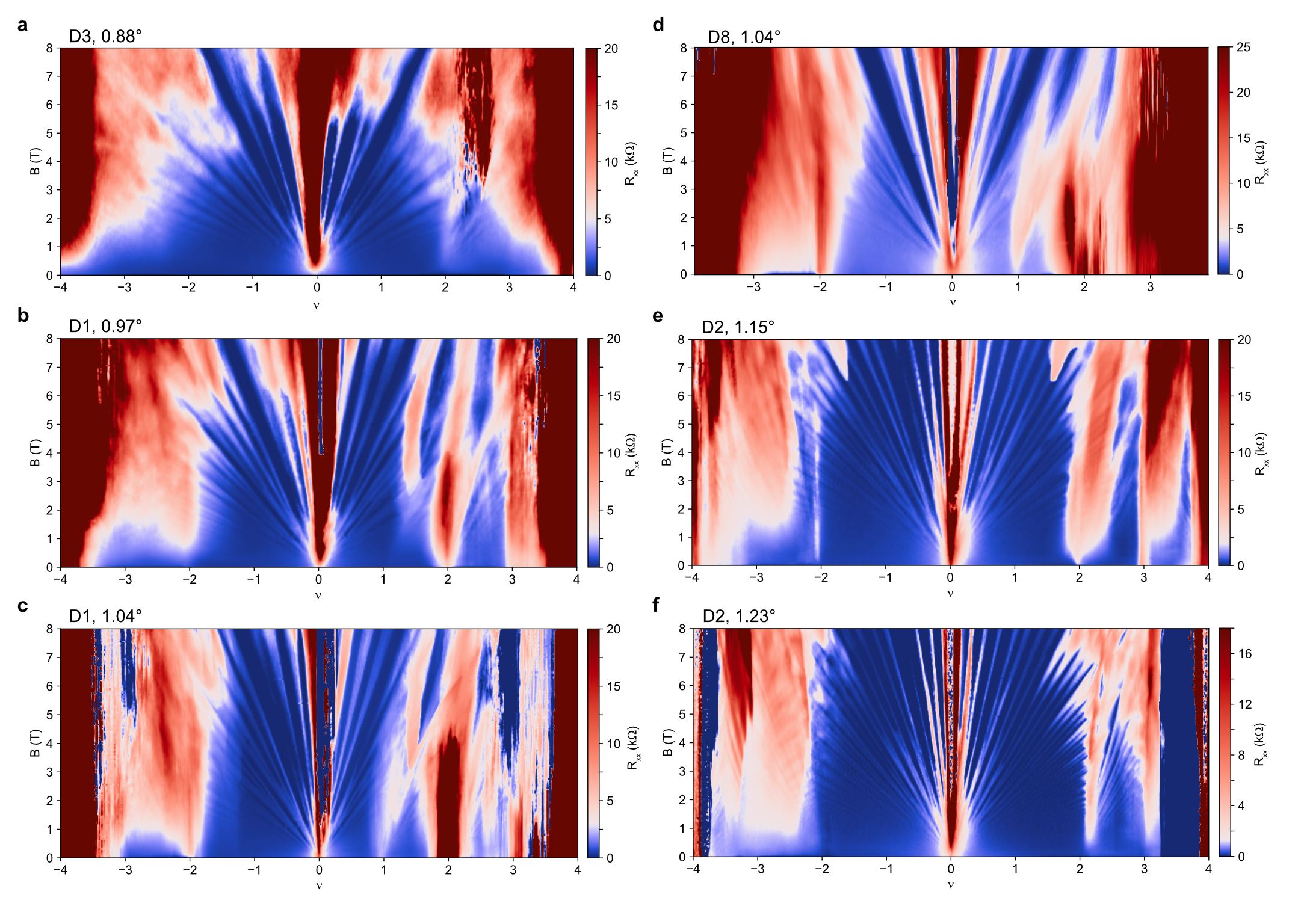}
    \centering
    \caption{{\bf Fan diagrams.} R$_{xx}$ measured as a function of magnetic
      field B and moir\' e filling factor $\nu$ for a collection of twist angles.
      One can see high-resistance states near $\nu$ = 2 just near the magic
      angle (0.97\degree ~to 1.15\degree) and near $\nu$ = 1 only in a very
      small range near 1.10\degree ~(by 1.04\degree, it is already disappearing,
      and it is gone at 1.15\degree ~here). On the edge of the magic angle,
      (such as at 0.97\degree) the $\nu$ = 2 state develops with a small
      magnetic field. The hole-side $\nu$ = --2 insulator is always smaller than
      the electron-side $\nu$ = 2 insulator, and it is not fully developed at 0
      magnetic field in these diagrams. The noisy features commonly seen for
      $|\nu|>$ 3 and occasionally for other values of $\nu$ at high field are
      likely due to contact/geometry effects near the insulating states.}
\label{exfig:multifandiagrams}
\end{figure}

Occasionally, we see evidence in transport of the presence of domains with slightly different 
twist angles (on the scale of the contact separation, or slightly smaller, a few hundred nm to a few µm),
resulting in multiple resistance features near $\nu$ = 2 (\prettyref{exfig:multifandiagrams}d), and 
a Landau fan diagram that is not as distinct as other devices. 
Other devices show variation in the twist angle over a range of contact pairs (length scales 2-20 µm, 
see \prettyref{extab:devicetable}) but still show less disorder in fan diagrams and strong correlations 
for each individual pair of contacts, owing likely to large domains. 

In the least disordered devices (in the context of transport measurements), 
the global twist-angle deviation between different 
contact pairs is below the detectable limit. For instance,
\prettyref{fig:Fig1}f shows a device with longitudinal resistance
traces that almost exactly match over four pairs of contacts. This device, made with
Au top and back gates and a monolayer of WSe\textsubscript{2} on top of the TBG, also shows
uniform resistance and superconducting features over the applicable electric
displacement field $D$ range (\prettyref{fig:Fig1}f, inset). This suggests a
low disorder as probed by $D$-field asymmetry that occurs in some
previous hBN encapsulated devices\cite{yankowitzTuningSuperconductivityTwisted2019}. 
The $D$-invariant superconductivity also implies that the substrate asymmetry introduced 
by WSe\textsubscript{2} is minimal, at least for twist angle of 0.88\degree. Moreover, we note that 
different contact pairs in this study are checked in a four-point measurement configuration (the standard used for 
measuring correlated states such as SC) and cross-checked with temperature dependence and magnetic 
field dependence for extraneous features, a process we find is more sensitive to disorder compared 
to the verification of twist angle disorder using two point measurements performed previously on 
hBN encapsulated devices. Further work (both in theory and experiment) is needed for more precise 
characterization of the role of disorder in TBG devices. 


%

\noindent {\bf Acknowledgments:} We acknowledge discussions with Cyprian
Lewandowski, Jason Alicea, and Alex Thomson. {\bf Funding:} This work has been
primarily supported by the DOE-QIS program (DE-SC0019166) and NSF-CAREER
(DMR-1753306). S.N-P. acknowledges support from the Sloan Foundation. G.R. and
S.N.-P. also acknowledge the support of the Institute for Quantum Information and
Matter, an NSF Physics Frontiers Center with support of the Gordon and Betty
Moore Foundation through Grant GBMF1250; Y.P. acknowledges support from the
startup fund from California State University, Northridge. F.v.O. is supported
by Deutsche Forschungsgemeinschaft within CRC 183 (project C02) as well as the project TWISTGRAPH.

\noindent {\bf Author Contribution:} R.P. and Y.Z. performed the measurements,
fabricated devices, and analyzed the data. H.P., Y.C., and H.K. helped with device fabrication and data analysis. Y. P. developed a theoretical model and performed 
model calculations in close collaboration with F.v.O. and G.R. K.W., and T.T. provides hBN 
crystals. S.N-P. supervised the project. R.P, Y.Z. Y.P., F.v.O. G.R. and S.N-P. wrote the
manuscript with the input of other authors. 

\noindent{\bf Competing interests:} The authors declare no competing interests.

\noindent {\bf Data availability:} The data that support the findings of this
study are available from the corresponding authors on reasonable
request. 

\noindent {\bf Code availability:} The code that support the findings of this
study are available from the corresponding authors on reasonable request.

\clearpage

\onecolumngrid  %
\beginsupplement
\section{Supplementary Information: Hierarchy of\\ Symmetry Breaking Correlated Phases\\ in Twisted Bilayer Graphene}

\centering
\noindent{\small Robert Polski, Yiran Zhang, Yang Peng,
  Harpreet Singh Arora, Youngjoon Choi, Hyunjin Kim,
  Kenji Watanabe, Takashi Taniguchi, Gil Refael, Felix von Oppen, and
  Stevan Nadj-Perge}
\flushleft

\subsection{Experiment: Supplementary figures and the overview table}

\FloatBarrier

\FloatBarrier

\FloatBarrier

\FloatBarrier

\FloatBarrier

\FloatBarrier

\FloatBarrier

\FloatBarrier

\FloatBarrier

\clearpage

\begin{table}[p]
  \centering
  \caption{{\bf Device table.} The superconducting transition temperatures (SC T$_c$), $\nu$ = 2 correlated
    insulator (CI) gap, and $\nu$ = 1 correlated state parameters measured for the
    devices used to plot the phase diagram in \prettyref{fig:Fig1}b,c  sorted by twist angle.}
\begin{tabular}{cccccc}
 \toprule
 Twist angle ($\pm$0.02) & Device & Hole SC T\textsubscript{c} (K) & Electron SC T\textsubscript{c} (K) & $\nu$=2 CI (meV) & $\nu$=1 state \\
 \colrule
 0.79  & D6 (M20) & N/A   & 0.382 & N/A   & N/A \\
 0.80   & D7 (W5) & N/A   & 0.54  & N/A   & N/A \\
 0.83  & D6 (M20) & N/A   & 0.702 & High-T peak & High-T peak \\
 0.87  & D6 (M20) & N/A   & 0.626 & High-T peak & High-T peak \\
 0.88  & D3 (S3) & 0.129 & 0.652 & High-T peak & High-T peak \\
 0.95 & D9 (M30) & 0.339 & 0.486 & $\Delta$=0.186 & High-T peak \\
 0.97  & D5 (M08) & 0.742 & 0.089 & $\Delta$=0.68 & High-T peak \\
 0.97  & D1 (S13) & 0.398 & 0.352 & $\Delta$=0.09 & High-T peak \\
 0.99 & D9 (M30) & 0.429 & N/A & $\Delta$=0.11 & Low-T peak, switching \\
 1.04  & D1 (S13) & 0.798 & N/A   & $\Delta$=0.89 & Low-T peak, hysteresis \\
 1.04  & D8 (M12) & 0.4   & 0.098 & $\Delta$=0.26 & Low-T peak \\
 1.10   & D1 (S13) & 1.59  & 0.083 & $\Delta$=0.84 & FM to 7K \\
 1.10   & D4 (W3) & 0.443 & N/A   & $\Delta$=0.27 & Low-T peak \\
 1.15  & D2 (S12) & 0.267 & 0.155 & $\Delta\sim$0.17 to SC & High-T peak \\
 1.23  & D2 (S12) & 0.317 & 0.128 & Disappearing   & N/A \\
 \botrule
 \label{extab:devicetable}
\end{tabular}
\end{table}

\FloatBarrier
\clearpage

\subsection{Pomeranchuk effect transition at $|\nu| \approx$ 1}
\setlength{\parskip}{15pt}

The phase transition at $|\nu| \approx$  1 has been previously explained as a 
transition from a fully flavor-symmetric Fermi liquid near charge-neutrality to a phase characterized 
by local free moments (approximately 1 per moire site) above $|\nu| \approx 1$ in 
analogy to the Pomeranchuk effect in He$^3$\cite{saitoIsospinPomeranchukEffect2021}. 
The free energy contribution due to the entropy of free spins is comparable with the 
contribution due to magnetic fields, so the phase transition line changes easily 
throughout the phase space of $\nu$, B, and T. When considering the grand canonical 
potential approach, allowing the filling factor $\nu$ to change, and setting 
the magnetic field to zero, the phase transition line follows\cite{rozenEntropicEvidencePomeranchuk2021}:

\begin{equation}
    \nu = \frac{1}{\Delta \mu}\left[ -\frac{1}{2}\Delta\gamma T^2 - \textrm{ln}(2)T + \Delta\epsilon \right]
    \label{eq:phasetransition}
\end{equation}

This quadratic equation fits to the $\nu$-T curves in Fig. \ref{fig:Fig2} of the main text and
provides the parameters plotted in panels \prettyref{fig:Fig2}c and \prettyref{fig:Fig2}d. $\Delta \mu$ 
represents the change in chemical potential between the Fermi liquid and isospin local 
moment phases, $\Delta \gamma$ is the change in the electronic specific heat (which is 
negative since the local moment phase has a smaller density of states), and $\Delta \epsilon$ 
is a free parameter related to a reference energies of the phases. A strong carrier 
density reset, and therefore strong correlations, are related to a strong $\Delta \mu$. Only the 
electron-side transitions were mapped in this study because they were detectable in a wider range of twist angles.

\subsection{Theory: Ten-band model with onsite interaction}

To model the interaction-induced symmetry breaking in twisted bilayer graphene (TBG), 
we take the non-interacting ten-band model for each flavor (spin, valley), and add 
interactions on top of it. Note that the correlation effects at charge neutrality for 
such system were modeled in a previous work,\cite{choiElectronicCorrelationsTwisted2019} 
based on a single-flavor ten-band model with onsite interactions.  
In this work, we consider the four flavors (two spins, two valleys) altogether.

\subsubsection{Single flavor}
\begin{figure}
\centering
\includegraphics[width=0.8\textwidth]{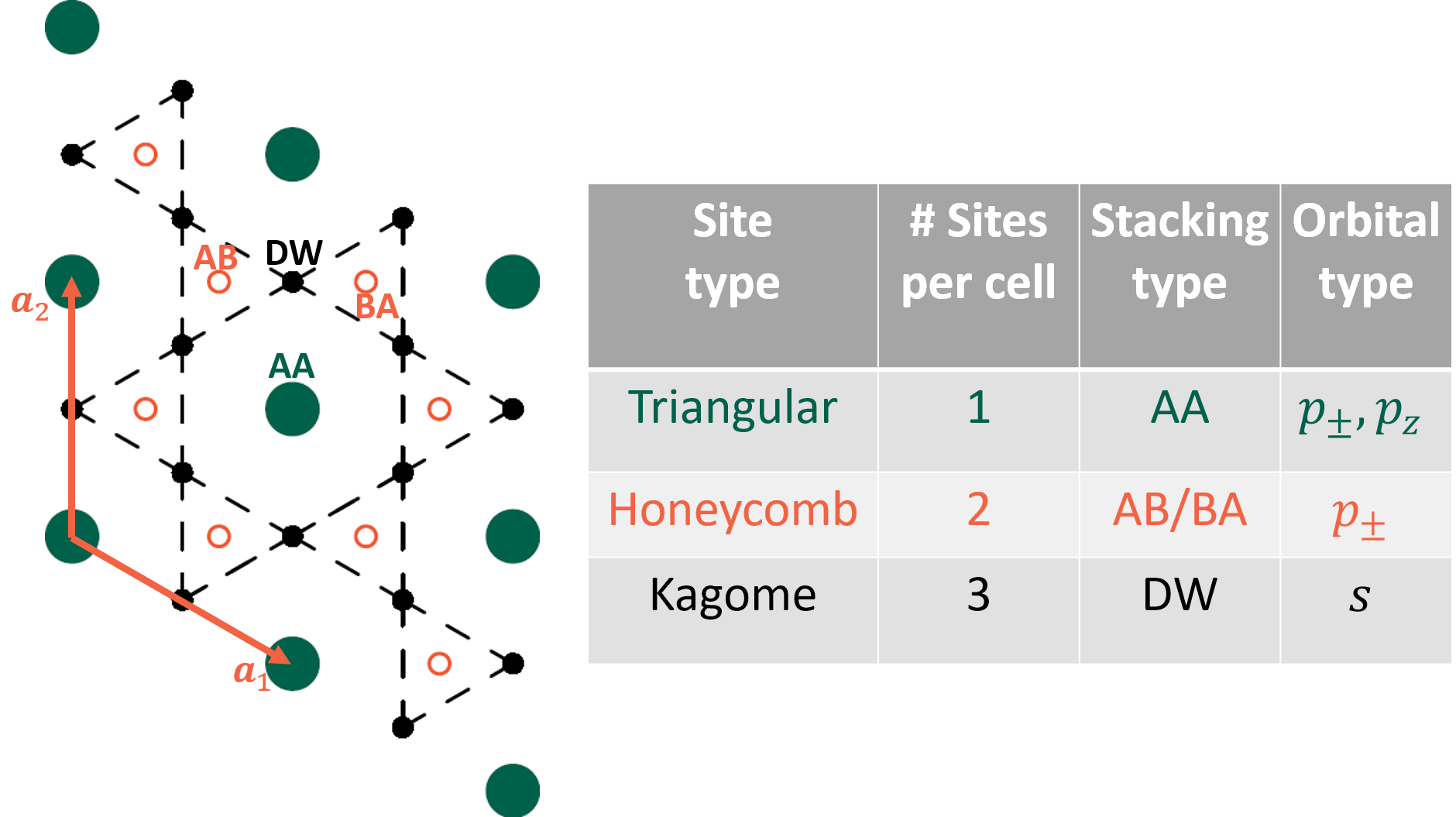}
\caption{\label{exfig:lattice} {\bf Lattice and orbitals for the ten-band model per flavor.} 
The green solid circles denote the triangular sites, corresponding to the AA stacking regions,
with $p_z$, $p_+$, and $p_-$ orbitals. 
The red empty circles indicate the honeycomb sites, either type A or type B,
corresponding to the AB/BA stacking regions,
with $p_+$ and $p_-$ orbitals on each of them.
The black solid circles stand for the three types of kagome sites, which are the domain wall regions
between the AB/BA regions. On each of these kagome sites, there is an $s$ orbital.
}
\end{figure}

Let us recall that the ten-band model for a single flavor of electrons in TBG 
is realized on a triangular lattice with basis vectors 
$\boldsymbol{a}_1 = (\sqrt{3}/2,-1/2)$ and $\boldsymbol{a}_2=(0,1)$. 
We write the Bravais lattice sites as 
$\boldsymbol{r}=r_1\boldsymbol{a}_1+r_2\boldsymbol{a}_2$ or 
simply as $\boldsymbol{r}=(r_1,r_2)$, where $r_{1,2}\in \mathbb{Z}$.
Within each unit cell, there are ten orbitals which are 
distributed on three different sites.
Explicitly, there are three orbitals, $p_z$, $p_+$, and $p_-$, 
on every triangular lattice site.
Each of the three kagome sites within a unit cell hosts an $s$ orbital.  
Finally, both A and B sublattices of the honeycomb sites 
have $p_+$ and $p_-$ orbitals. 
These are indicated and summarized in \prettyref{exfig:lattice}.

Note that these ten orbitals should be regarded as Wannier orbitals which are 
able to faithfully produce the ten Bloch states closest to charge neutrality, 
including the two flat bands, while satisfying the same symmetry constraints 
of the original TBG system. 

Throughout this work, we order the ten orbitals as 
\begin{equation}
c_{\boldsymbol{r}} =\left( \tau_{z,\boldsymbol{r}},\tau_{+,\boldsymbol{r}},
\tau_{-,\boldsymbol{r}},\kappa_{1,\boldsymbol{r}},\kappa_{2,\boldsymbol{r}},\kappa_{2,\boldsymbol{r}},
\eta_{A+,\boldsymbol{r}},\eta_{A-,\boldsymbol{r}},\eta_{B+,\boldsymbol{r}},
\eta_{B-,\boldsymbol{r}} \right)^T,
\end{equation}
where $\tau$, $\kappa$, and $\eta$  denote operators on the triangular, kagome, and honeycomb sites respectively.

For the Hamiltonian of a single flavor (e.g. say $H_{K\uparrow}$,
spin up in valley $K$), we took the same parameters as the ones in Ref.\cite{choiElectronicCorrelationsTwisted2019}

\subsubsection{Including spin and valley}
The full non-interacting Hamiltonian including all flavors (spins and valleys) can be obtained in the following way. 
First, noting that the Hamiltonian does not explicitly depend on spin, in other words the Hamiltonians in both valleys have spin $SU(2)$ symmetry, 
we can, for each valley, simply duplicate the single-spin Hamiltonian to obtain the Hamiltonian for the two spin components. 
For example, the Hamiltonian for valley $K$ is $H_K = H _{K\uparrow}\oplus H_{K\downarrow} $, with $H_{K\downarrow} = H_{K\uparrow}$.

Second, just as in monolayer graphene, the Hamiltonians at opposite valleys are related by the spinless time-reversal operation $\mathcal{T}$. Thus, we have the full Hamiltonian
\begin{equation}
    H = H_K \oplus \mathcal{T}H_K\mathcal{T}^{-1}.
\end{equation}

\subsubsection{Including interactions}
We start from the general electron-electron interaction 
\begin{align}
     H_{int}&=\frac{1}{2}\sum_{\eta\eta'}\sum_{\sigma\sigma'}\int d\boldsymbol{r}d\boldsymbol{r}'\,\psi_{\eta\sigma}^{\dagger}(\boldsymbol{r})\psi_{\eta'\sigma'}^{\dagger}(\boldsymbol{r}')V(\boldsymbol{r}-\boldsymbol{r}')\psi_{\eta'\sigma'}(\boldsymbol{r})\psi_{\eta\sigma}(\boldsymbol{r}).
\end{align}
By projecting the interaction onto the subspace spanned by the ten (Wannier) orbitals of the ten-band model, and keeping only the onsite interaction on the triangular lattice sites, we obtained
\begin{align}
    H_{int} \simeq \sum_{\boldsymbol{r}}\sum_{\eta\eta'\sigma\sigma'}\sum_{m_1,m_2,m_3,m_4} 
    V^{\eta\eta'}_{m_1m_2m_3m_4} (\tau^{(\eta\sigma)}_{m_1,\boldsymbol{r}})^\dagger
    (\tau_{m_2,\boldsymbol{r}}^{(\eta'\sigma')})^\dagger
    \tau_{m_3,\boldsymbol{r}}^{(\eta'\sigma')}
    \tau_{m_4,\boldsymbol{r}}^{(\eta\sigma)},
\end{align}
where $\eta,\eta'$ are the valley indices, $\sigma,\sigma'$ denote the spin, and  $m_1, m_2,\ldots =z,+,-$ label the three types of p orbitals on the triangular sites. The interaction matrix elements
\begin{equation}
   V^{\eta\eta'}_{m_1m_2m_3m_4}  = \int d\boldsymbol{r} d\boldsymbol{r}' V(\boldsymbol{r}-\boldsymbol{r}')
   \phi_{m_{1}}^{(\eta)*}(\boldsymbol{r})\phi_{m_{2}}^{(\eta')*}(\boldsymbol{r}')\phi_{m_{3}}^{(\eta')}(\boldsymbol{r}')\phi_{m_{4}}^{(\eta)}(\boldsymbol{r}),
\end{equation}
where $\phi_{m}^{(\eta)} (\boldsymbol{r})$ is the wave function for the $m$-type p orbital for valley $\eta$. Here, we have assumed that these wave functions do not depend on spin, but they do depend on the valley indices according to 
\begin{equation}
    \phi_{\pm}^{(\eta)}(\boldsymbol{r}) = (\phi_{\mp}^{(-\eta)}(\boldsymbol{r}))^* , \quad 
    \phi_{z}^{(\eta)}(\boldsymbol{r}) = (\phi_{z}^{(-\eta)}(\boldsymbol{r}))^*.
\end{equation}

Because of the rotational symmetry of the p orbitals, the intra-valley interaction matrix elements can be parametrized in terms of two independent parameters $U$ and $J$, as
\begin{equation}
    V^{++}_{m_1m_2m_3m_4}=U\delta_{m_1 m_4}\delta_{m_2 m_3 }+J\left[\delta_{m_{1}m_{3}}\delta_{m_{2}m_{4}}+(-1)^{m_{1}+m_{4}}\delta_{-m_{1}m_{2}}\delta_{-m_{4}m_{3}}\right] = V^{--}_{m_1m_2m_3m_4},
\end{equation}
where for $p_z$ orbitals $m=0$. On the other hand, the inter-valley interaction matrix elements can be parametrized by the same two parameters, as
\begin{equation}
    V^{+-}_{m_1m_2m_3m_4} = U\delta_{m_{1}m_{4}}\delta_{m_{2}m_{3}}+J\left[\delta_{-m_{1}m_{3}}\delta_{m_{2},-m_{4}}+(-1)^{m_{1}+m_{4}}\delta_{m_{1}m_{2}}\delta_{m_{4}m_{3}}\right] = V^{-+}_{m_1m_2m_3m_4}.
\end{equation}

In the current work, we set $J=0$ for simplicity in the following.
The interaction Hamiltonian in momentum space can be written as
\begin{align}
   H_{int} &=\frac{U}{2V}\sum_{\eta}\sum_{\sigma\sigma'}\sum_{m_{1}m_{2}}\sum_{\boldsymbol{k}_{1}\boldsymbol{k}_{2}\boldsymbol{q}}\left\{ \tau_{m_{1}}^{(\eta\sigma)\dagger}(\boldsymbol{k}_{1})\tau_{m_{2}}^{(\eta\sigma')\dagger}(\boldsymbol{k}_{2})\tau_{m_{2}}^{(\eta\sigma')}(\boldsymbol{k}_{2}+\boldsymbol{q})\tau_{m_{1}}^{(\eta\sigma)}(\boldsymbol{k}_{1}-\boldsymbol{q})\right. \nonumber \\
   & \left.
  +\tau_{m_{1}}^{(\eta\sigma)\dagger}(\boldsymbol{k}_{1})\tau_{m_{2}}^{(-\eta\sigma')\dagger}(\boldsymbol{k}_{2})\tau_{m_{2}}^{(-\eta\sigma')}(\boldsymbol{k}_{2}+\boldsymbol{q})\tau_{m_{1}}^{(\eta\sigma)}(\boldsymbol{k}_{1}-\boldsymbol{q})\right\},  
\end{align}
where $V$ is the volume of the system size.
Now we apply the Hartree-Fock approximiation. Up to a constant, we obtain the mean-field Hamiltonian
\begin{equation}
    H_{int}^{HF}=\sum_{\eta\sigma}\sum_{\boldsymbol{k}}\sum_{m_{1}m_{2}}\tau_{m_{1}}^{(\eta\sigma)\dagger}(\boldsymbol{k})\tilde{W}_{m_{1}m_{2}}^{\eta,\sigma}\tau_{m_{2}}^{(\eta\sigma)}(\boldsymbol{k}),
\end{equation}
with
\begin{align}
    \tilde{W}_{m_{1}m_{2}}^{\eta,\sigma} &=U\left(\sum_{\eta'\sigma'}\mathrm{Tr} P^{\eta',\sigma'}\delta_{m_{1}m_{2}}-P_{m_{1}m_{2}}^{\eta,\sigma}\right)
    \label{eq:W_mat}
\end{align}
and
\begin{equation}
    P_{m_{1}m_{2}}^{\eta,\sigma}=\frac{1}{V}\sum_{\boldsymbol{k'}}\langle\tau_{m_{2}}^{(\eta\sigma)\dagger}(\boldsymbol{k}')\tau_{m_{1}}^{(\eta\sigma)}(\boldsymbol{k}')\rangle
\end{equation}
It can be seen that the two terms in Eq.~(\ref{eq:W_mat}) can be identified as Hartree and Fock contributions, respectively.

In addition to the interaction, we assume there is a background positive charge distribution such that at charge neutrality the system is neutral. This potential can be modeled as
\begin{equation}
    H_{bg}=-U\sum_{\boldsymbol{k}}\sum_{\eta'\sigma'}\mathrm{Tr} P_{CN}^{\eta,\sigma}
    \sum_{m\eta\sigma}\tau_{m}^{(\eta\sigma)\dagger}(\boldsymbol{k})
    \tau_{m}^{(\eta\sigma)}(\boldsymbol{k}),
\end{equation}
where $P_{CN}^{\eta,\sigma}$ is computed from the non-interacting Hamiltonian
at charge neutrality. 
Thus, the total mean-field Hamiltonian can be written as
\begin{equation}
    H = H_{nonint} + \sum_{\eta\sigma}\sum_{\boldsymbol{k}}\sum_{m_{1}m_{2}}\tau_{m_{1}}^{(\eta\sigma)\dagger}(\boldsymbol{k})\tilde{W}_{m_{1}m_{2}}^{\eta,\sigma}\tau_{m_{2}}^{(\eta\sigma)}(\boldsymbol{k})
\end{equation}
with $H_{nonint}$ the non-interacting ten-band model and
\begin{align}
W_{m_{1}m_{2}}^{\eta,\sigma} & =\left(\sum_{\eta'\sigma'}U_{H}\mathrm{Tr}(P^{\eta',\sigma'}-P_{CN}^{\eta',\sigma'})\delta_{m_{1}m_{2}}-U_{F}P_{m_{1}m_{2}}^{\eta,\sigma}\right),
\end{align}
which has to be determined self-consistently by diagonalizing $H$.

Note that here we have generalized the mean-field Hamiltonian by choosing two independent parameters $U_H$ and $U_F$ for the contributions due to the Hartree and Fock terms. These two parameters can be regarded as phenomenological parameters, similar to the Landau parameters in the Fermi liquid theory. 

The energy per electron can be computed as
\begin{equation}
    E = \sum_{\eta\sigma}\left\{\frac{1}{V}\sum_{\boldsymbol{k}}\epsilon_{\eta,\sigma}(\boldsymbol{k}) - \frac{1}{2}\mathrm{Tr}\left[\tilde{W}^{\eta,\sigma}P^{\eta,\sigma}\right] \right\},
\end{equation}
with 
\begin{equation}
    \tilde{W}^{\eta,\sigma} = U_H\sum_{\eta'\sigma'}\mathrm{Tr} P^{\eta',\sigma'}\delta_{m_{1}m_{2}}-U_F P_{m_{1}m_{2}}^{\eta,\sigma},
\end{equation}
which generalizes Eq.~(\ref{eq:W_mat}) with independent Hartree and Fock contributions.

\subsection{Competing states near $\nu$ = 1}

\begin{figure}
    \centering
    \includegraphics[width=0.7\textwidth]{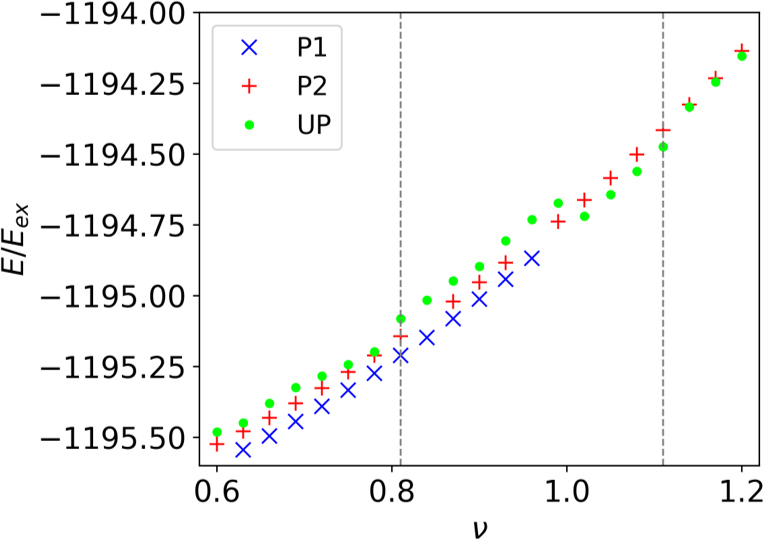}
    \caption{{\bf Ground state energies with filling.} Ground state energy per electron for different competing phases, near $\nu$ = 1. The two dashed vertical lines in indicate where the filling $\nu$ is 0.81 and 1.11, see also \prettyref{exfig:C3_broken} and \prettyref{exfig:nu_1_states}.}
    \label{exfig:energies}
\end{figure}

In this section, we show several possible phases that may exist near $\nu$ = 1, based on the interacting 
ten-band model. In \prettyref{exfig:energies}, we show the ground state energy per electron for different 
phases. 

Here, P1 and P2 states have maximal flavor polarization. Namely, the resulting state will be close to 
the configuration where each flavor, labeled by $\eta$,$\sigma$, is filled (depleted) sequentially above (below) 
charge neutrality. This is done by first adding this constraint to the self-consistent iterations until 
convergence is met. Then one relaxes the constraint and performs one additional iteration. The resulting state 
may not have the complete polarization on the maximally polarized flavor.

The UP state can also be a cascaded state.
In this state, particularly, there are $[|\nu|]$ filled (depleted) flavors if $\nu>0$ ($\nu<0$), and the rest of the flavors 
will be filled (depleted) equally from charge neutrality (Here $[\cdot]$ denotes the truncated integer part of 
$\nu$). In other words, the fractionally filled flavors will have equal filling fractions.

Energetically, from the 
\prettyref{exfig:energies}a, we see that the system prefers to be in the 
gapped P1 state, although the gapless P2 state is close in energy when $\nu <1$, 
and it prefers to be in the UP state when $\nu > 1$. This coincide with the 
cascade picture previously proposed\cite{zondinerCascadePhaseTransitions2020}.

It is worth mentioning that due to the onsite nature of the interaction, there exist degeneracies when we consider different permutations of the filling fractions for different flavors. This means one can arbitrarily 
choose the flavors which are maximally polarized.

\begin{figure}
    \centering
    \includegraphics[width=0.9\textwidth]{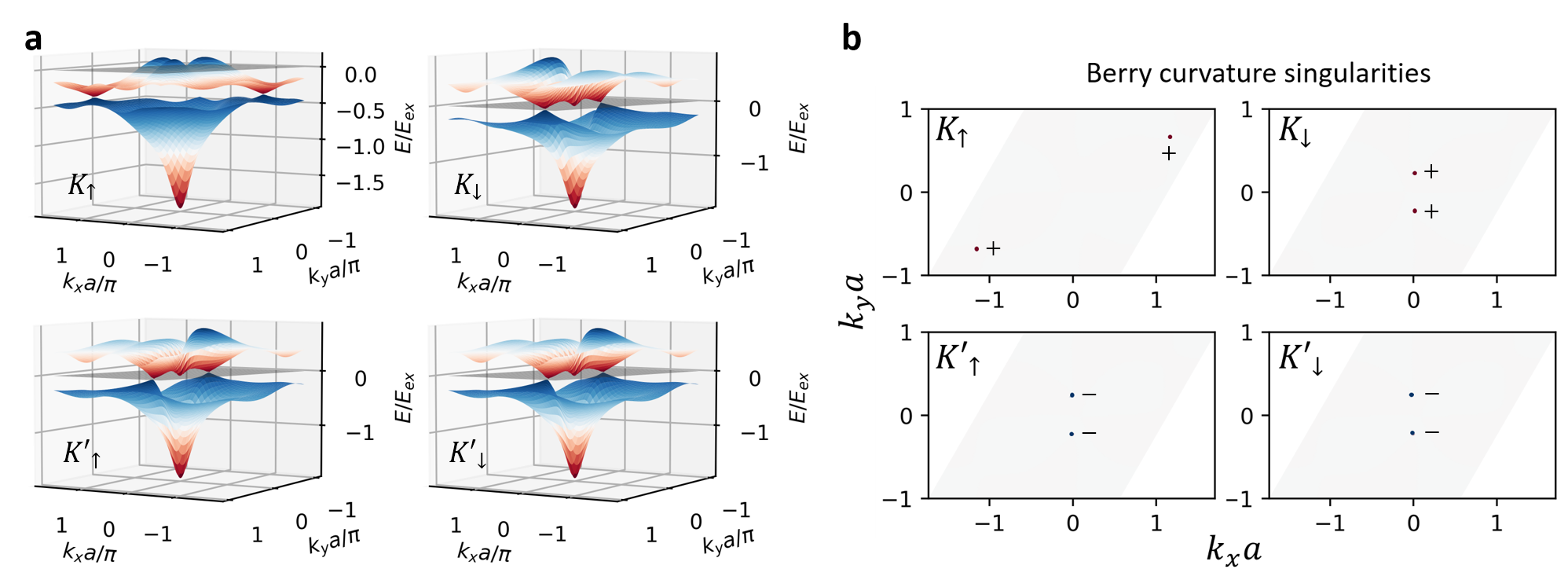}
    \caption{{\bf Theory of the gapless (P2) state.} {\bf a}, Mean-field band structure
    obtained from the ten-band model for the case of broken $C_3\mathcal{T}$ symmetry at $\nu$ =  0.81.
    {\bf b}, Singularities of the Berry curvature $\Omega_{k_x,k_y}$ for the conduction flat band are located at the gapless points,  as in {\bf a}. The $+/-$ signs denotes the signs of the singularities, which are the same as the sign of the gapless Dirac nodes for each flavor.}
    \label{exfig:C3_broken}
\end{figure}

\begin{figure}
    \centering
    \includegraphics[width=0.8\textwidth]{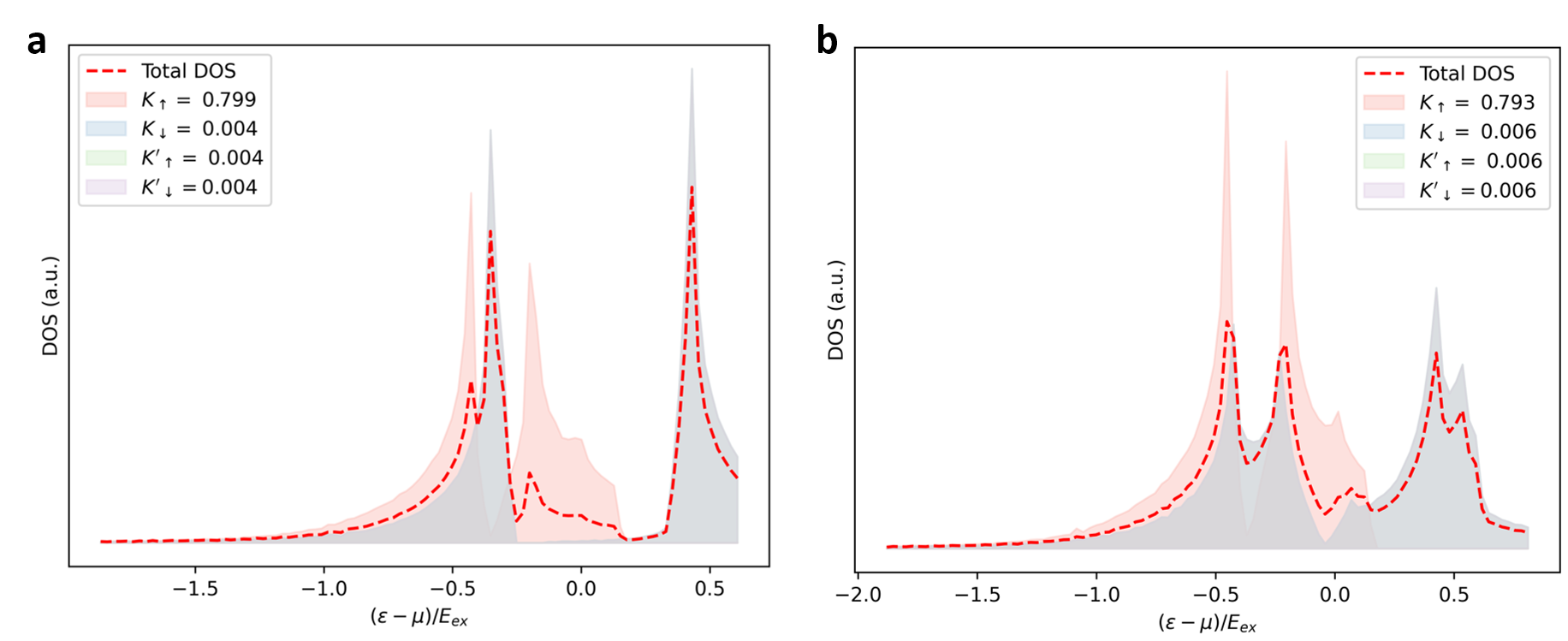}
    \caption{{\bf Total and flavor-resolved density of states, at $\nu$ = 0.81.} {\bf a}, the $C_{2}\mathcal{T}$-broken (P1) state. {\bf b}, The $C_3$-broken (P2) state. The fill colors each represent a spin/valley flavor, the gray color results from multiple flavors overlapping, and the red line is the normalized total density of states. The legends indicate the individual filling factor of each flavor.}
    \label{exfig:DOS_C2T_C3}
\end{figure}

The P1 state has $C_2\mathcal{T}$-broken gaps in the three flavors near charge neutrality. 
The band structures, as well as the Berry curvature, for this state at $\nu$ =  0.81 are shown 
in \prettyref{fig:Fig4} of the main text. 
The P2 state is gapless in all flavors. Moreover, 
for the fractionally filled flavors, $C_3$ symmetry is broken.  
The band structure of different 
flavors for the P2 state at $\nu$ = 0.81 is shown in \prettyref{exfig:C3_broken}, alongside the 
positions of the Dirac points, where the Berry curvature singularities are located. 
Because of this, the P2 state will not exhibit anomalous Hall effects, whereas the P1 state carries almost quantized Hall conductance since the Berry curvature is mostly distributed around the $\Gamma$ point where the upper band bottom is located.   
\prettyref{exfig:DOS_C2T_C3}, shows density of states for the P1 and P2 states.

\begin{figure}
    \centering
    \includegraphics[width=0.95\textwidth]{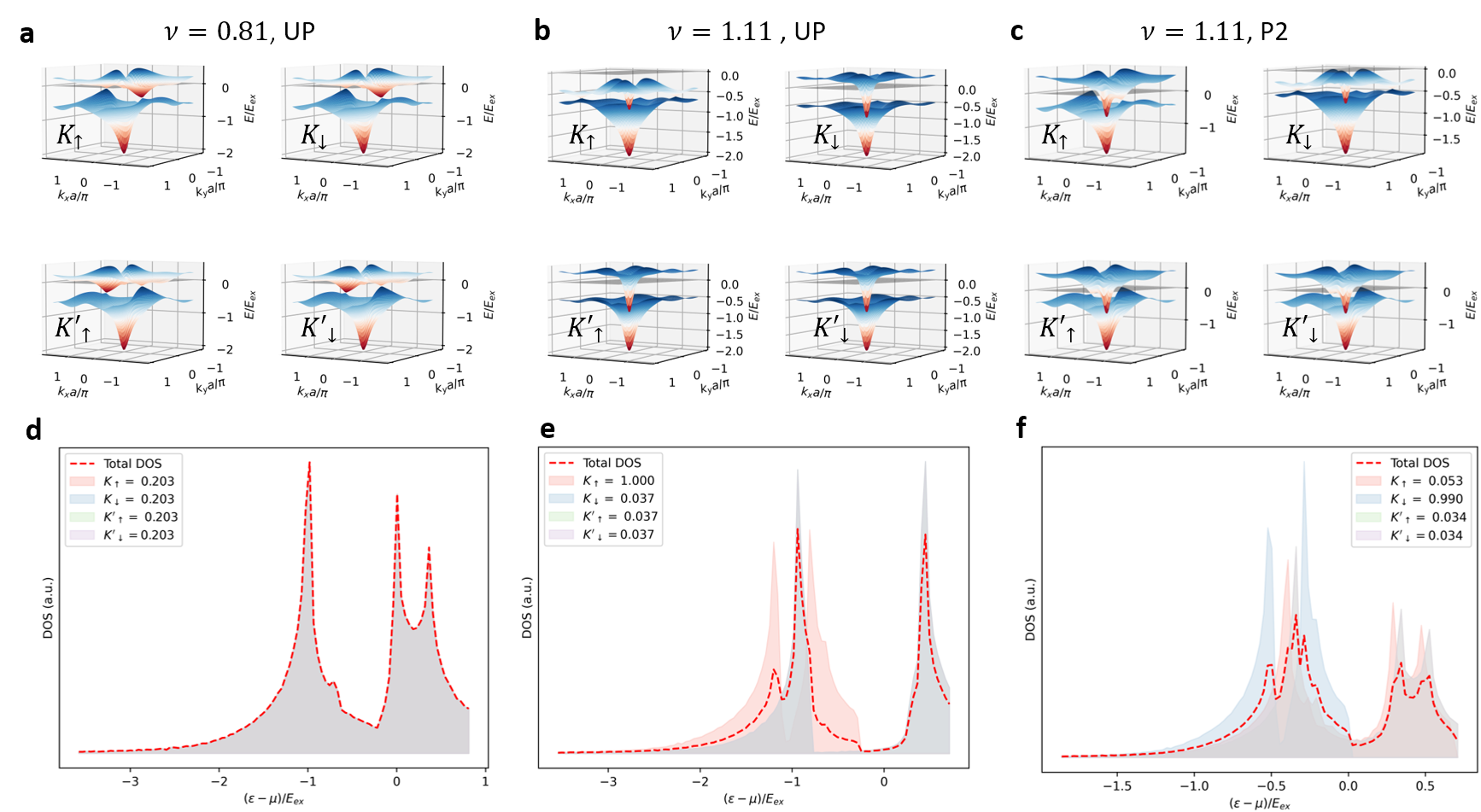}
    \caption{{\bf Possible ground states near $\nu$ = 1.} {\bf a}-{\bf c}, band 
    structures. {\bf d}-{\bf f}, density of states. UP state at $\nu$ = 0.81 ({\bf a, 
    d}). UP state at $\nu$ = 1.11 ({\bf b, e}). P2 state at $\nu$ = 1.11 ({\bf c, f}).}
    \label{exfig:nu_1_states}
\end{figure}

In \prettyref{exfig:nu_1_states}, we show additional band structures and density of
states, for UP states at $\nu=0.81, 1.11$, and P2 state at $\nu=1.11$. 
Here, the UP state at $\nu=0.81$ has no flavor polarization and carries zero Hall conductance. 
On the other hand, the UP state at $\nu=1.11$ is after cascade, with one flavor almost 
filled and the rest three equally and partially filled. When the latter three flavors 
are filled slightly near the $\Gamma$ pocket, they can still contribute to a finite 
Hall conductance for the reason similar to the one for the P1 state above.
Finally, the P2 state for $\nu=1.11$ cannot contribute to a nonzero Hall conductance.

\begin{thebibliography}{49}%
\makeatletter
\providecommand \@ifxundefined [1]{%
 \@ifx{#1\undefined}
}%
\providecommand \@ifnum [1]{%
 \ifnum #1\expandafter \@firstoftwo
 \else \expandafter \@secondoftwo
 \fi
}%
\providecommand \@ifx [1]{%
 \ifx #1\expandafter \@firstoftwo
 \else \expandafter \@secondoftwo
 \fi
}%
\providecommand \natexlab [1]{#1}%
\providecommand \enquote  [1]{``#1''}%
\providecommand \bibnamefont  [1]{#1}%
\providecommand \bibfnamefont [1]{#1}%
\providecommand \citenamefont [1]{#1}%
\providecommand \href@noop [0]{\@secondoftwo}%
\providecommand \href [0]{\begingroup \@sanitize@url \@href}%
\providecommand \@href[1]{\@@startlink{#1}\@@href}%
\providecommand \@@href[1]{\endgroup#1\@@endlink}%
\providecommand \@sanitize@url [0]{\catcode `\\12\catcode `\$12\catcode
  `\&12\catcode `\#12\catcode `\^12\catcode `\_12\catcode `\%12\relax}%
\providecommand \@@startlink[1]{}%
\providecommand \@@endlink[0]{}%
\providecommand \url  [0]{\begingroup\@sanitize@url \@url }%
\providecommand \@url [1]{\endgroup\@href {#1}{\urlprefix }}%
\providecommand \urlprefix  [0]{URL }%
\providecommand \Eprint [0]{\href }%
\providecommand \doibase [0]{https://doi.org/}%
\providecommand \selectlanguage [0]{\@gobble}%
\providecommand \bibinfo  [0]{\@secondoftwo}%
\providecommand \bibfield  [0]{\@secondoftwo}%
\providecommand \translation [1]{[#1]}%
\providecommand \BibitemOpen [0]{}%
\providecommand \bibitemStop [0]{}%
\providecommand \bibitemNoStop [0]{.\EOS\space}%
\providecommand \EOS [0]{\spacefactor3000\relax}%
\providecommand \BibitemShut  [1]{\csname bibitem#1\endcsname}%
\let\auto@bib@innerbib\@empty
\bibitem [{\citenamefont {Cao}\ \emph {et~al.}(2018{\natexlab{a}})\citenamefont
  {Cao}, \citenamefont {Fatemi}, \citenamefont {Demir}, \citenamefont {Fang},
  \citenamefont {Tomarken}, \citenamefont {Luo}, \citenamefont
  {{Sanchez-Yamagishi}}, \citenamefont {Watanabe}, \citenamefont {Taniguchi},
  \citenamefont {Kaxiras}, \citenamefont {Ashoori},\ and\ \citenamefont
  {{Jarillo-Herrero}}}]{caoCorrelatedInsulatorBehaviour2018}%
  \BibitemOpen
  \bibfield  {author} {\bibinfo {author} {\bibfnamefont {Y.}~\bibnamefont
  {Cao}}, \bibinfo {author} {\bibfnamefont {V.}~\bibnamefont {Fatemi}},
  \bibinfo {author} {\bibfnamefont {A.}~\bibnamefont {Demir}}, \bibinfo
  {author} {\bibfnamefont {S.}~\bibnamefont {Fang}}, \bibinfo {author}
  {\bibfnamefont {S.~L.}\ \bibnamefont {Tomarken}}, \bibinfo {author}
  {\bibfnamefont {J.~Y.}\ \bibnamefont {Luo}}, \bibinfo {author} {\bibfnamefont
  {J.~D.}\ \bibnamefont {{Sanchez-Yamagishi}}}, \bibinfo {author}
  {\bibfnamefont {K.}~\bibnamefont {Watanabe}}, \bibinfo {author}
  {\bibfnamefont {T.}~\bibnamefont {Taniguchi}}, \bibinfo {author}
  {\bibfnamefont {E.}~\bibnamefont {Kaxiras}}, \bibinfo {author} {\bibfnamefont
  {R.~C.}\ \bibnamefont {Ashoori}},\ and\ \bibinfo {author} {\bibfnamefont
  {P.}~\bibnamefont {{Jarillo-Herrero}}},\ }\href
  {https://doi.org/10.1038/nature26154} {\bibfield  {journal} {\bibinfo
  {journal} {Nature}\ }\textbf {\bibinfo {volume} {556}},\ \bibinfo {pages}
  {80} (\bibinfo {year} {2018}{\natexlab{a}})}\BibitemShut {NoStop}%
\bibitem [{\citenamefont {Cao}\ \emph {et~al.}(2018{\natexlab{b}})\citenamefont
  {Cao}, \citenamefont {Fatemi}, \citenamefont {Fang}, \citenamefont
  {Watanabe}, \citenamefont {Taniguchi}, \citenamefont {Kaxiras},\ and\
  \citenamefont
  {{Jarillo-Herrero}}}]{caoUnconventionalSuperconductivityMagicangle2018}%
  \BibitemOpen
  \bibfield  {author} {\bibinfo {author} {\bibfnamefont {Y.}~\bibnamefont
  {Cao}}, \bibinfo {author} {\bibfnamefont {V.}~\bibnamefont {Fatemi}},
  \bibinfo {author} {\bibfnamefont {S.}~\bibnamefont {Fang}}, \bibinfo {author}
  {\bibfnamefont {K.}~\bibnamefont {Watanabe}}, \bibinfo {author}
  {\bibfnamefont {T.}~\bibnamefont {Taniguchi}}, \bibinfo {author}
  {\bibfnamefont {E.}~\bibnamefont {Kaxiras}},\ and\ \bibinfo {author}
  {\bibfnamefont {P.}~\bibnamefont {{Jarillo-Herrero}}},\ }\href
  {https://doi.org/10.1038/nature26160} {\bibfield  {journal} {\bibinfo
  {journal} {Nature}\ }\textbf {\bibinfo {volume} {556}},\ \bibinfo {pages}
  {43} (\bibinfo {year} {2018}{\natexlab{b}})}\BibitemShut {NoStop}%
\bibitem [{\citenamefont {Yankowitz}\ \emph {et~al.}(2019)\citenamefont
  {Yankowitz}, \citenamefont {Chen}, \citenamefont {Polshyn}, \citenamefont
  {Zhang}, \citenamefont {Watanabe}, \citenamefont {Taniguchi}, \citenamefont
  {Graf}, \citenamefont {Young},\ and\ \citenamefont
  {Dean}}]{yankowitzTuningSuperconductivityTwisted2019}%
  \BibitemOpen
  \bibfield  {author} {\bibinfo {author} {\bibfnamefont {M.}~\bibnamefont
  {Yankowitz}}, \bibinfo {author} {\bibfnamefont {S.}~\bibnamefont {Chen}},
  \bibinfo {author} {\bibfnamefont {H.}~\bibnamefont {Polshyn}}, \bibinfo
  {author} {\bibfnamefont {Y.}~\bibnamefont {Zhang}}, \bibinfo {author}
  {\bibfnamefont {K.}~\bibnamefont {Watanabe}}, \bibinfo {author}
  {\bibfnamefont {T.}~\bibnamefont {Taniguchi}}, \bibinfo {author}
  {\bibfnamefont {D.}~\bibnamefont {Graf}}, \bibinfo {author} {\bibfnamefont
  {A.~F.}\ \bibnamefont {Young}},\ and\ \bibinfo {author} {\bibfnamefont
  {C.~R.}\ \bibnamefont {Dean}},\ }\href
  {https://doi.org/10.1126/science.aav1910} {\bibfield  {journal} {\bibinfo
  {journal} {Science}\ }\textbf {\bibinfo {volume} {363}},\ \bibinfo {pages}
  {1059} (\bibinfo {year} {2019})}\BibitemShut {NoStop}%
\bibitem [{\citenamefont {Lu}\ \emph {et~al.}(2019)\citenamefont {Lu},
  \citenamefont {Stepanov}, \citenamefont {Yang}, \citenamefont {Xie},
  \citenamefont {Aamir}, \citenamefont {Das}, \citenamefont {Urgell},
  \citenamefont {Watanabe}, \citenamefont {Taniguchi}, \citenamefont {Zhang},
  \citenamefont {Bachtold}, \citenamefont {MacDonald},\ and\ \citenamefont
  {Efetov}}]{luSuperconductorsOrbitalMagnets2019}%
  \BibitemOpen
  \bibfield  {author} {\bibinfo {author} {\bibfnamefont {X.}~\bibnamefont
  {Lu}}, \bibinfo {author} {\bibfnamefont {P.}~\bibnamefont {Stepanov}},
  \bibinfo {author} {\bibfnamefont {W.}~\bibnamefont {Yang}}, \bibinfo {author}
  {\bibfnamefont {M.}~\bibnamefont {Xie}}, \bibinfo {author} {\bibfnamefont
  {M.~A.}\ \bibnamefont {Aamir}}, \bibinfo {author} {\bibfnamefont
  {I.}~\bibnamefont {Das}}, \bibinfo {author} {\bibfnamefont {C.}~\bibnamefont
  {Urgell}}, \bibinfo {author} {\bibfnamefont {K.}~\bibnamefont {Watanabe}},
  \bibinfo {author} {\bibfnamefont {T.}~\bibnamefont {Taniguchi}}, \bibinfo
  {author} {\bibfnamefont {G.}~\bibnamefont {Zhang}}, \bibinfo {author}
  {\bibfnamefont {A.}~\bibnamefont {Bachtold}}, \bibinfo {author}
  {\bibfnamefont {A.~H.}\ \bibnamefont {MacDonald}},\ and\ \bibinfo {author}
  {\bibfnamefont {D.~K.}\ \bibnamefont {Efetov}},\ }\href
  {https://doi.org/10.1038/s41586-019-1695-0} {\bibfield  {journal} {\bibinfo
  {journal} {Nature}\ }\textbf {\bibinfo {volume} {574}},\ \bibinfo {pages}
  {653} (\bibinfo {year} {2019})}\BibitemShut {NoStop}%
\bibitem [{\citenamefont {Arora}\ \emph {et~al.}(2020)\citenamefont {Arora},
  \citenamefont {Polski}, \citenamefont {Zhang}, \citenamefont {Thomson},
  \citenamefont {Choi}, \citenamefont {Kim}, \citenamefont {Lin}, \citenamefont
  {Wilson}, \citenamefont {Xu}, \citenamefont {Chu}, \citenamefont {Watanabe},
  \citenamefont {Taniguchi}, \citenamefont {Alicea},\ and\ \citenamefont
  {{Nadj-Perge}}}]{aroraSuperconductivityMetallicTwisted2020}%
  \BibitemOpen
  \bibfield  {author} {\bibinfo {author} {\bibfnamefont {H.~S.}\ \bibnamefont
  {Arora}}, \bibinfo {author} {\bibfnamefont {R.}~\bibnamefont {Polski}},
  \bibinfo {author} {\bibfnamefont {Y.}~\bibnamefont {Zhang}}, \bibinfo
  {author} {\bibfnamefont {A.}~\bibnamefont {Thomson}}, \bibinfo {author}
  {\bibfnamefont {Y.}~\bibnamefont {Choi}}, \bibinfo {author} {\bibfnamefont
  {H.}~\bibnamefont {Kim}}, \bibinfo {author} {\bibfnamefont {Z.}~\bibnamefont
  {Lin}}, \bibinfo {author} {\bibfnamefont {I.~Z.}\ \bibnamefont {Wilson}},
  \bibinfo {author} {\bibfnamefont {X.}~\bibnamefont {Xu}}, \bibinfo {author}
  {\bibfnamefont {J.-H.}\ \bibnamefont {Chu}}, \bibinfo {author} {\bibfnamefont
  {K.}~\bibnamefont {Watanabe}}, \bibinfo {author} {\bibfnamefont
  {T.}~\bibnamefont {Taniguchi}}, \bibinfo {author} {\bibfnamefont
  {J.}~\bibnamefont {Alicea}},\ and\ \bibinfo {author} {\bibfnamefont
  {S.}~\bibnamefont {{Nadj-Perge}}},\ }\href
  {https://doi.org/10.1038/s41586-020-2473-8} {\bibfield  {journal} {\bibinfo
  {journal} {Nature}\ }\textbf {\bibinfo {volume} {583}},\ \bibinfo {pages}
  {379} (\bibinfo {year} {2020})}\BibitemShut {NoStop}%
\bibitem [{\citenamefont {Serlin}\ \emph {et~al.}(2019)\citenamefont {Serlin},
  \citenamefont {Tschirhart}, \citenamefont {Polshyn}, \citenamefont {Zhang},
  \citenamefont {Zhu}, \citenamefont {Watanabe}, \citenamefont {Taniguchi},
  \citenamefont {Balents},\ and\ \citenamefont
  {Young}}]{serlinIntrinsicQuantizedAnomalous2019}%
  \BibitemOpen
  \bibfield  {author} {\bibinfo {author} {\bibfnamefont {M.}~\bibnamefont
  {Serlin}}, \bibinfo {author} {\bibfnamefont {C.~L.}\ \bibnamefont
  {Tschirhart}}, \bibinfo {author} {\bibfnamefont {H.}~\bibnamefont {Polshyn}},
  \bibinfo {author} {\bibfnamefont {Y.}~\bibnamefont {Zhang}}, \bibinfo
  {author} {\bibfnamefont {J.}~\bibnamefont {Zhu}}, \bibinfo {author}
  {\bibfnamefont {K.}~\bibnamefont {Watanabe}}, \bibinfo {author}
  {\bibfnamefont {T.}~\bibnamefont {Taniguchi}}, \bibinfo {author}
  {\bibfnamefont {L.}~\bibnamefont {Balents}},\ and\ \bibinfo {author}
  {\bibfnamefont {A.~F.}\ \bibnamefont {Young}},\ }\href
  {https://doi.org/10.1126/science.aay5533} {\bibfield  {journal} {\bibinfo
  {journal} {Science}\ }\textbf {\bibinfo {volume} {367}},\ \bibinfo {pages}
  {900} (\bibinfo {year} {2019})}\BibitemShut {NoStop}%
\bibitem [{\citenamefont {Sharpe}\ \emph {et~al.}(2019)\citenamefont {Sharpe},
  \citenamefont {Fox}, \citenamefont {Barnard}, \citenamefont {Finney},
  \citenamefont {Watanabe}, \citenamefont {Taniguchi}, \citenamefont
  {Kastner},\ and\ \citenamefont
  {{Goldhaber-Gordon}}}]{sharpeEmergentFerromagnetismThreequarters2019}%
  \BibitemOpen
  \bibfield  {author} {\bibinfo {author} {\bibfnamefont {A.~L.}\ \bibnamefont
  {Sharpe}}, \bibinfo {author} {\bibfnamefont {E.~J.}\ \bibnamefont {Fox}},
  \bibinfo {author} {\bibfnamefont {A.~W.}\ \bibnamefont {Barnard}}, \bibinfo
  {author} {\bibfnamefont {J.}~\bibnamefont {Finney}}, \bibinfo {author}
  {\bibfnamefont {K.}~\bibnamefont {Watanabe}}, \bibinfo {author}
  {\bibfnamefont {T.}~\bibnamefont {Taniguchi}}, \bibinfo {author}
  {\bibfnamefont {M.~A.}\ \bibnamefont {Kastner}},\ and\ \bibinfo {author}
  {\bibfnamefont {D.}~\bibnamefont {{Goldhaber-Gordon}}},\ }\href
  {https://doi.org/10.1126/science.aaw3780} {\bibfield  {journal} {\bibinfo
  {journal} {Science}\ }\textbf {\bibinfo {volume} {365}},\ \bibinfo {pages}
  {605} (\bibinfo {year} {2019})}\BibitemShut {NoStop}%
\bibitem [{\citenamefont {Cao}\ \emph {et~al.}(2020)\citenamefont {Cao},
  \citenamefont {Chowdhury}, \citenamefont {{Rodan-Legrain}}, \citenamefont
  {{Rubies-Bigorda}}, \citenamefont {Watanabe}, \citenamefont {Taniguchi},
  \citenamefont {Senthil},\ and\ \citenamefont
  {{Jarillo-Herrero}}}]{caoStrangeMetalMagicAngle2020}%
  \BibitemOpen
  \bibfield  {author} {\bibinfo {author} {\bibfnamefont {Y.}~\bibnamefont
  {Cao}}, \bibinfo {author} {\bibfnamefont {D.}~\bibnamefont {Chowdhury}},
  \bibinfo {author} {\bibfnamefont {D.}~\bibnamefont {{Rodan-Legrain}}},
  \bibinfo {author} {\bibfnamefont {O.}~\bibnamefont {{Rubies-Bigorda}}},
  \bibinfo {author} {\bibfnamefont {K.}~\bibnamefont {Watanabe}}, \bibinfo
  {author} {\bibfnamefont {T.}~\bibnamefont {Taniguchi}}, \bibinfo {author}
  {\bibfnamefont {T.}~\bibnamefont {Senthil}},\ and\ \bibinfo {author}
  {\bibfnamefont {P.}~\bibnamefont {{Jarillo-Herrero}}},\ }\href
  {https://doi.org/10.1103/PhysRevLett.124.076801} {\bibfield  {journal}
  {\bibinfo  {journal} {Physical Review Letters}\ }\textbf {\bibinfo {volume}
  {124}},\ \bibinfo {pages} {076801} (\bibinfo {year} {2020})},\ \Eprint
  {https://arxiv.org/abs/1901.03710} {arXiv:1901.03710} \BibitemShut {NoStop}%
\bibitem [{\citenamefont {Stepanov}\ \emph
  {et~al.}(2020{\natexlab{a}})\citenamefont {Stepanov}, \citenamefont {Xie},
  \citenamefont {Taniguchi}, \citenamefont {Watanabe}, \citenamefont {Lu},
  \citenamefont {MacDonald}, \citenamefont {Bernevig},\ and\ \citenamefont
  {Efetov}}]{stepanovCompetingZerofieldChern2020}%
  \BibitemOpen
  \bibfield  {author} {\bibinfo {author} {\bibfnamefont {P.}~\bibnamefont
  {Stepanov}}, \bibinfo {author} {\bibfnamefont {M.}~\bibnamefont {Xie}},
  \bibinfo {author} {\bibfnamefont {T.}~\bibnamefont {Taniguchi}}, \bibinfo
  {author} {\bibfnamefont {K.}~\bibnamefont {Watanabe}}, \bibinfo {author}
  {\bibfnamefont {X.}~\bibnamefont {Lu}}, \bibinfo {author} {\bibfnamefont
  {A.~H.}\ \bibnamefont {MacDonald}}, \bibinfo {author} {\bibfnamefont {B.~A.}\
  \bibnamefont {Bernevig}},\ and\ \bibinfo {author} {\bibfnamefont {D.~K.}\
  \bibnamefont {Efetov}},\ }\href@noop {} {\bibfield  {journal} {\bibinfo
  {journal} {arXiv:2012.15126 [cond-mat]}\ } (\bibinfo {year}
  {2020}{\natexlab{a}})},\ \Eprint {https://arxiv.org/abs/2012.15126}
  {arXiv:2012.15126 [cond-mat]} \BibitemShut {NoStop}%
\bibitem [{\citenamefont {Bistritzer}\ and\ \citenamefont
  {MacDonald}(2011)}]{bistritzerMoireBandsTwisted2011}%
  \BibitemOpen
  \bibfield  {author} {\bibinfo {author} {\bibfnamefont {R.}~\bibnamefont
  {Bistritzer}}\ and\ \bibinfo {author} {\bibfnamefont {A.~H.}\ \bibnamefont
  {MacDonald}},\ }\href {https://doi.org/10.1073/pnas.1108174108} {\bibfield
  {journal} {\bibinfo  {journal} {Proceedings of the National Academy of
  Sciences}\ }\textbf {\bibinfo {volume} {108}},\ \bibinfo {pages} {12233}
  (\bibinfo {year} {2011})}\BibitemShut {NoStop}%
\bibitem [{\citenamefont {Zondiner}\ \emph {et~al.}(2020)\citenamefont
  {Zondiner}, \citenamefont {Rozen}, \citenamefont {{Rodan-Legrain}},
  \citenamefont {Cao}, \citenamefont {Queiroz}, \citenamefont {Taniguchi},
  \citenamefont {Watanabe}, \citenamefont {Oreg}, \citenamefont {{von Oppen}},
  \citenamefont {Stern}, \citenamefont {Berg}, \citenamefont
  {{Jarillo-Herrero}},\ and\ \citenamefont
  {Ilani}}]{zondinerCascadePhaseTransitions2020}%
  \BibitemOpen
  \bibfield  {author} {\bibinfo {author} {\bibfnamefont {U.}~\bibnamefont
  {Zondiner}}, \bibinfo {author} {\bibfnamefont {A.}~\bibnamefont {Rozen}},
  \bibinfo {author} {\bibfnamefont {D.}~\bibnamefont {{Rodan-Legrain}}},
  \bibinfo {author} {\bibfnamefont {Y.}~\bibnamefont {Cao}}, \bibinfo {author}
  {\bibfnamefont {R.}~\bibnamefont {Queiroz}}, \bibinfo {author} {\bibfnamefont
  {T.}~\bibnamefont {Taniguchi}}, \bibinfo {author} {\bibfnamefont
  {K.}~\bibnamefont {Watanabe}}, \bibinfo {author} {\bibfnamefont
  {Y.}~\bibnamefont {Oreg}}, \bibinfo {author} {\bibfnamefont {F.}~\bibnamefont
  {{von Oppen}}}, \bibinfo {author} {\bibfnamefont {A.}~\bibnamefont {Stern}},
  \bibinfo {author} {\bibfnamefont {E.}~\bibnamefont {Berg}}, \bibinfo {author}
  {\bibfnamefont {P.}~\bibnamefont {{Jarillo-Herrero}}},\ and\ \bibinfo
  {author} {\bibfnamefont {S.}~\bibnamefont {Ilani}},\ }\href
  {https://doi.org/10.1038/s41586-020-2373-y} {\bibfield  {journal} {\bibinfo
  {journal} {Nature}\ }\textbf {\bibinfo {volume} {582}},\ \bibinfo {pages}
  {203} (\bibinfo {year} {2020})}\BibitemShut {NoStop}%
\bibitem [{\citenamefont {Wong}\ \emph {et~al.}(2020)\citenamefont {Wong},
  \citenamefont {Nuckolls}, \citenamefont {Oh}, \citenamefont {Lian},
  \citenamefont {Xie}, \citenamefont {Jeon}, \citenamefont {Watanabe},
  \citenamefont {Taniguchi}, \citenamefont {Bernevig},\ and\ \citenamefont
  {Yazdani}}]{wongCascadeElectronicTransitions2020}%
  \BibitemOpen
  \bibfield  {author} {\bibinfo {author} {\bibfnamefont {D.}~\bibnamefont
  {Wong}}, \bibinfo {author} {\bibfnamefont {K.~P.}\ \bibnamefont {Nuckolls}},
  \bibinfo {author} {\bibfnamefont {M.}~\bibnamefont {Oh}}, \bibinfo {author}
  {\bibfnamefont {B.}~\bibnamefont {Lian}}, \bibinfo {author} {\bibfnamefont
  {Y.}~\bibnamefont {Xie}}, \bibinfo {author} {\bibfnamefont {S.}~\bibnamefont
  {Jeon}}, \bibinfo {author} {\bibfnamefont {K.}~\bibnamefont {Watanabe}},
  \bibinfo {author} {\bibfnamefont {T.}~\bibnamefont {Taniguchi}}, \bibinfo
  {author} {\bibfnamefont {B.~A.}\ \bibnamefont {Bernevig}},\ and\ \bibinfo
  {author} {\bibfnamefont {A.}~\bibnamefont {Yazdani}},\ }\href
  {https://doi.org/10.1038/s41586-020-2339-0} {\bibfield  {journal} {\bibinfo
  {journal} {Nature}\ }\textbf {\bibinfo {volume} {582}},\ \bibinfo {pages}
  {198} (\bibinfo {year} {2020})}\BibitemShut {NoStop}%
\bibitem [{\citenamefont {Xie}\ and\ \citenamefont
  {MacDonald}(2020)}]{xieWeakfieldHallResistivity2020}%
  \BibitemOpen
  \bibfield  {author} {\bibinfo {author} {\bibfnamefont {M.}~\bibnamefont
  {Xie}}\ and\ \bibinfo {author} {\bibfnamefont {A.~H.}\ \bibnamefont
  {MacDonald}},\ }\href@noop {} {\bibfield  {journal} {\bibinfo  {journal}
  {arXiv:2010.07928 [cond-mat]}\ } (\bibinfo {year} {2020})},\ \Eprint
  {https://arxiv.org/abs/2010.07928} {arXiv:2010.07928 [cond-mat]} \BibitemShut
  {NoStop}%
\bibitem [{\citenamefont {Qin}\ \emph {et~al.}(2021)\citenamefont {Qin},
  \citenamefont {Zou},\ and\ \citenamefont
  {MacDonald}}]{qinCriticalMagneticFields2021}%
  \BibitemOpen
  \bibfield  {author} {\bibinfo {author} {\bibfnamefont {W.}~\bibnamefont
  {Qin}}, \bibinfo {author} {\bibfnamefont {B.}~\bibnamefont {Zou}},\ and\
  \bibinfo {author} {\bibfnamefont {A.~H.}\ \bibnamefont {MacDonald}},\
  }\href@noop {} {\bibfield  {journal} {\bibinfo  {journal} {arXiv:2102.10504
  [cond-mat]}\ } (\bibinfo {year} {2021})},\ \Eprint
  {https://arxiv.org/abs/2102.10504} {arXiv:2102.10504 [cond-mat]} \BibitemShut
  {NoStop}%
\bibitem [{\citenamefont {Cherng}\ \emph {et~al.}(2007)\citenamefont {Cherng},
  \citenamefont {Refael},\ and\ \citenamefont
  {Demler}}]{cherngSuperfluidityMagnetismMulticomponent2007}%
  \BibitemOpen
  \bibfield  {author} {\bibinfo {author} {\bibfnamefont {R.~W.}\ \bibnamefont
  {Cherng}}, \bibinfo {author} {\bibfnamefont {G.}~\bibnamefont {Refael}},\
  and\ \bibinfo {author} {\bibfnamefont {E.}~\bibnamefont {Demler}},\ }\href
  {https://doi.org/10.1103/PhysRevLett.99.130406} {\bibfield  {journal}
  {\bibinfo  {journal} {Physical Review Letters}\ }\textbf {\bibinfo {volume}
  {99}},\ \bibinfo {pages} {130406} (\bibinfo {year} {2007})}\BibitemShut
  {NoStop}%
\bibitem [{\citenamefont {You}\ and\ \citenamefont
  {Vishwanath}(2019)}]{youSuperconductivityValleyFluctuations2019}%
  \BibitemOpen
  \bibfield  {author} {\bibinfo {author} {\bibfnamefont {Y.-Z.}\ \bibnamefont
  {You}}\ and\ \bibinfo {author} {\bibfnamefont {A.}~\bibnamefont
  {Vishwanath}},\ }\href {https://doi.org/10.1038/s41535-019-0153-4} {\bibfield
   {journal} {\bibinfo  {journal} {npj Quantum Materials}\ }\textbf {\bibinfo
  {volume} {4}},\ \bibinfo {pages} {1} (\bibinfo {year} {2019})}\BibitemShut
  {NoStop}%
\bibitem [{\citenamefont {Gonz{\'a}lez}\ and\ \citenamefont
  {Stauber}(2019)}]{gonzalezKohnLuttingerSuperconductivityTwisted2019}%
  \BibitemOpen
  \bibfield  {author} {\bibinfo {author} {\bibfnamefont {J.}~\bibnamefont
  {Gonz{\'a}lez}}\ and\ \bibinfo {author} {\bibfnamefont {T.}~\bibnamefont
  {Stauber}},\ }\href {https://doi.org/10.1103/PhysRevLett.122.026801}
  {\bibfield  {journal} {\bibinfo  {journal} {Physical Review Letters}\
  }\textbf {\bibinfo {volume} {122}},\ \bibinfo {pages} {026801} (\bibinfo
  {year} {2019})}\BibitemShut {NoStop}%
\bibitem [{\citenamefont {Saito}\ \emph {et~al.}(2021)\citenamefont {Saito},
  \citenamefont {Yang}, \citenamefont {Ge}, \citenamefont {Liu}, \citenamefont
  {Taniguchi}, \citenamefont {Watanabe}, \citenamefont {Li}, \citenamefont
  {Berg},\ and\ \citenamefont {Young}}]{saitoIsospinPomeranchukEffect2021}%
  \BibitemOpen
  \bibfield  {author} {\bibinfo {author} {\bibfnamefont {Y.}~\bibnamefont
  {Saito}}, \bibinfo {author} {\bibfnamefont {F.}~\bibnamefont {Yang}},
  \bibinfo {author} {\bibfnamefont {J.}~\bibnamefont {Ge}}, \bibinfo {author}
  {\bibfnamefont {X.}~\bibnamefont {Liu}}, \bibinfo {author} {\bibfnamefont
  {T.}~\bibnamefont {Taniguchi}}, \bibinfo {author} {\bibfnamefont
  {K.}~\bibnamefont {Watanabe}}, \bibinfo {author} {\bibfnamefont {J.~I.~A.}\
  \bibnamefont {Li}}, \bibinfo {author} {\bibfnamefont {E.}~\bibnamefont
  {Berg}},\ and\ \bibinfo {author} {\bibfnamefont {A.~F.}\ \bibnamefont
  {Young}},\ }\href {https://doi.org/10.1038/s41586-021-03409-2} {\bibfield
  {journal} {\bibinfo  {journal} {Nature}\ }\textbf {\bibinfo {volume} {592}},\
  \bibinfo {pages} {220} (\bibinfo {year} {2021})}\BibitemShut {NoStop}%
\bibitem [{\citenamefont {Rozen}\ \emph {et~al.}(2021)\citenamefont {Rozen},
  \citenamefont {Park}, \citenamefont {Zondiner}, \citenamefont {Cao},
  \citenamefont {{Rodan-Legrain}}, \citenamefont {Taniguchi}, \citenamefont
  {Watanabe}, \citenamefont {Oreg}, \citenamefont {Stern}, \citenamefont
  {Berg}, \citenamefont {{Jarillo-Herrero}},\ and\ \citenamefont
  {Ilani}}]{rozenEntropicEvidencePomeranchuk2021}%
  \BibitemOpen
  \bibfield  {author} {\bibinfo {author} {\bibfnamefont {A.}~\bibnamefont
  {Rozen}}, \bibinfo {author} {\bibfnamefont {J.~M.}\ \bibnamefont {Park}},
  \bibinfo {author} {\bibfnamefont {U.}~\bibnamefont {Zondiner}}, \bibinfo
  {author} {\bibfnamefont {Y.}~\bibnamefont {Cao}}, \bibinfo {author}
  {\bibfnamefont {D.}~\bibnamefont {{Rodan-Legrain}}}, \bibinfo {author}
  {\bibfnamefont {T.}~\bibnamefont {Taniguchi}}, \bibinfo {author}
  {\bibfnamefont {K.}~\bibnamefont {Watanabe}}, \bibinfo {author}
  {\bibfnamefont {Y.}~\bibnamefont {Oreg}}, \bibinfo {author} {\bibfnamefont
  {A.}~\bibnamefont {Stern}}, \bibinfo {author} {\bibfnamefont
  {E.}~\bibnamefont {Berg}}, \bibinfo {author} {\bibfnamefont {P.}~\bibnamefont
  {{Jarillo-Herrero}}},\ and\ \bibinfo {author} {\bibfnamefont
  {S.}~\bibnamefont {Ilani}},\ }\href
  {https://doi.org/10.1038/s41586-021-03319-3} {\bibfield  {journal} {\bibinfo
  {journal} {Nature}\ }\textbf {\bibinfo {volume} {592}},\ \bibinfo {pages}
  {214} (\bibinfo {year} {2021})},\ \Eprint {https://arxiv.org/abs/2009.01836}
  {arXiv:2009.01836} \BibitemShut {NoStop}%
\bibitem [{\citenamefont {Song}\ and\ \citenamefont
  {Bernevig}(2021)}]{songMATBGTopologicalHeavy2021}%
  \BibitemOpen
  \bibfield  {author} {\bibinfo {author} {\bibfnamefont {Z.-D.}\ \bibnamefont
  {Song}}\ and\ \bibinfo {author} {\bibfnamefont {B.~A.}\ \bibnamefont
  {Bernevig}},\ }\href@noop {} {\bibfield  {journal} {\bibinfo  {journal}
  {arXiv:2111.05865 [cond-mat]}\ } (\bibinfo {year} {2021})},\ \Eprint
  {https://arxiv.org/abs/2111.05865} {arXiv:2111.05865 [cond-mat]} \BibitemShut
  {NoStop}%
\bibitem [{\citenamefont {Stewart}(1984)}]{stewartHeavyfermionSystems1984}%
  \BibitemOpen
  \bibfield  {author} {\bibinfo {author} {\bibfnamefont {G.~R.}\ \bibnamefont
  {Stewart}},\ }\href {https://doi.org/10.1103/RevModPhys.56.755} {\bibfield
  {journal} {\bibinfo  {journal} {Reviews of Modern Physics}\ }\textbf
  {\bibinfo {volume} {56}},\ \bibinfo {pages} {755} (\bibinfo {year}
  {1984})}\BibitemShut {NoStop}%
\bibitem [{\citenamefont {Lonzarich}\ \emph {et~al.}(2016)\citenamefont
  {Lonzarich}, \citenamefont {Pines},\ and\ \citenamefont
  {Yang}}]{lonzarichNewMicroscopicFramework2016}%
  \BibitemOpen
  \bibfield  {author} {\bibinfo {author} {\bibfnamefont {G.}~\bibnamefont
  {Lonzarich}}, \bibinfo {author} {\bibfnamefont {D.}~\bibnamefont {Pines}},\
  and\ \bibinfo {author} {\bibfnamefont {Y.-f.}\ \bibnamefont {Yang}},\ }\href
  {https://doi.org/10.1088/1361-6633/80/2/024501} {\bibfield  {journal}
  {\bibinfo  {journal} {Reports on Progress in Physics}\ }\textbf {\bibinfo
  {volume} {80}},\ \bibinfo {pages} {024501} (\bibinfo {year}
  {2016})}\BibitemShut {NoStop}%
\bibitem [{\citenamefont {Continentino}\ and\ \citenamefont
  {Ferreira}(2004)}]{continentinoPomeranchukEffectUnstable2004}%
  \BibitemOpen
  \bibfield  {author} {\bibinfo {author} {\bibfnamefont {M.~A.}\ \bibnamefont
  {Continentino}}\ and\ \bibinfo {author} {\bibfnamefont {A.~S.}\ \bibnamefont
  {Ferreira}},\ }\href {https://doi.org/10.1103/PhysRevB.69.233104} {\bibfield
  {journal} {\bibinfo  {journal} {Physical Review B}\ }\textbf {\bibinfo
  {volume} {69}},\ \bibinfo {pages} {233104} (\bibinfo {year}
  {2004})}\BibitemShut {NoStop}%
\bibitem [{\citenamefont {Wu}\ \emph {et~al.}(2019)\citenamefont {Wu},
  \citenamefont {Hwang},\ and\ \citenamefont
  {Das~Sarma}}]{wuPhononinducedGiantLinearin2019}%
  \BibitemOpen
  \bibfield  {author} {\bibinfo {author} {\bibfnamefont {F.}~\bibnamefont
  {Wu}}, \bibinfo {author} {\bibfnamefont {E.}~\bibnamefont {Hwang}},\ and\
  \bibinfo {author} {\bibfnamefont {S.}~\bibnamefont {Das~Sarma}},\ }\href
  {https://doi.org/10.1103/PhysRevB.99.165112} {\bibfield  {journal} {\bibinfo
  {journal} {Physical Review B}\ }\textbf {\bibinfo {volume} {99}},\ \bibinfo
  {pages} {165112} (\bibinfo {year} {2019})}\BibitemShut {NoStop}%
\bibitem [{\citenamefont {Polshyn}\ \emph {et~al.}(2019)\citenamefont
  {Polshyn}, \citenamefont {Yankowitz}, \citenamefont {Chen}, \citenamefont
  {Zhang}, \citenamefont {Watanabe}, \citenamefont {Taniguchi}, \citenamefont
  {Dean},\ and\ \citenamefont
  {Young}}]{polshynLargeLinearintemperatureResistivity2019}%
  \BibitemOpen
  \bibfield  {author} {\bibinfo {author} {\bibfnamefont {H.}~\bibnamefont
  {Polshyn}}, \bibinfo {author} {\bibfnamefont {M.}~\bibnamefont {Yankowitz}},
  \bibinfo {author} {\bibfnamefont {S.}~\bibnamefont {Chen}}, \bibinfo {author}
  {\bibfnamefont {Y.}~\bibnamefont {Zhang}}, \bibinfo {author} {\bibfnamefont
  {K.}~\bibnamefont {Watanabe}}, \bibinfo {author} {\bibfnamefont
  {T.}~\bibnamefont {Taniguchi}}, \bibinfo {author} {\bibfnamefont {C.~R.}\
  \bibnamefont {Dean}},\ and\ \bibinfo {author} {\bibfnamefont {A.~F.}\
  \bibnamefont {Young}},\ }\href {https://doi.org/10.1038/s41567-019-0596-3}
  {\bibfield  {journal} {\bibinfo  {journal} {Nature Physics}\ }\textbf
  {\bibinfo {volume} {15}},\ \bibinfo {pages} {1011} (\bibinfo {year}
  {2019})}\BibitemShut {NoStop}%
\bibitem [{\citenamefont {Sarma}\ and\ \citenamefont
  {Wu}(2022)}]{sarmaStrangeMetallicityMoir2022}%
  \BibitemOpen
  \bibfield  {author} {\bibinfo {author} {\bibfnamefont {S.~D.}\ \bibnamefont
  {Sarma}}\ and\ \bibinfo {author} {\bibfnamefont {F.}~\bibnamefont {Wu}},\
  }\href@noop {} {\bibfield  {journal} {\bibinfo  {journal} {arXiv:2201.10270
  [cond-mat]}\ } (\bibinfo {year} {2022})},\ \Eprint
  {https://arxiv.org/abs/2201.10270} {arXiv:2201.10270 [cond-mat]} \BibitemShut
  {NoStop}%
\bibitem [{\citenamefont {Jaoui}\ \emph {et~al.}(2022)\citenamefont {Jaoui},
  \citenamefont {Das}, \citenamefont {Di~Battista}, \citenamefont
  {{D{\'i}ez-M{\'e}rida}}, \citenamefont {Lu}, \citenamefont {Watanabe},
  \citenamefont {Taniguchi}, \citenamefont {Ishizuka}, \citenamefont
  {Levitov},\ and\ \citenamefont {Efetov}}]{jaouiQuantumCriticalBehavior2022}%
  \BibitemOpen
  \bibfield  {author} {\bibinfo {author} {\bibfnamefont {A.}~\bibnamefont
  {Jaoui}}, \bibinfo {author} {\bibfnamefont {I.}~\bibnamefont {Das}}, \bibinfo
  {author} {\bibfnamefont {G.}~\bibnamefont {Di~Battista}}, \bibinfo {author}
  {\bibfnamefont {J.}~\bibnamefont {{D{\'i}ez-M{\'e}rida}}}, \bibinfo {author}
  {\bibfnamefont {X.}~\bibnamefont {Lu}}, \bibinfo {author} {\bibfnamefont
  {K.}~\bibnamefont {Watanabe}}, \bibinfo {author} {\bibfnamefont
  {T.}~\bibnamefont {Taniguchi}}, \bibinfo {author} {\bibfnamefont
  {H.}~\bibnamefont {Ishizuka}}, \bibinfo {author} {\bibfnamefont
  {L.}~\bibnamefont {Levitov}},\ and\ \bibinfo {author} {\bibfnamefont {D.~K.}\
  \bibnamefont {Efetov}},\ }\href@noop {} {\bibfield  {journal} {\bibinfo
  {journal} {arXiv:2108.07753 [cond-mat]}\ } (\bibinfo {year} {2022})},\
  \Eprint {https://arxiv.org/abs/2108.07753} {arXiv:2108.07753 [cond-mat]}
  \BibitemShut {NoStop}%
\bibitem [{\citenamefont {Goodwin}\ \emph {et~al.}(2020)\citenamefont
  {Goodwin}, \citenamefont {Vitale}, \citenamefont {Liang}, \citenamefont
  {Mostofi},\ and\ \citenamefont
  {Lischner}}]{goodwinHartreeTheoryCalculations2020}%
  \BibitemOpen
  \bibfield  {author} {\bibinfo {author} {\bibfnamefont {Z.~A.~H.}\
  \bibnamefont {Goodwin}}, \bibinfo {author} {\bibfnamefont {V.}~\bibnamefont
  {Vitale}}, \bibinfo {author} {\bibfnamefont {X.}~\bibnamefont {Liang}},
  \bibinfo {author} {\bibfnamefont {A.~A.}\ \bibnamefont {Mostofi}},\ and\
  \bibinfo {author} {\bibfnamefont {J.}~\bibnamefont {Lischner}},\ }\href
  {https://doi.org/10.1088/2516-1075/ab9f94} {\bibfield  {journal} {\bibinfo
  {journal} {Electronic Structure}\ }\textbf {\bibinfo {volume} {2}},\ \bibinfo
  {pages} {034001} (\bibinfo {year} {2020})},\ \Eprint
  {https://arxiv.org/abs/2004.14784} {arXiv:2004.14784} \BibitemShut {NoStop}%
\bibitem [{\citenamefont {Saito}\ \emph {et~al.}(2020)\citenamefont {Saito},
  \citenamefont {Ge}, \citenamefont {Watanabe}, \citenamefont {Taniguchi},\
  and\ \citenamefont {Young}}]{saitoIndependentSuperconductorsCorrelated2020}%
  \BibitemOpen
  \bibfield  {author} {\bibinfo {author} {\bibfnamefont {Y.}~\bibnamefont
  {Saito}}, \bibinfo {author} {\bibfnamefont {J.}~\bibnamefont {Ge}}, \bibinfo
  {author} {\bibfnamefont {K.}~\bibnamefont {Watanabe}}, \bibinfo {author}
  {\bibfnamefont {T.}~\bibnamefont {Taniguchi}},\ and\ \bibinfo {author}
  {\bibfnamefont {A.~F.}\ \bibnamefont {Young}},\ }\href
  {https://doi.org/10.1038/s41567-020-0928-3} {\bibfield  {journal} {\bibinfo
  {journal} {Nature Physics}\ }\textbf {\bibinfo {volume} {16}},\ \bibinfo
  {pages} {926} (\bibinfo {year} {2020})}\BibitemShut {NoStop}%
\bibitem [{\citenamefont {Stepanov}\ \emph
  {et~al.}(2020{\natexlab{b}})\citenamefont {Stepanov}, \citenamefont {Das},
  \citenamefont {Lu}, \citenamefont {Fahimniya}, \citenamefont {Watanabe},
  \citenamefont {Taniguchi}, \citenamefont {Koppens}, \citenamefont {Lischner},
  \citenamefont {Levitov},\ and\ \citenamefont
  {Efetov}}]{stepanovUntyingInsulatingSuperconducting2020}%
  \BibitemOpen
  \bibfield  {author} {\bibinfo {author} {\bibfnamefont {P.}~\bibnamefont
  {Stepanov}}, \bibinfo {author} {\bibfnamefont {I.}~\bibnamefont {Das}},
  \bibinfo {author} {\bibfnamefont {X.}~\bibnamefont {Lu}}, \bibinfo {author}
  {\bibfnamefont {A.}~\bibnamefont {Fahimniya}}, \bibinfo {author}
  {\bibfnamefont {K.}~\bibnamefont {Watanabe}}, \bibinfo {author}
  {\bibfnamefont {T.}~\bibnamefont {Taniguchi}}, \bibinfo {author}
  {\bibfnamefont {F.~H.~L.}\ \bibnamefont {Koppens}}, \bibinfo {author}
  {\bibfnamefont {J.}~\bibnamefont {Lischner}}, \bibinfo {author}
  {\bibfnamefont {L.}~\bibnamefont {Levitov}},\ and\ \bibinfo {author}
  {\bibfnamefont {D.~K.}\ \bibnamefont {Efetov}},\ }\href
  {https://doi.org/10.1038/s41586-020-2459-6} {\bibfield  {journal} {\bibinfo
  {journal} {Nature}\ }\textbf {\bibinfo {volume} {583}},\ \bibinfo {pages}
  {375} (\bibinfo {year} {2020}{\natexlab{b}})}\BibitemShut {NoStop}%
\bibitem [{\citenamefont {Zhu}\ \emph {et~al.}(2020)\citenamefont {Zhu},
  \citenamefont {Su},\ and\ \citenamefont
  {MacDonald}}]{zhuVoltageControlledMagneticReversal2020}%
  \BibitemOpen
  \bibfield  {author} {\bibinfo {author} {\bibfnamefont {J.}~\bibnamefont
  {Zhu}}, \bibinfo {author} {\bibfnamefont {J.-J.}\ \bibnamefont {Su}},\ and\
  \bibinfo {author} {\bibfnamefont {A.~H.}\ \bibnamefont {MacDonald}},\ }\href
  {https://doi.org/10.1103/PhysRevLett.125.227702} {\bibfield  {journal}
  {\bibinfo  {journal} {Physical Review Letters}\ }\textbf {\bibinfo {volume}
  {125}},\ \bibinfo {pages} {227702} (\bibinfo {year} {2020})}\BibitemShut
  {NoStop}%
\bibitem [{\citenamefont {Polshyn}\ \emph {et~al.}(2020)\citenamefont
  {Polshyn}, \citenamefont {Zhu}, \citenamefont {Kumar}, \citenamefont {Zhang},
  \citenamefont {Yang}, \citenamefont {Tschirhart}, \citenamefont {Serlin},
  \citenamefont {Watanabe}, \citenamefont {Taniguchi}, \citenamefont
  {MacDonald},\ and\ \citenamefont
  {Young}}]{polshynElectricalSwitchingMagnetic2020}%
  \BibitemOpen
  \bibfield  {author} {\bibinfo {author} {\bibfnamefont {H.}~\bibnamefont
  {Polshyn}}, \bibinfo {author} {\bibfnamefont {J.}~\bibnamefont {Zhu}},
  \bibinfo {author} {\bibfnamefont {M.~A.}\ \bibnamefont {Kumar}}, \bibinfo
  {author} {\bibfnamefont {Y.}~\bibnamefont {Zhang}}, \bibinfo {author}
  {\bibfnamefont {F.}~\bibnamefont {Yang}}, \bibinfo {author} {\bibfnamefont
  {C.~L.}\ \bibnamefont {Tschirhart}}, \bibinfo {author} {\bibfnamefont
  {M.}~\bibnamefont {Serlin}}, \bibinfo {author} {\bibfnamefont
  {K.}~\bibnamefont {Watanabe}}, \bibinfo {author} {\bibfnamefont
  {T.}~\bibnamefont {Taniguchi}}, \bibinfo {author} {\bibfnamefont {A.~H.}\
  \bibnamefont {MacDonald}},\ and\ \bibinfo {author} {\bibfnamefont {A.~F.}\
  \bibnamefont {Young}},\ }\href {https://doi.org/10.1038/s41586-020-2963-8}
  {\bibfield  {journal} {\bibinfo  {journal} {Nature}\ }\textbf {\bibinfo
  {volume} {588}},\ \bibinfo {pages} {66} (\bibinfo {year} {2020})}\BibitemShut
  {NoStop}%
\bibitem [{\citenamefont {Lian}\ \emph {et~al.}(2021)\citenamefont {Lian},
  \citenamefont {Song}, \citenamefont {Regnault}, \citenamefont {Efetov},
  \citenamefont {Yazdani},\ and\ \citenamefont
  {Bernevig}}]{lianTwistedBilayerGraphene2021}%
  \BibitemOpen
  \bibfield  {author} {\bibinfo {author} {\bibfnamefont {B.}~\bibnamefont
  {Lian}}, \bibinfo {author} {\bibfnamefont {Z.-D.}\ \bibnamefont {Song}},
  \bibinfo {author} {\bibfnamefont {N.}~\bibnamefont {Regnault}}, \bibinfo
  {author} {\bibfnamefont {D.~K.}\ \bibnamefont {Efetov}}, \bibinfo {author}
  {\bibfnamefont {A.}~\bibnamefont {Yazdani}},\ and\ \bibinfo {author}
  {\bibfnamefont {B.~A.}\ \bibnamefont {Bernevig}},\ }\href
  {https://doi.org/10.1103/PhysRevB.103.205414} {\bibfield  {journal} {\bibinfo
   {journal} {Physical Review B}\ }\textbf {\bibinfo {volume} {103}},\ \bibinfo
  {pages} {205414} (\bibinfo {year} {2021})}\BibitemShut {NoStop}%
\bibitem [{\citenamefont {Lin}\ \emph {et~al.}(2021)\citenamefont {Lin},
  \citenamefont {Zhang}, \citenamefont {Morissette}, \citenamefont {Wang},
  \citenamefont {Liu}, \citenamefont {Rhodes}, \citenamefont {Watanabe},
  \citenamefont {Taniguchi}, \citenamefont {Hone},\ and\ \citenamefont
  {Li}}]{linProximityinducedSpinorbitCoupling2021}%
  \BibitemOpen
  \bibfield  {author} {\bibinfo {author} {\bibfnamefont {J.-X.}\ \bibnamefont
  {Lin}}, \bibinfo {author} {\bibfnamefont {Y.-H.}\ \bibnamefont {Zhang}},
  \bibinfo {author} {\bibfnamefont {E.}~\bibnamefont {Morissette}}, \bibinfo
  {author} {\bibfnamefont {Z.}~\bibnamefont {Wang}}, \bibinfo {author}
  {\bibfnamefont {S.}~\bibnamefont {Liu}}, \bibinfo {author} {\bibfnamefont
  {D.}~\bibnamefont {Rhodes}}, \bibinfo {author} {\bibfnamefont
  {K.}~\bibnamefont {Watanabe}}, \bibinfo {author} {\bibfnamefont
  {T.}~\bibnamefont {Taniguchi}}, \bibinfo {author} {\bibfnamefont
  {J.}~\bibnamefont {Hone}},\ and\ \bibinfo {author} {\bibfnamefont {J.~I.~A.}\
  \bibnamefont {Li}},\ }\href@noop {} {\bibfield  {journal} {\bibinfo
  {journal} {arXiv:2102.06566 [cond-mat]}\ } (\bibinfo {year} {2021})},\
  \Eprint {https://arxiv.org/abs/2102.06566} {arXiv:2102.06566 [cond-mat]}
  \BibitemShut {NoStop}%
\bibitem [{\citenamefont {Grover}\ \emph {et~al.}(2022)\citenamefont {Grover},
  \citenamefont {Bocarsly}, \citenamefont {Uri}, \citenamefont {Stepanov},
  \citenamefont {Di~Battista}, \citenamefont {Roy}, \citenamefont {Xiao},
  \citenamefont {Meltzer}, \citenamefont {Myasoedov}, \citenamefont {Pareek},
  \citenamefont {Watanabe}, \citenamefont {Taniguchi}, \citenamefont {Yan},
  \citenamefont {Stern}, \citenamefont {Berg}, \citenamefont {Efetov},\ and\
  \citenamefont {Zeldov}}]{groverImagingChernMosaic2022}%
  \BibitemOpen
  \bibfield  {author} {\bibinfo {author} {\bibfnamefont {S.}~\bibnamefont
  {Grover}}, \bibinfo {author} {\bibfnamefont {M.}~\bibnamefont {Bocarsly}},
  \bibinfo {author} {\bibfnamefont {A.}~\bibnamefont {Uri}}, \bibinfo {author}
  {\bibfnamefont {P.}~\bibnamefont {Stepanov}}, \bibinfo {author}
  {\bibfnamefont {G.}~\bibnamefont {Di~Battista}}, \bibinfo {author}
  {\bibfnamefont {I.}~\bibnamefont {Roy}}, \bibinfo {author} {\bibfnamefont
  {J.}~\bibnamefont {Xiao}}, \bibinfo {author} {\bibfnamefont {A.~Y.}\
  \bibnamefont {Meltzer}}, \bibinfo {author} {\bibfnamefont {Y.}~\bibnamefont
  {Myasoedov}}, \bibinfo {author} {\bibfnamefont {K.}~\bibnamefont {Pareek}},
  \bibinfo {author} {\bibfnamefont {K.}~\bibnamefont {Watanabe}}, \bibinfo
  {author} {\bibfnamefont {T.}~\bibnamefont {Taniguchi}}, \bibinfo {author}
  {\bibfnamefont {B.}~\bibnamefont {Yan}}, \bibinfo {author} {\bibfnamefont
  {A.}~\bibnamefont {Stern}}, \bibinfo {author} {\bibfnamefont
  {E.}~\bibnamefont {Berg}}, \bibinfo {author} {\bibfnamefont {D.~K.}\
  \bibnamefont {Efetov}},\ and\ \bibinfo {author} {\bibfnamefont
  {E.}~\bibnamefont {Zeldov}},\ }\href@noop {} {\bibfield  {journal} {\bibinfo
  {journal} {arXiv:2201.06901 [cond-mat]}\ } (\bibinfo {year} {2022})},\
  \Eprint {https://arxiv.org/abs/2201.06901} {arXiv:2201.06901 [cond-mat]}
  \BibitemShut {NoStop}%
\bibitem [{\citenamefont {Polshyn}\ \emph {et~al.}(2021)\citenamefont
  {Polshyn}, \citenamefont {Zhang}, \citenamefont {Kumar}, \citenamefont
  {Soejima}, \citenamefont {Ledwith}, \citenamefont {Watanabe}, \citenamefont
  {Taniguchi}, \citenamefont {Vishwanath}, \citenamefont {Zaletel},\ and\
  \citenamefont {Young}}]{polshynTopologicalChargeDensity2021}%
  \BibitemOpen
  \bibfield  {author} {\bibinfo {author} {\bibfnamefont {H.}~\bibnamefont
  {Polshyn}}, \bibinfo {author} {\bibfnamefont {Y.}~\bibnamefont {Zhang}},
  \bibinfo {author} {\bibfnamefont {M.~A.}\ \bibnamefont {Kumar}}, \bibinfo
  {author} {\bibfnamefont {T.}~\bibnamefont {Soejima}}, \bibinfo {author}
  {\bibfnamefont {P.}~\bibnamefont {Ledwith}}, \bibinfo {author} {\bibfnamefont
  {K.}~\bibnamefont {Watanabe}}, \bibinfo {author} {\bibfnamefont
  {T.}~\bibnamefont {Taniguchi}}, \bibinfo {author} {\bibfnamefont
  {A.}~\bibnamefont {Vishwanath}}, \bibinfo {author} {\bibfnamefont {M.~P.}\
  \bibnamefont {Zaletel}},\ and\ \bibinfo {author} {\bibfnamefont {A.~F.}\
  \bibnamefont {Young}},\ }\href@noop {} {\bibfield  {journal} {\bibinfo
  {journal} {arXiv:2104.01178 [cond-mat]}\ } (\bibinfo {year} {2021})},\
  \Eprint {https://arxiv.org/abs/2104.01178} {arXiv:2104.01178 [cond-mat]}
  \BibitemShut {NoStop}%
\bibitem [{\citenamefont {Yu}\ \emph {et~al.}(2021)\citenamefont {Yu},
  \citenamefont {Foutty}, \citenamefont {Han}, \citenamefont {Barber},
  \citenamefont {Schattner}, \citenamefont {Watanabe}, \citenamefont
  {Taniguchi}, \citenamefont {Phillips}, \citenamefont {Shen}, \citenamefont
  {Kivelson},\ and\ \citenamefont
  {Feldman}}]{yuCorrelatedHofstadterSpectrum2021}%
  \BibitemOpen
  \bibfield  {author} {\bibinfo {author} {\bibfnamefont {J.}~\bibnamefont
  {Yu}}, \bibinfo {author} {\bibfnamefont {B.~A.}\ \bibnamefont {Foutty}},
  \bibinfo {author} {\bibfnamefont {Z.}~\bibnamefont {Han}}, \bibinfo {author}
  {\bibfnamefont {M.~E.}\ \bibnamefont {Barber}}, \bibinfo {author}
  {\bibfnamefont {Y.}~\bibnamefont {Schattner}}, \bibinfo {author}
  {\bibfnamefont {K.}~\bibnamefont {Watanabe}}, \bibinfo {author}
  {\bibfnamefont {T.}~\bibnamefont {Taniguchi}}, \bibinfo {author}
  {\bibfnamefont {P.}~\bibnamefont {Phillips}}, \bibinfo {author}
  {\bibfnamefont {Z.-X.}\ \bibnamefont {Shen}}, \bibinfo {author}
  {\bibfnamefont {S.~A.}\ \bibnamefont {Kivelson}},\ and\ \bibinfo {author}
  {\bibfnamefont {B.~E.}\ \bibnamefont {Feldman}},\ }\href@noop {} {\bibfield
  {journal} {\bibinfo  {journal} {arXiv:2108.00009 [cond-mat]}\ } (\bibinfo
  {year} {2021})},\ \Eprint {https://arxiv.org/abs/2108.00009}
  {arXiv:2108.00009 [cond-mat]} \BibitemShut {NoStop}%
\bibitem [{\citenamefont {Nuckolls}\ \emph {et~al.}(2020)\citenamefont
  {Nuckolls}, \citenamefont {Oh}, \citenamefont {Wong}, \citenamefont {Lian},
  \citenamefont {Watanabe}, \citenamefont {Taniguchi}, \citenamefont
  {Bernevig},\ and\ \citenamefont
  {Yazdani}}]{nuckollsStronglyCorrelatedChern2020}%
  \BibitemOpen
  \bibfield  {author} {\bibinfo {author} {\bibfnamefont {K.~P.}\ \bibnamefont
  {Nuckolls}}, \bibinfo {author} {\bibfnamefont {M.}~\bibnamefont {Oh}},
  \bibinfo {author} {\bibfnamefont {D.}~\bibnamefont {Wong}}, \bibinfo {author}
  {\bibfnamefont {B.}~\bibnamefont {Lian}}, \bibinfo {author} {\bibfnamefont
  {K.}~\bibnamefont {Watanabe}}, \bibinfo {author} {\bibfnamefont
  {T.}~\bibnamefont {Taniguchi}}, \bibinfo {author} {\bibfnamefont {B.~A.}\
  \bibnamefont {Bernevig}},\ and\ \bibinfo {author} {\bibfnamefont
  {A.}~\bibnamefont {Yazdani}},\ }\href
  {https://doi.org/10.1038/s41586-020-3028-8} {\bibfield  {journal} {\bibinfo
  {journal} {Nature}\ }\textbf {\bibinfo {volume} {588}},\ \bibinfo {pages}
  {610} (\bibinfo {year} {2020})}\BibitemShut {NoStop}%
\bibitem [{\citenamefont {Choi}\ \emph
  {et~al.}(2021{\natexlab{a}})\citenamefont {Choi}, \citenamefont {Kim},
  \citenamefont {Peng}, \citenamefont {Thomson}, \citenamefont {Lewandowski},
  \citenamefont {Polski}, \citenamefont {Zhang}, \citenamefont {Arora},
  \citenamefont {Watanabe}, \citenamefont {Taniguchi}, \citenamefont {Alicea},\
  and\ \citenamefont
  {{Nadj-Perge}}}]{choiCorrelationdrivenTopologicalPhases2021}%
  \BibitemOpen
  \bibfield  {author} {\bibinfo {author} {\bibfnamefont {Y.}~\bibnamefont
  {Choi}}, \bibinfo {author} {\bibfnamefont {H.}~\bibnamefont {Kim}}, \bibinfo
  {author} {\bibfnamefont {Y.}~\bibnamefont {Peng}}, \bibinfo {author}
  {\bibfnamefont {A.}~\bibnamefont {Thomson}}, \bibinfo {author} {\bibfnamefont
  {C.}~\bibnamefont {Lewandowski}}, \bibinfo {author} {\bibfnamefont
  {R.}~\bibnamefont {Polski}}, \bibinfo {author} {\bibfnamefont
  {Y.}~\bibnamefont {Zhang}}, \bibinfo {author} {\bibfnamefont {H.~S.}\
  \bibnamefont {Arora}}, \bibinfo {author} {\bibfnamefont {K.}~\bibnamefont
  {Watanabe}}, \bibinfo {author} {\bibfnamefont {T.}~\bibnamefont {Taniguchi}},
  \bibinfo {author} {\bibfnamefont {J.}~\bibnamefont {Alicea}},\ and\ \bibinfo
  {author} {\bibfnamefont {S.}~\bibnamefont {{Nadj-Perge}}},\ }\href
  {https://doi.org/10.1038/s41586-020-03159-7} {\bibfield  {journal} {\bibinfo
  {journal} {Nature}\ }\textbf {\bibinfo {volume} {589}},\ \bibinfo {pages}
  {536} (\bibinfo {year} {2021}{\natexlab{a}})}\BibitemShut {NoStop}%
\bibitem [{\citenamefont {Choi}\ \emph
  {et~al.}(2021{\natexlab{b}})\citenamefont {Choi}, \citenamefont {Kim},
  \citenamefont {Lewandowski}, \citenamefont {Peng}, \citenamefont {Thomson},
  \citenamefont {Polski}, \citenamefont {Zhang}, \citenamefont {Watanabe},
  \citenamefont {Taniguchi}, \citenamefont {Alicea},\ and\ \citenamefont
  {{Nadj-Perge}}}]{choiInteractiondrivenBandFlattening2021}%
  \BibitemOpen
  \bibfield  {author} {\bibinfo {author} {\bibfnamefont {Y.}~\bibnamefont
  {Choi}}, \bibinfo {author} {\bibfnamefont {H.}~\bibnamefont {Kim}}, \bibinfo
  {author} {\bibfnamefont {C.}~\bibnamefont {Lewandowski}}, \bibinfo {author}
  {\bibfnamefont {Y.}~\bibnamefont {Peng}}, \bibinfo {author} {\bibfnamefont
  {A.}~\bibnamefont {Thomson}}, \bibinfo {author} {\bibfnamefont
  {R.}~\bibnamefont {Polski}}, \bibinfo {author} {\bibfnamefont
  {Y.}~\bibnamefont {Zhang}}, \bibinfo {author} {\bibfnamefont
  {K.}~\bibnamefont {Watanabe}}, \bibinfo {author} {\bibfnamefont
  {T.}~\bibnamefont {Taniguchi}}, \bibinfo {author} {\bibfnamefont
  {J.}~\bibnamefont {Alicea}},\ and\ \bibinfo {author} {\bibfnamefont
  {S.}~\bibnamefont {{Nadj-Perge}}},\ }\href
  {https://doi.org/10.1038/s41567-021-01359-0} {\bibfield  {journal} {\bibinfo
  {journal} {Nature Physics}\ }\textbf {\bibinfo {volume} {17}},\ \bibinfo
  {pages} {1375} (\bibinfo {year} {2021}{\natexlab{b}})}\BibitemShut {NoStop}%
\bibitem [{\citenamefont {Po}\ \emph {et~al.}(2019)\citenamefont {Po},
  \citenamefont {Zou}, \citenamefont {Senthil},\ and\ \citenamefont
  {Vishwanath}}]{poFaithfulTightbindingModels2019}%
  \BibitemOpen
  \bibfield  {author} {\bibinfo {author} {\bibfnamefont {H.~C.}\ \bibnamefont
  {Po}}, \bibinfo {author} {\bibfnamefont {L.}~\bibnamefont {Zou}}, \bibinfo
  {author} {\bibfnamefont {T.}~\bibnamefont {Senthil}},\ and\ \bibinfo {author}
  {\bibfnamefont {A.}~\bibnamefont {Vishwanath}},\ }\href
  {https://doi.org/10.1103/PhysRevB.99.195455} {\bibfield  {journal} {\bibinfo
  {journal} {Physical Review B}\ }\textbf {\bibinfo {volume} {99}},\ \bibinfo
  {pages} {195455} (\bibinfo {year} {2019})}\BibitemShut {NoStop}%
\bibitem [{\citenamefont {Choi}\ \emph {et~al.}(2019)\citenamefont {Choi},
  \citenamefont {Kemmer}, \citenamefont {Peng}, \citenamefont {Thomson},
  \citenamefont {Arora}, \citenamefont {Polski}, \citenamefont {Zhang},
  \citenamefont {Ren}, \citenamefont {Alicea}, \citenamefont {Refael},
  \citenamefont {{von Oppen}}, \citenamefont {Watanabe}, \citenamefont
  {Taniguchi},\ and\ \citenamefont
  {{Nadj-Perge}}}]{choiElectronicCorrelationsTwisted2019}%
  \BibitemOpen
  \bibfield  {author} {\bibinfo {author} {\bibfnamefont {Y.}~\bibnamefont
  {Choi}}, \bibinfo {author} {\bibfnamefont {J.}~\bibnamefont {Kemmer}},
  \bibinfo {author} {\bibfnamefont {Y.}~\bibnamefont {Peng}}, \bibinfo {author}
  {\bibfnamefont {A.}~\bibnamefont {Thomson}}, \bibinfo {author} {\bibfnamefont
  {H.}~\bibnamefont {Arora}}, \bibinfo {author} {\bibfnamefont
  {R.}~\bibnamefont {Polski}}, \bibinfo {author} {\bibfnamefont
  {Y.}~\bibnamefont {Zhang}}, \bibinfo {author} {\bibfnamefont
  {H.}~\bibnamefont {Ren}}, \bibinfo {author} {\bibfnamefont {J.}~\bibnamefont
  {Alicea}}, \bibinfo {author} {\bibfnamefont {G.}~\bibnamefont {Refael}},
  \bibinfo {author} {\bibfnamefont {F.}~\bibnamefont {{von Oppen}}}, \bibinfo
  {author} {\bibfnamefont {K.}~\bibnamefont {Watanabe}}, \bibinfo {author}
  {\bibfnamefont {T.}~\bibnamefont {Taniguchi}},\ and\ \bibinfo {author}
  {\bibfnamefont {S.}~\bibnamefont {{Nadj-Perge}}},\ }\href
  {https://doi.org/10.1038/s41567-019-0606-5} {\bibfield  {journal} {\bibinfo
  {journal} {Nature Physics}\ }\textbf {\bibinfo {volume} {15}},\ \bibinfo
  {pages} {1174} (\bibinfo {year} {2019})}\BibitemShut {NoStop}%
\bibitem [{\citenamefont {Cea}\ and\ \citenamefont
  {Guinea}(2020)}]{ceaBandStructureInsulating2020}%
  \BibitemOpen
  \bibfield  {author} {\bibinfo {author} {\bibfnamefont {T.}~\bibnamefont
  {Cea}}\ and\ \bibinfo {author} {\bibfnamefont {F.}~\bibnamefont {Guinea}},\
  }\href {https://doi.org/10.1103/PhysRevB.102.045107} {\bibfield  {journal}
  {\bibinfo  {journal} {Physical Review B}\ }\textbf {\bibinfo {volume}
  {102}},\ \bibinfo {pages} {045107} (\bibinfo {year} {2020})}\BibitemShut
  {NoStop}%
\bibitem [{\citenamefont {Kwan}\ \emph {et~al.}(2021)\citenamefont {Kwan},
  \citenamefont {Wagner}, \citenamefont {Soejima}, \citenamefont {Zaletel},
  \citenamefont {Simon}, \citenamefont {Parameswaran},\ and\ \citenamefont
  {Bultinck}}]{kwanKekulSpiralOrder2021}%
  \BibitemOpen
  \bibfield  {author} {\bibinfo {author} {\bibfnamefont {Y.~H.}\ \bibnamefont
  {Kwan}}, \bibinfo {author} {\bibfnamefont {G.}~\bibnamefont {Wagner}},
  \bibinfo {author} {\bibfnamefont {T.}~\bibnamefont {Soejima}}, \bibinfo
  {author} {\bibfnamefont {M.~P.}\ \bibnamefont {Zaletel}}, \bibinfo {author}
  {\bibfnamefont {S.~H.}\ \bibnamefont {Simon}}, \bibinfo {author}
  {\bibfnamefont {S.~A.}\ \bibnamefont {Parameswaran}},\ and\ \bibinfo {author}
  {\bibfnamefont {N.}~\bibnamefont {Bultinck}},\ }\href@noop {} {\bibfield
  {journal} {\bibinfo  {journal} {arXiv:2105.05857 [cond-mat]}\ } (\bibinfo
  {year} {2021})},\ \Eprint {https://arxiv.org/abs/2105.05857}
  {arXiv:2105.05857 [cond-mat]} \BibitemShut {NoStop}%
\bibitem [{\citenamefont {Shavit}\ \emph {et~al.}(2021)\citenamefont {Shavit},
  \citenamefont {Berg}, \citenamefont {Stern},\ and\ \citenamefont
  {Oreg}}]{shavitTheoryCorrelatedInsulators2021}%
  \BibitemOpen
  \bibfield  {author} {\bibinfo {author} {\bibfnamefont {G.}~\bibnamefont
  {Shavit}}, \bibinfo {author} {\bibfnamefont {E.}~\bibnamefont {Berg}},
  \bibinfo {author} {\bibfnamefont {A.}~\bibnamefont {Stern}},\ and\ \bibinfo
  {author} {\bibfnamefont {Y.}~\bibnamefont {Oreg}},\ }\href@noop {} {\bibfield
   {journal} {\bibinfo  {journal} {arXiv:2107.08486 [cond-mat]}\ } (\bibinfo
  {year} {2021})},\ \Eprint {https://arxiv.org/abs/2107.08486}
  {arXiv:2107.08486 [cond-mat]} \BibitemShut {NoStop}%
\bibitem [{\citenamefont {Tseng}\ \emph {et~al.}(2022)\citenamefont {Tseng},
  \citenamefont {Ma}, \citenamefont {Liu}, \citenamefont {Watanabe},
  \citenamefont {Taniguchi}, \citenamefont {Chu},\ and\ \citenamefont
  {Yankowitz}}]{tsengAnomalousHallEffect2022}%
  \BibitemOpen
  \bibfield  {author} {\bibinfo {author} {\bibfnamefont {C.-C.}\ \bibnamefont
  {Tseng}}, \bibinfo {author} {\bibfnamefont {X.}~\bibnamefont {Ma}}, \bibinfo
  {author} {\bibfnamefont {Z.}~\bibnamefont {Liu}}, \bibinfo {author}
  {\bibfnamefont {K.}~\bibnamefont {Watanabe}}, \bibinfo {author}
  {\bibfnamefont {T.}~\bibnamefont {Taniguchi}}, \bibinfo {author}
  {\bibfnamefont {J.-H.}\ \bibnamefont {Chu}},\ and\ \bibinfo {author}
  {\bibfnamefont {M.}~\bibnamefont {Yankowitz}},\ }\href@noop {} {\bibfield
  {journal} {\bibinfo  {journal} {arXiv:2202.01734 [cond-mat]}\ } (\bibinfo
  {year} {2022})},\ \Eprint {https://arxiv.org/abs/2202.01734}
  {arXiv:2202.01734 [cond-mat]} \BibitemShut {NoStop}%
\bibitem [{\citenamefont {Lin}\ \emph {et~al.}(2022)\citenamefont {Lin},
  \citenamefont {Zhang}, \citenamefont {Morissette}, \citenamefont {Wang},
  \citenamefont {Liu}, \citenamefont {Rhodes}, \citenamefont {Watanabe},
  \citenamefont {Taniguchi}, \citenamefont {Hone},\ and\ \citenamefont
  {Li}}]{linSpinorbitDrivenFerromagnetism2022}%
  \BibitemOpen
  \bibfield  {author} {\bibinfo {author} {\bibfnamefont {J.-X.}\ \bibnamefont
  {Lin}}, \bibinfo {author} {\bibfnamefont {Y.-H.}\ \bibnamefont {Zhang}},
  \bibinfo {author} {\bibfnamefont {E.}~\bibnamefont {Morissette}}, \bibinfo
  {author} {\bibfnamefont {Z.}~\bibnamefont {Wang}}, \bibinfo {author}
  {\bibfnamefont {S.}~\bibnamefont {Liu}}, \bibinfo {author} {\bibfnamefont
  {D.}~\bibnamefont {Rhodes}}, \bibinfo {author} {\bibfnamefont
  {K.}~\bibnamefont {Watanabe}}, \bibinfo {author} {\bibfnamefont
  {T.}~\bibnamefont {Taniguchi}}, \bibinfo {author} {\bibfnamefont
  {J.}~\bibnamefont {Hone}},\ and\ \bibinfo {author} {\bibfnamefont {J.~I.~A.}\
  \bibnamefont {Li}},\ }\href {https://doi.org/10.1126/science.abh2889}
  {\bibfield  {journal} {\bibinfo  {journal} {Science}\ }\textbf {\bibinfo
  {volume} {375}},\ \bibinfo {pages} {437} (\bibinfo {year}
  {2022})}\BibitemShut {NoStop}%
\bibitem [{\citenamefont {Nam}\ and\ \citenamefont
  {Koshino}(2017)}]{namLatticeRelaxationEnergy2017}%
  \BibitemOpen
  \bibfield  {author} {\bibinfo {author} {\bibfnamefont {N.~N.~T.}\
  \bibnamefont {Nam}}\ and\ \bibinfo {author} {\bibfnamefont {M.}~\bibnamefont
  {Koshino}},\ }\href {https://doi.org/10.1103/PhysRevB.96.075311} {\bibfield
  {journal} {\bibinfo  {journal} {Physical Review B}\ }\textbf {\bibinfo
  {volume} {96}},\ \bibinfo {pages} {075311} (\bibinfo {year}
  {2017})}\BibitemShut {NoStop}%
\bibitem [{\citenamefont {Uri}\ \emph {et~al.}(2020)\citenamefont {Uri},
  \citenamefont {Grover}, \citenamefont {Cao}, \citenamefont {Crosse},
  \citenamefont {Bagani}, \citenamefont {{Rodan-Legrain}}, \citenamefont
  {Myasoedov}, \citenamefont {Watanabe}, \citenamefont {Taniguchi},
  \citenamefont {Moon}, \citenamefont {Koshino}, \citenamefont
  {{Jarillo-Herrero}},\ and\ \citenamefont
  {Zeldov}}]{uriMappingTwistangleDisorder2020}%
  \BibitemOpen
  \bibfield  {author} {\bibinfo {author} {\bibfnamefont {A.}~\bibnamefont
  {Uri}}, \bibinfo {author} {\bibfnamefont {S.}~\bibnamefont {Grover}},
  \bibinfo {author} {\bibfnamefont {Y.}~\bibnamefont {Cao}}, \bibinfo {author}
  {\bibfnamefont {J.~A.}\ \bibnamefont {Crosse}}, \bibinfo {author}
  {\bibfnamefont {K.}~\bibnamefont {Bagani}}, \bibinfo {author} {\bibfnamefont
  {D.}~\bibnamefont {{Rodan-Legrain}}}, \bibinfo {author} {\bibfnamefont
  {Y.}~\bibnamefont {Myasoedov}}, \bibinfo {author} {\bibfnamefont
  {K.}~\bibnamefont {Watanabe}}, \bibinfo {author} {\bibfnamefont
  {T.}~\bibnamefont {Taniguchi}}, \bibinfo {author} {\bibfnamefont
  {P.}~\bibnamefont {Moon}}, \bibinfo {author} {\bibfnamefont {M.}~\bibnamefont
  {Koshino}}, \bibinfo {author} {\bibfnamefont {P.}~\bibnamefont
  {{Jarillo-Herrero}}},\ and\ \bibinfo {author} {\bibfnamefont
  {E.}~\bibnamefont {Zeldov}},\ }\href
  {https://doi.org/10.1038/s41586-020-2255-3} {\bibfield  {journal} {\bibinfo
  {journal} {Nature}\ }\textbf {\bibinfo {volume} {581}},\ \bibinfo {pages}
  {47} (\bibinfo {year} {2020})}\BibitemShut {NoStop}%
\end{thebibliography}
\end{document}